\documentclass[preprint,12pt,sort&compress]{elsarticle}

\usepackage{epsfig}

\usepackage{amssymb}
\usepackage[utf8]{inputenc}

\journal{Progress in Particle and Nuclear Physics}

\newcommand{\bc}{\begin{center}}
\newcommand{\ec}{\end{center}}
\newcommand{\be}{\begin{equation}}
\newcommand{\ee}{\end{equation}}
\newcommand{\bea}{\begin{eqnarray}}
\newcommand{\eea}{\end{eqnarray}}
\newcommand{\ba}{\begin{array}}
\newcommand{\ea}{\end{array}}
\newcommand{\lb}{\label}
\newcommand{\rf}{\ref}
\newcommand{\bfg}{\begin{figure}[htbp]}
\newcommand{\efg}{\end{figure}}

\begin{document}

\begin{frontmatter}



\title{Tetraquarks in large-$N_c^{}$ QCD}


\author[1]{Wolfgang Lucha}\ead{wolfgang.lucha@oeaw.ac.at}

\author[2,3,4]{Dmitri Melikhov}\ead{dmitri\_melikhov@gmx.de}

\author[5]{Hagop Sazdjian\corref{cor1}}\ead{sazdjian@ijclab.in2p3.fr}

\cortext[cor1]{Corresponding author}

\address[1]{Institute for High Energy Physics, Austrian Academy of 
Sciences, Nikolsdorfergasse 18, A-1050 Vienna, Austria}

\address[2]{D.~V.~Skobeltsyn Institute of Nuclear Physics,
M.~V.~Lomonosov Moscow State University, 119991 Moscow, Russia}

\address[3]{Joint Institute for Nuclear Research, 141980 Dubna,
Russia}

\address[4]{Faculty of Physics, University of Vienna,
Boltzmanngasse 5, A-1090 Vienna, Austria}

\address[5]{Universit\'e Paris-Saclay, CNRS/IN2P3, IJCLab, 91405
Orsay, France}

\begin{abstract}
The generalization of the color gauge group SU(3) to SU$(N_c)$,
with $N_c$ taking arbitrarily large values, as had been proposed
and developed by 't Hooft, has allowed for a decisive progress in
the understanding of many qualitative, as well as quantitative,
aspects of QCD in its nonperturbative regime. In particular, the
notion of valence quarks receives there a precise meaning.
The present work reviews the various aspects of the extension of
this approach to the case of tetraquark states, which are a category
of the general class of exotic states, also called multiquark states,
whose internal valence-quark structure does not match with that of
ordinary hadrons, and which have received, in recent years, many
experimental confirmations.
The primary question of describing, or probing, on theoretical grounds,
multiquark states is first examined. The signature of such states
inside Feynman diagrams in relation with their singularities is
highlighted. The main mechanisms of formation of tetraquark states,
provided by the diquark model and the molecular scheme, are considered
together with their specific implications.
The properties of tetraquark states at large $N_c$ are
analyzed through the Feynman diagrams that describe two-meson
scattering amplitudes. It turns out that, in that limit, the possible
formation of tetraquark states is mainly due to the mutual interactions
of their internal mesonic clusters. These essentially
arise from the quark-rearrangement, or quark-interchange, mechanism.
In coupled-channel meson-meson scattering amplitudes, one may expect
the occurrence of two independent tetraquark states, each having
priviledged couplings with the two mesons of their dominant channel.
The question of the energy balance of various schemes in the static
limit is also analyzed. The clarification of the mechanisms that are
at work in the formation of tetraquarks is the main outcome from the
large-$N_c$ approach to this problem.
\par

\end{abstract}



\end{frontmatter}


\tableofcontents


\section{Introduction} \lb{s1}

The possibility of the existence of exotic hadrons, containing more
valence quarks and antiquarks than a quark-antiquark pair for mesons
and three quarks for baryons, had been considered since the early
days of the quark model \cite{GellMann:1964nj,Zweig:1964jf}.
The lack of experimental data about these hypothetical objects during
the following decades has pushed them for a long time into a marginal 
situation. However, with the advent of Quantum Chromodynamics (QCD)
as a theory of strong interactions, with the fundamental properties
of asymptotic freedom and confinement
\cite{Fritzsch:1973pi,Gross:1973id,Politzer:1973fx}, the road was
open for detailed theoretical investigations on the subject
\cite{Jaffe:1976ig,Jaffe:1976ih}.
\par
Progress has been achieved for the last two decades when
several experiments have been able to detect, with sufficient
precision, many candidates to represent exotic hadrons, the
latter not being matched with the quantum numbers or the
constituent content of the usual quark model
\cite{Choi:2003ue,Aubert:2003fg,Besson:2003cp,Aubert:2005rm,
Ablikim:2013mio,Liu:2013dau,Ablikim:2013wzq,Aaij:2014jqa,Aaij:2015tga,
Aaij:2020fnh,Aaij:2020ypa,Wu:2020hmk}.
These discoveries stimulated in turn vast theoretical investigations
in order to understand and interpret the detailed dynamics that give
rise to their existence. General accounts of them can be found in
recent review articles
\cite{Chen:2016qju,Hosaka:2016pey,Lebed:2016hpi,Esposito:2016noz,
Ali:2017jda,Guo:2017jvc,Olsen:2017bmm,Karliner:2017qhf,
Albuquerque:2018jkn,Liu:2019zoy,Ali:2019roi,Brambilla:2019esw}.
\par
While in the framework of QCD theory one explains in a satisfactory
way the spectrum and transition properties of ordinary hadrons
\cite{Chodos:1974je,Chodos:1974pn,DeGrand:1975cf,DeRujula:1975qlm,
Appelquist:1975ya,Eichten:1978tg,Eichten:1979ms,Isgur:1978xj,
Godfrey:1985xj,LeYaouanc:1984ntu,Lucha:1991vn}, the problem of exotic
hadrons comparatively encounters additional difficulties.
In QCD, which is a non-Abelian gauge theory, observable quantities
must be color gauge invariant. This naturally explains why quarks and
gluons, the building blocks of the theory, are not individually
observed in nature; only color-singlet objects, actually represented
by the ordinary hadrons, which are bound states of quarks and gluons,
are detected as free particles. This property is commonly depicted
by the term ``confinement of quarks and gluons''. It should be
emphasized, however, that the term confinement has here a stronger
meaning than the terms ``color screening'', also used in the
literature. The bound states of quarks and gluons do not resemble
the positronium- or hydrogen-like bound states of Quantum
Electrodynamics (QED). Hadrons, which, in principle, are infinite in
number, have masses squared that increase linearly with respect to
their spin, lying along ``Regge trajectories''
\cite{Regge:1959mz,Regge:1960zc,Veneziano:1968yb,Eden:1971jm,
Gribov:2009zz}. This is due to the
fact that the potential energy of the binding forces increases with
respect to the mutual distances between quarks and/or gluons
\cite{Wilson:1974sk,Brown:1979ya}. However, the confining forces are
only operative between colored objects or within color-singlet
objects. There are no confining or long-range van der Waals forces
that would operate between hadrons. The latter mutually interact by
means of short-range forces, represented by contact terms or
meson-exchange terms\footnote{The terms ``long-range forces'' and
``short-range forces'' are often used with different meanings in
the literature. At the level of quark and gluon interactions,
the long-range forces are represented by the confining forces
and their residual van der Waals forces, while the short-range
forces are those that operate by means of hadron exchanges or by
contact interactions between hadrons. In the effective theories
of hadrons, long-range forces are those that operate by meson
exchanges (Yukawa-type forces), mostly by pion exchanges, while
short-range forces are those resulting from contact interactions.
We use, in the present work, the convention related to the
quark and gluon interactions.}.
\par
The color gauge invariance principle should also apply to the
constitution of exotic hadrons, also called ``multiquark states''.
However, one realizes here that color-singlet multiquark states
can easily be generated by products or combinations of products
of free or interacting ordinary hadron states \cite{Jaffe:2008zz}.
Since ordinary hadrons mutually interact by means of short-range
forces (cf. footnote above), one should not find in that case
spectra similar to that
of the confining interactions found earlier in QCD. Rather, one
would have loosely bound states, or corresponding resonances,
typical of molecular or nuclear states, whose illustrated
representative is the deuteron
\cite{Weinberg:1965zz,Voloshin:1976ap,DeRujula:1976zlg,
Tornqvist:1993ng}.
\par
The existence of molecular-type exotic hadrons does not 
\textit{a priori} exclude the formation of exotic hadrons by means
of confining forces acting directly on the quarks and gluons.
In such a case, one would expect to obtain bound states of compact
size, due to the confining nature of all the operative forces.
Such states are also designated as ``compact exotic hadrons'',
in contradistinction to the ``molecular states''.
\par
The eventual existence of compact multiquark states would implicitly
mean that they are color or cluster irreducible, in the sense that
they cannot be decomposed as combinations of products of simpler color
gauge invariant states, which are the usual hadronic states.
This basic property does not seem, however, realizable. When the
multiquark generating operators or sources are expressed in local
form, they can always be reexpressed, by means of Fierz
rearrangements, as combinations of products of ordinary hadronic
currents \cite{Jaffe:2008zz}. More generally, even when the multiquark
operators take a multilocal form with the aid of gauge links,
the color or cluster reducibility phenomenon continues to occur
\cite{Lucha:2019cdc}. This means that, on formal grounds, compact
multiquark states are not the most favored outcome of the exotic
hadron construction. To ensure their existence on theoretical
grounds, one should be able to find underlying dynamical mechanisms
that ensure their stability against the natural forces of
dislocation, represented by the formation of internal hadronic
clusters.
\par
The primary mechanism that might operate for the formation of
compact multiquark states is the ``diquark'' one
\cite{Jaffe:1976ih,Anselmino:1992vg}, which has been first
utilized in the study of baryon spectroscopy
\cite{Lichtenberg:1982jp,Fleck:1988vm,Schafer:1993ra}. It is based on
the observation that in SU(3), the color gauge group underlying QCD,
the fundamental representation to which the quarks are belonging is
the triplet $\mathbf{3}$; therefore, two quarks belong either to the
antitriplet representation $\mathbf{\overline 3}$, which is
antisymmetric, or to the sextet representation $\mathbf{6}$, which is
symmetric. The forces that act between the quarks are generally
attractive in the $\mathbf{\overline 3}$ representation and will have
the tendancy to form bound states of diquarks, which, in the case of
compact sizes, will behave as nearly pointlike antiquarks, thus in
turn being attracted by a third quark.
\par
The idea of the diquark mechanism has naturally been extended to the
case of exotic hadrons \cite{Jaffe:2003sg,Shuryak:2003zi,Maiani:2004vq,
Maiani:2005pe}, giving rise to detailed investigations
\cite{Ebert:2007rn,Ebert:2010af,Maiani:2015vwa,Wu:2016vtq,Lebed:2017min,
Ali:2017wsf,Ali:2019clg,Giron:2020wpx,Faustov:2020qfm}.
This model, associated with approximate flavor symmetry and flavor
independence of the confining forces, generally predicts, in the case
of existence of exotic hadrons, several flavor multiplets of such states.
\par
On the other hand, the effective forces that operate for the formation
of molecular-type states are more dependent on the quark flavors and
hence may not predict as many flavor multiplets as the diquark
mechanism. One of the salient features of the molecular picture is
its tendancy to predict bound states lying near the two-hadron
threshold.
\par
The difficulties encountered in clearly predicting the domains or
conditions of existence of each category of exotic hadrons, compact
or molecular, are intimately related to the fact that QCD theory is
characterized by having a nonperturbative regime at large distances,
which is not yet analytically solved.
Efficient tools are provided through the recognition of approximate
symmetries and the use of related effective field theories, among
which one may quote chiral perturbation theory
\cite{Weinberg:1978kz,Gasser:1983yg,Gasser:1984gg,Manohar:1996cq},
heavy-quark effective theories
\cite{Georgi:1990um,Isgur:1991wq,Wise:1992hn,Neubert:1993mb,
Casalbuoni:1996pg} and effective theories involving nucleons
\cite{Gasser:1987rb,Weinberg:1991um,Ordonez:1995rz,Kaplan:1996xu,
Epelbaum:1998ka,Epelbaum:1999dj,Bedaque:2002mn,Hammer:2019poc}.
Nonrelativistic effective field theories, aimed at studying
bound-state properties, have been first developed in Ref.
\cite{Caswell:1985ui} and later completed under the name of
``potential nonrelativistic QCD'' in Ref. \cite{Brambilla:2004jw}.
Other analytic approaches are forced to hinge on simplifying models,
which in turn are subjected to theoretical debates.
\par
For the time being, lattice theory remains one of the most powerful
tools for the solution of QCD in its nonperturbative
regime. It is essentially based on a numerical approach, discretizing
the continuum of spacetime over a euclidean finite-volume lattice
\cite{Wilson:1974sk,Kogut:1982ds}. Lattice theory has also made
decisive progress in the calculation of scattering amplitudes
\cite{Luscher:1990ux,Briceno:2017max}. Concerning the exotic
hadrons, it has already provided many results, mainly
in sectors containing two heavy quarks
\cite{Prelovsek:2013cra,Prelovsek:2014swa,Ikeda:2013vwa,Ikeda:2016zwx,
Padmanath:2015era,Bicudo:2015vta,Bicudo:2017usw,Bicudo:2017szl,
Francis:2016hui,Francis:2018jyb,Cheung:2017tnt,Hughes:2017xie,
Junnarkar:2018twb,Liu:2012zya,Lang:2014yfa}.  
However, they are not all conclusive concerning the existence of
tetraquarks. No evidence seems to be found, in general, in sectors
with hidden or doubly-charm flavors \cite{Prelovsek:2014swa,
Ikeda:2013vwa,Ikeda:2016zwx,Padmanath:2015era,Bicudo:2015vta,
Cheung:2017tnt}, except in \cite{Prelovsek:2013cra,Junnarkar:2018twb}.
Open charm states are found in \cite{Liu:2012zya,Lang:2014yfa}.
The analysis of $bb\bar b\bar b$ systems does not seem conclusive
\cite{Bicudo:2017usw,Hughes:2017xie}. Evidence for deeply-bound
states in doubly-bottom and open bottom-charm sectors with light
quarks is found in \cite{Francis:2016hui,Francis:2018jyb,
Junnarkar:2018twb,Bicudo:2015vta,Bicudo:2017szl}.
(Other considerations for lattice theory calculations can be found
in \cite{Guo:2013nja}.)
\par  
Although the analyses of the internal structure of tetraquark
candidates, found in lattice theory calculations, are not yet
complete, a first-level information may be obtained from the
overlap coefficients between the energy levels and the interpolating
operators that are implemented. The fact that there is little
overlap with tetraquark-like operators might suggest a larger size
for the state. More precise information about the size of the
observed states could be obtained from the calculation of appropriate
form factors.
\par
Among the approaches for the analysis of the nonperturbative
properties of QCD, the large-$N_c$ limiting method, originally
introduced by 't Hooft
\cite{'tHooft:1973jz,'tHooft:1974hx,Callan:1975ps},
has been proven to be one of the most fruitful ones. It consists
in generalizing the color gauge group SU(3) of QCD to the more
general case of SU($N_c$), with $N_c$ considered as a parameter.
It turns out that the large-$N_c$ limit of the theory, associated
with a scaling of the coupling constant as $g=O(1/\sqrt{N_c})$, has
more simplifying features than in the finite-$N_c$ case.
The color-singlet parts of Feynman diagrams can then be classified
according to their topological properties: it is the ``planar''
diagrams that are the dominant ones, while the other types of diagram
can systematically be classified, according to their more complicated
topology, within a perturbative expansion in $1/N_c$, as providing
nonleading contributions. 
\par
This approach does not solve the theory, but allows, assuming that
the large-$N_c$ limit is a smooth one, a better understanding of
some of its salient features. In this limit, the spectrum of the
theory is essentially made of an infinite tower of free stable mesons,
their mutual interactions occurring through nonleading effects in
$1/N_c$ \cite{Witten:1979kh,Witten:1979pi}.
This clearly shows that, in the hadronic world, the $\rho$-meson,
for instance, is as elementary as the pion and could not be
considered as a composite object of two pions
\cite{Jaffe:2008zz}. Another outcome is a natural
explanation of the OZI (Okubo, Zweig, Iizuka) rule
\cite{Zweig:1964jf,Okubo:1963fa,Iizuka:1966fk},
which asserts that leading strong interaction processes involving
hadrons are those that have nonzero connecting quark lines
between the initial and final states. Processes not satisfying this
rule occur in nondominant orders of the $1/N_c$ expansion and are
naturally subleading \cite{Witten:1979kh,Witten:1979pi}.
The large-$N_c^{}$ limit brings also a natural explanation of the
absence, at leading order, of the quark-antiquark sea inside hadrons
and a theoretical support to Regge phenomenology
\cite{Regge:1959mz,Regge:1960zc,Veneziano:1968yb,Eden:1971jm,
Gribov:2009zz}, 
in which, in first approximation, hadronic processes are well
described  by tree diagrams with hadron exchanges
\cite{Veneziano:1968yb}. 
\par
The large-$N_c$ approach to QCD has received much attention during
the past decades in many phenomenological calculations related to 
hadronic physics. Its main virtue is to provide a theoretical
basis for qualitative simplifications and for the understanding
of the data.
\par
The large-$N_c^{}$ limiting procedure has also been, over the
last two decades, a decisive tool for new investigations in
the search for possibly existing duality relations between
gauge and string theories \cite{Maldacena:1997re,Maldacena:1998im,
Witten:1998qj,Witten:1998zw,Gubser:1998bc,Aharony:1999ti,
Maldacena:1999fi}. 
\par
The purpose of the present article is to present a
review of the main properties of the large-$N_c$ approach, with
emphasis put on its applications to exotic hadrons. Rather than 
focusing on particular candidates or particular data, our aim is
to introduce the general aspects of the method, which might be
applicable to a wide variety of situations.
\par
The large-$N_c$ analysis plays a crucial role in the recognition
of those QCD Feynman diagrams that might contribute to the formation
of exotic hadrons. It is in the large-$N_c$ limit that the counting
of the quark content of a hadronic state takes a systematic
mathematical meaning, associated with the order of expansion with
respect to $1/N_c$.
However, a straightforward transposition to the case of
exotic hadrons of results known from the large-$N_c$ approach to
ordinary hadrons might lead, in some cases, to wrong predictions.
It is here that the analysis of the singularities of Feynman diagrams
with respect to the quark content becomes of primary importance.
This is usually done with the help of the Landau equations
\cite{Landau:1959fi,Eden:1966dnq,Itzykson:1980rh}.
\par
In summary, the large-$N_c$ approach may be considered as one of the
basic tools for a systematic investigation of the nonperturbative
regime of QCD, with the objective of gaining complementary
information with respect to other well-established approaches.
\par
The paper is organized as follows. In Sec.~\rf{s2}, we present the
general aspects of the large-$N_c$ approach. Section \rf{s3} is
devoted to a review of the various descriptions of exotic states
by means of interpolating currents or multilocal operators.
In Sec.~\rf{s4}, emphasis is put on the singularities of Feynman
diagrams and the Landau equations for the recognition of a possible
presence of tetraquark states. Section \rf{s5} studies the properties
of tetraquarks through the Feynman diagrams of meson-meson scattering
amplitudes in terms of quark and gluon lines. Various cases of quark
flavor contents are considered. In Sec.~\rf{s6}, some salient
features of the molecular scheme, in relation with effective theories,
are reviewed. In Sec.~\rf{s7}, the question of the reducibility of
multiquark operators is considered and the energy balance of various
configurations is studied in the static limit. The notion of
geometric partitioning is introduced. A summary and concluding
remarks follow in Sec.~\rf{s8}.
\par

\section{Large-$N_c$ limit} \lb{s2}

\subsection{General aspects} \lb{s21}

Quantum Chromodynamics is a non-Abelian gauge theory with the
color gauge group SU(3), with three quark fields $\psi^a$ ($a=1,2,3$),
belonging to the fundamental representation $\mathbf{3}$, three
antiquark fields $\overline{\psi}_b$ ($b=1,2,3$), belonging to the
antifundamental representation $\mathbf{\bar{3}}$ and eight gluon
fields $A_{\mu}^B$ ($B=1,\ldots,8$), belonging to the adjoint
representation
$\mathbf{8}$ \cite{Fritzsch:1973pi,Gross:1973id,Politzer:1973fx}.
The primary, CP conserving, Lagrangian density, written in matrix
form in color space and with $N_f$ different quark flavors, reads
\be \lb{2e1}
\mathcal{L}=-\frac{1}{2}\mathrm{tr}_cF_{\mu\nu}F^{\mu\nu}+
\sum_{j=1}^{N_f}\overline{\psi_j}(iD_{\mu}\gamma^{\mu}-m_j)\psi_j,
\ee
where $D$ is the covariant derivative,
$D_{\mu}=1\partial_{\mu}-igT^BA_{\mu}^B$,
$F$ is the field strength,
$F_{\mu\nu}=(i/g)[D_{\mu},D_{\nu}]=T^BF_{\mu\nu}^B$,
$g$ is the coupling constant
and $T^B$ are the generators of the gauge group in the fundamental
representation, with normalization
$\mathrm{tr}(T^AT^B)=(1/2)\delta^{AB}$.
The quantization of the theory requires the introduction into the
previous Lagrangian density of a gauge-fixing term together with
a part containing auxiliary anticommuting scalar fields, the
so-called Faddeev-Popov ghosts \cite{Faddeev:1967fc,Abers:1973qs,
Itzykson:1980rh,Cheng:1985bj,Kaku:1993ym}.
\par
QCD has the property of asymptotic freedom, which asserts
that the theory becomes almost free at short distances, or at high
energies, while it becomes unstable at large distances, or at low
energies. This is interpreted as the sign of a new regime,
characterized by the confinement of the fundamental particles of
the theory. One striking feature of the theory is that the coupling
constant, which appears in the primary Lagrangian density (\rf{2e1}),
is not a free parameter: it is absorbed into the definition of the
mass scale, usually denoted by $\Lambda_{\mathrm{QCD}}$
\cite{Buras:1977qg,Zyla:2020zbs}, a phenomenon called
``dimensional (or mass) transmutation''.
Actually, the mass of the proton is mainly determined by
$\Lambda_{\mathrm{QCD}}$ and not by the masses of the quarks that enter
into its constitution; even if the quarks were massless, the
proton would continue having a mass of the order of its physical
mass. This is in contrast to the behavior in the electroweak sector
of the Standard Model, where the Higgs mechanism is at the origin of
the mass scales. Therefore, the masses of the light quarks $u,\ d,\ s$
do not play a major role in the theory and could, in many cases, be
taken as zero.
\par
The presence of a free parameter in a theory allows one to search for
approximate solutions for some particular values of the parameter
and then to apply perturbation theory around those values
\cite{Nielsen:1973cs}. QED and the short-distance regime of QCD
provide particular examples of this procedure, the expansion being
realized around the free theory. 't Hooft observed that QCD possesses
a hidden free parameter, represented by the dimension of the color
gauge group SU(3), provided one considers it as part of the general
class of non-Abelian gauge theories SU($N_c$), with the particular
physical value $N_c=3$ of the parameter $N_c$ \cite{'tHooft:1973jz}.
He studied the limit of large values of $N_c$, with the quark fields
belonging to the fundamental representation, which is of dimension
$N_c$, and the gluon fields belonging to the adjoint representation,
which is of dimension $(N_c^2-1)$, and showed that the
theory in its nonperturbative regime becomes simplified in many
instances.
\par
A first glance of the effect of this limit can be taken at the level
of the $\beta$-function, which displays the implicit mass-scale
dependence of the coupling constant
\cite{Gross:1973id,Politzer:1973fx,Politzer:1974fr,Gross:1975vu}
and which is a gauge and renormalization-group independent quantity
up to two loops. At one-loop order it reads
\be \lb{2e2}
\beta(g)\equiv \mu\frac{\partial g}{\partial\mu}=
-\frac{1}{16\pi^2}\Big(\frac{11}{3}N_c-\frac{2}{3}N_f\Big)g^3,
\ee
where $\mu$ is the mass scale at which renormalization has been
defined. Asymptotic freedom is realized for a negative value of
$\beta$; this is indeed the case with the physical values $N_c=3$
and $N_f=6$. Taking now large values of $N_c$, while keeping $N_f$ 
fixed, one notices that the negativity of $\beta$ is 
strengthened. To ensure, however, a smooth limit, so that
$\Lambda_{\mathrm{QCD}}$ remains independent of $N_c^{}$, one should
admit that at the same time the coupling constant $g$ scales with
$N_c$ like $N_c^{-1/2}$; the product $g^2N_c$ then remains constant
with respect to $N_c^{}$:
\be \lb{2e3}
g^2N_c\equiv \lambda=O(N_c^0).
\ee
The corresponding $\beta$-function is, for large $N_c$,
\be \lb{2e4}
\beta(\lambda)\equiv\mu\frac{\partial \lambda}{\partial\mu}=
-\frac{11}{24\pi^2}\lambda^2+O\Big(\frac{1}{N_c}\Big).
\ee
A similar conclusion is also obtained at the two-loop level.
\par
Generally, the quark flavor number, $N_f$, manifests itself through
quark-loop contributions. Equation (\rf{2e4}) shows that in the
large-$N_c$ limit, with fixed $N_f$, the quark loop contributions
are expected to be of nonleading order. This is one of the
important qualitative simplifications that occur on practical
grounds in the large-$N_c$ limit.
\par
Other types of large-$N_c$ limits have also been considered in the
past and presently for various purposes.
The simultaneous limits of large values of
$N_c$ and $N_f$, with $N_c/N_f$ fixed, has been considered by
Veneziano \cite{Veneziano:1976wm}; it is evident, from the previous
observation, that in that case the quark loops continue contributing
to leading order. In another limiting procedure, one assumes that
the quark fields belong to the second-rank antisymmetric tensor
representation
\cite{Corrigan:1979xf,Armoni:2003fb,Armoni:2004uu,Cherman:2006iy};
for $N_c=3$, this representation is equivalent to the antitriplet
one and therefore the physical content of actual QCD is
not modified. For general $N_c$, the dimension of that
representation is $N_c(N_c-1)/2$ and hence the number of degrees of
freedom of the quark fields scale as $N_c^2$; this prevents the
quark loops from disappearing from the leading order.
\par
The above variants of the large-$N_c$ limit have their own
phenomenological advantages. We shall stick, however, in the
present review, to the more traditional scheme developed by
't Hooft, because of its greater simplicity. Reviews about the
large-$N_c$ limit can be found in
\cite{Coleman:1985rnk,Migdal:1984gj,
Manohar:1998xv,Lebed:1998st,Makeenko:1999hq,Jenkins:2009wm,
Lucini:2012gg,Lucini:2013qja,Esposito:2016noz,Hernandez:2020tbc}.
\par

\subsection{Topological properties in color space} \lb{s22}

To study in more detail the properties of the theory in the
large-$N_c$ limit, it is advantageous to use a color two-index
notation for the gluon fields \cite{'tHooft:1973jz}. Since they
belong to the adjoint representation and the latter is contained
in the direct product of the fundamental and antifundamental
representations, one may represent the gluon fields with the
notation $A_{\ b,\mu}^a$, its relationship with the conventional
notation being the following:
\be \lb{2e5}
A_{\ b,\mu}^a=(A_{\mu}^BT^B)_{\ b}^a,\ \ \ \ \
A_{\ b,\mu}^{a\dagger}=A_{\ a,\mu}^b,\ \ \ \ \ A_{\ a,\mu}^a=0,
\ \ \ \ \ a,b=1,\ldots,N_c,
\ee
the third equation being a consequence of the property of the $T$'s
tracelessness. A similar notation can also be adopted for the ghost
fields, which are introduced together with the gauge-fixing term
in order to quantize the theory; however, for the simplicity of
presentation, we shall not explicitly write down ghost fields in
the remaining part of the paper, nor shall we draw the corresponding
Feynman diagrams; their presence does not modify the main qualitative
features that are drawn from the gluon fields.
\par
With the above convention, the color contents of the quark and
gluon propagators are
\be \lb{2e6}
\langle\psi_{i,\alpha}^a(x)\overline\psi_{b,j,\beta}(y)\rangle=
\delta_{ij}^{}\delta_{\ b}^a S_{\alpha\beta}(x-y),
\ee
where $i$ and $j$ are flavor indices, $\alpha$ and $\beta$ spinor
indices, and $S$ is the color-independent Dirac field propagator,
\be \lb{2e7}
\langle A_{\ b,\mu}^a(x)A_{\ d,\nu}^c(y)\rangle=
\Big(\delta_{\ d}^a\delta_{\ b}^c-\frac{1}{N_c}
\delta_{\ b}^a\delta_{\ d}^c\Big)D_{\mu\nu}(x-y),
\ee
where $D$ is the color-independent part of the gluon propagator.
The term proportional to $1/N_c$ in the last equation ensures the
traceless property of the gluon field. However, because of the
factor $1/N_c$, it could be neglected in leading-order calculations;
this amounts to replacing the gauge group SU($N_c$) by U($N_c$) and
the $(N_c^2-1)$ gluon fields by $N_c^2$ ones. The
corresponding approximation is of order $1/N_c^2$
(cf.~Ref.~\cite{Coleman:1985rnk}, Appendix C). This considerably
simplifies the diagrammatic representation of the gluon propagator:
as far as the color indices are concerned, the gluon propagates as
a quark-antiquark pair, which suggests a double-line representation
for the gluon propagator. Figure \rf{2f1} depicts, in two columns,
the correspondence between the conventional and the double-line
representations.
\par
\bfg 
\vspace*{1 cm}
\bc
\epsfig{file=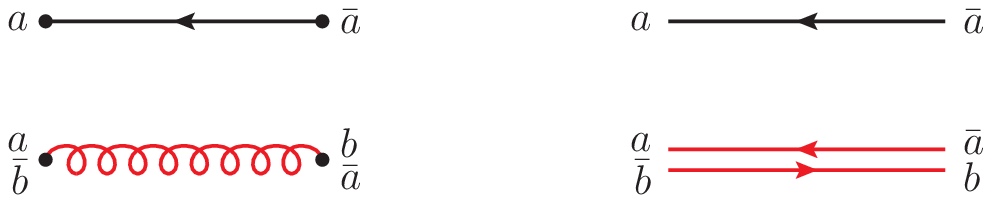,scale=1.}
\caption{The quark and gluon propagators in the conventional and
the double-line representations, first and second columns,
respectively. Lower indices are distinguished by bars.}  
\lb{2f1}
\ec
\efg
\par
Concerning the interaction parts, they have structures of the
following types: $\overline\psi_a\gamma^{\mu}A_{\ b,\mu}^a\psi^b$,\-
$A_{\ b,\mu}^aA_{\ c,\nu}^b\partial^{\mu}A_{\ a}^{c,\nu}$,\-
$A_{\ b,\mu}^aA_{\ c,\nu}^bA_{\ d}^{c,\mu}A_{\ a}^{d,\nu}$,\- quark
flavor being conserved. One notices that
a lower color index is always contracted with the upper index of a
neighboring field and this ensures the continuity of lines arriving
at a vertex and departing from it. The corresponding vertex diagrams
are presented in Fig.~\rf{2f2} in both representations.
\par
\bfg 
\bc
\epsfig{file=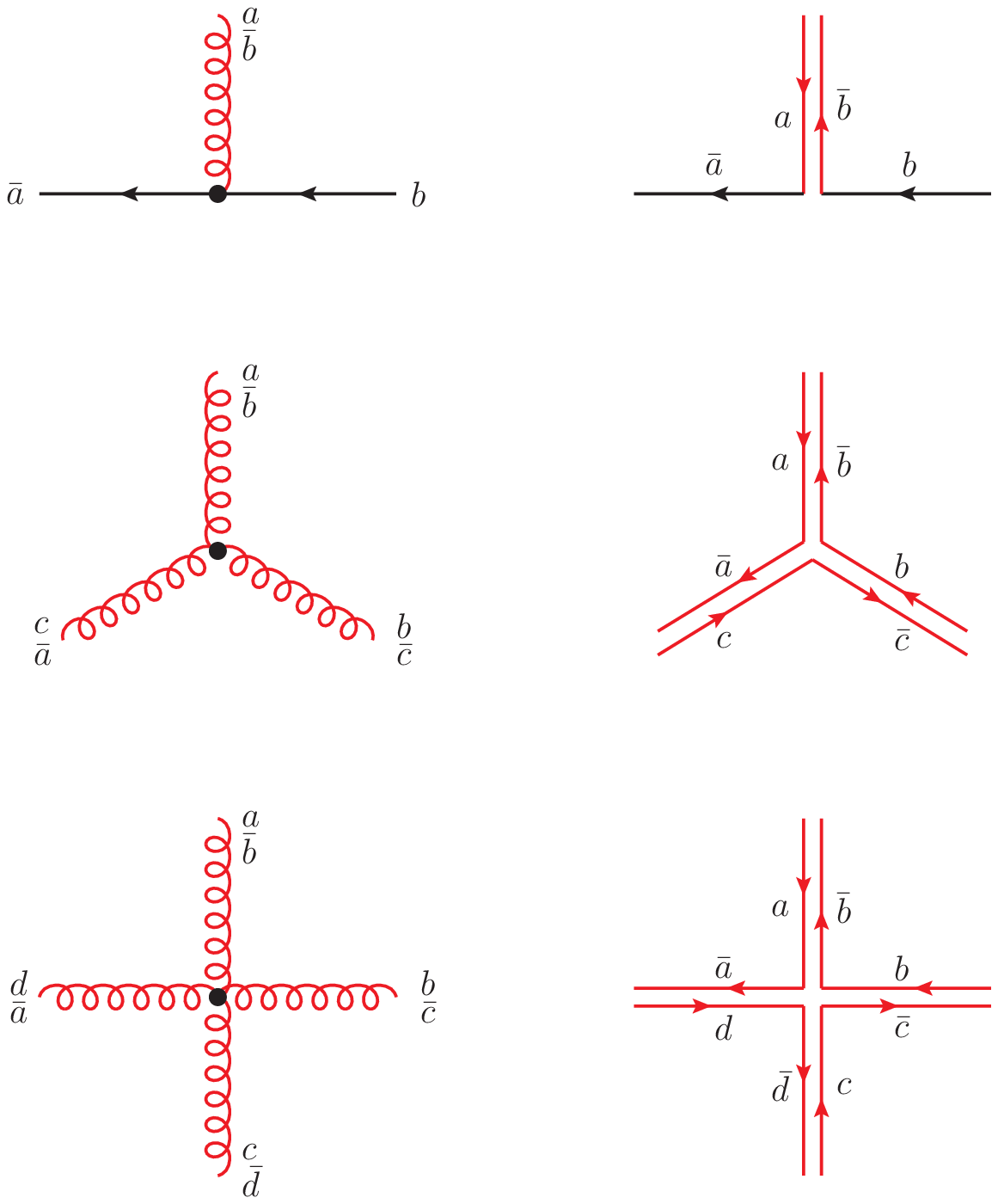,scale=0.8} 
\caption{The quark-quark-gluon, three-gluon and four-gluon
vertices in the conventional and the double-line representations,
first and second columns, respectively. Lower indices are
distinguished by bars.}  
\lb{2f2}
\ec
\efg
\par
The double-line representation allows a better control of the
color flow inside Feynman diagrams. To have a first glance of
it, we consider the two lowest-order contributions to the gluon
self-energy, represented by a quark loop (Fig.~\rf{2f3}a)
and a gluon loop (Fig.~\rf{2f3}b), respectively. Each color
loop produces a factor $N_c^{}$. The external gluon field color
indices being fixed, diagram (a) does not have such a loop;
furthermore, it contains two quark-quark-gluon vertices, each
involving a coupling constant $g$; taking into account the
large-$N_c^{}$ behavior of the latter [Eq.~(\rf{2e3})], one finds
that the large-$N_c^{}$ behavior of diagram (a) is $O(N_c^{-1})$.
On the other hand, diagram (b) contains one internal color loop,
providing an additional factor $N_c^{}$ with respect to the previous
diagram. Therefore, the large-$N_c^{}$ behavior of diagram (b) is
$O(N_c^{0})$. Thus, among the two diagrams of Fig.~\rf{2f3}, it is
diagram (b) which contributes to the leading-order behavior.
\par
\bfg 
\bc
\epsfig{file=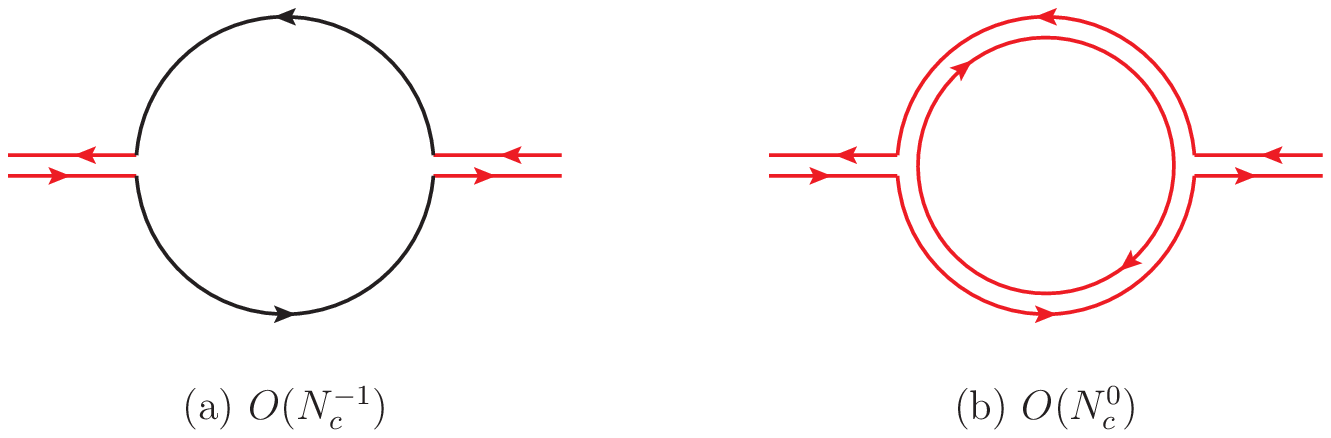,scale=0.7}
\caption{The lowest-order gluon self-energy contributions: (a) quark
loop; (b) gluon loop. The corresponding orders in large-$N_c$
behavior are also indicated.}  
\lb{2f3}
\ec
\efg
\par
The origin of the difference of contributions between the two
diagrams can easily be understood: it is related to the distinct
representations of the color group to which the quark and the
gluon fields belong: one has $N_c^{}$ quarks in the fundamental
representation against $(N_c^2-1)$ gluons in the adjoint  
representation. At leading orders, internal gluon lines produce
$N_c^{}$ times more contributions than quark lines, a feature that
the double-line representation displays explicitly through the
supplementary color loops. This means that, at leading orders, the
quark loops, which actually are in $N_f^{}$ duplicates, can be
neglected altogether, except when they appear as contractions of
external quark lines. This is one of the main advantages of the
large-$N_c{}$ limiting procedure adopted by 't Hooft.
\par
As a second example of the large-$N_c^{}$ counting rules, we consider,
still in the gluon self-energy part, one-gluon exchange diagrams
containing either a quark loop or a gluon loop (Fig.~\rf{2f4}).
Diagram (a) contains two color loops, producing a factor $N_c^2$,
together with six vertices, producing a factor $N_c^{-3}$
[Eq.~(\rf{2e3})]. Its global behavior is therefore $O(N_c^{-1})$.
Diagram (b) has one additional color loop and thus its behavior
is $O(N_c^{0})$. One again verifies the general property of the
nonleading character of the internal quark loops.
\bfg 
\bc
\epsfig{file=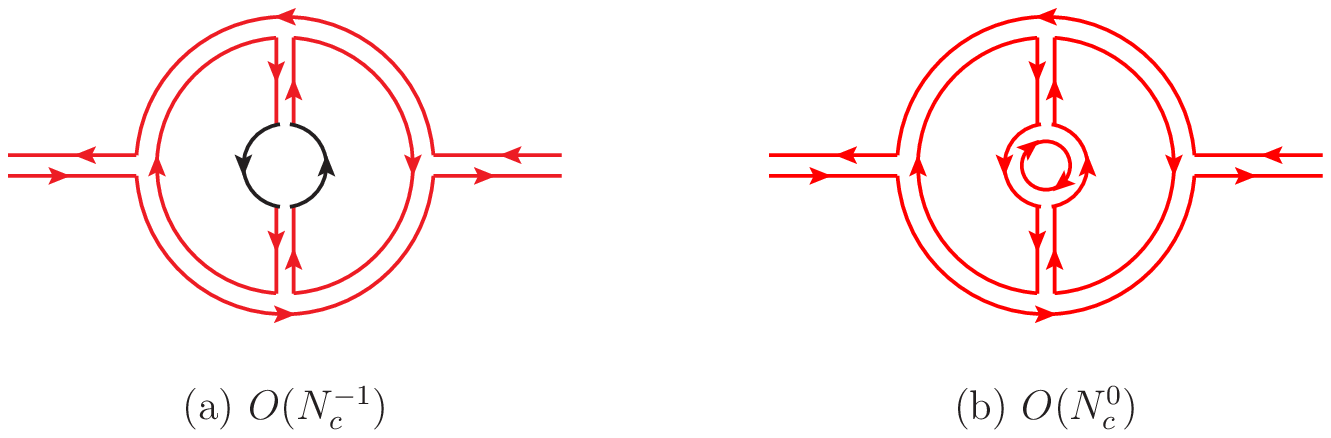,scale=0.8}
\caption{One-gluon exchange inside the gluon self-energy diagram:
(a) including a quark loop; (b) including a gluon loop.
The corresponding orders in large-$N_c$ behavior are also indicated.}
\lb{2f4}
\ec
\efg
\par
This second example outlines some other general features of the
large-$N_c^{}$ behavior, that we would like to emphasize. The leading
$N_c^{}$-behavior of Fig.~\rf{2f4} is the same as in Fig.~\rf{2f3}.
This means that the large-$N_c^{}$ behavior does not follow the
usual perturbative expansion in the coupling constant. The number
of color loops may balance the occurrence of new vertices. The
example of Fig.~\rf{2f4} may easily be generalized to more
complicated types of diagram, where one finds again the same
leading large-$N_c^{}$ behavior. The common feature of all these
diagrams is that they can be mapped on a plane, and more generally
on a two-dimensional surface, without allowing crossings of color
lines. For this reason, they are called \textit{planar diagrams},
which can be considered as belonging to a particular topological
class in color space. It is to be emphasized that they are infinite
in number.
\par
On the other hand, the nonleading diagrams of Figs.~\rf{2f3} and
\rf{2f4}, which contain the quark loops, could also be considered
as planar. They, however, display an additional topological property,
which is associated with the notion of a \textit{hole}. Comparing
diagrams (a) and (b) of both figures, one may distinguish figures
(a) from figures (b) by the occurrence in (a) of a hole in place of
the internal color loop that exists inside the gluon loop in (b).
Therefore, diagrams (a) above can be considered as planar, but
with one hole inside the plane. It is evident that each occurrence
of a hole produces a factor $N_c^{-1}$ in the large-$N_c^{}$ counting
rules.
\par
The diagrams that do not fulfill the planarity condition are called
\textit{nonplanar}. They occur when, after their projection on a
plane, some of the color lines cross each other. An example of
such a case is provided by the two-gluon crossed-exchange diagram
between two quark lines. Figure \rf{2f5} displays a few Feynman
diagrams occurring in the perturbative expansion of the two-point
function of the color-singlet bilinear current $j_{\bar k\ell}$,
$\langle j_{\bar k\ell}(x)j_{\bar k\ell}^{\dagger}(y)\rangle$,
defined as
\be \lb{2e8}
j_{\bar k\ell}^{}=\overline\psi_{a,k}^{}\psi_{\ell}^a,
\ee
where $a$ is a color index and $k$ and $\ell$ are fixed flavor
indices; Dirac matrices and spinor indices have been omitted, as
not being of primary importance in the present evaluation. Diagrams
(a), (b) and (c) are planar and provide the leading large-$N_c$
behavior. Diagram (d), representing the two-gluon crossed-ladder
diagram, is nonplanar. One observes that it contains only one
color loop, against the three color loops of the two-gluon ladder
diagram (c). Its large-$N_c^{}$ behavior is therefore $O(N_c^{-1})$,
against the $O(N_c^{1})$ behavior of the three planar diagrams (a),
(b) and (c).
\par
\bfg
\bc
\epsfig{file=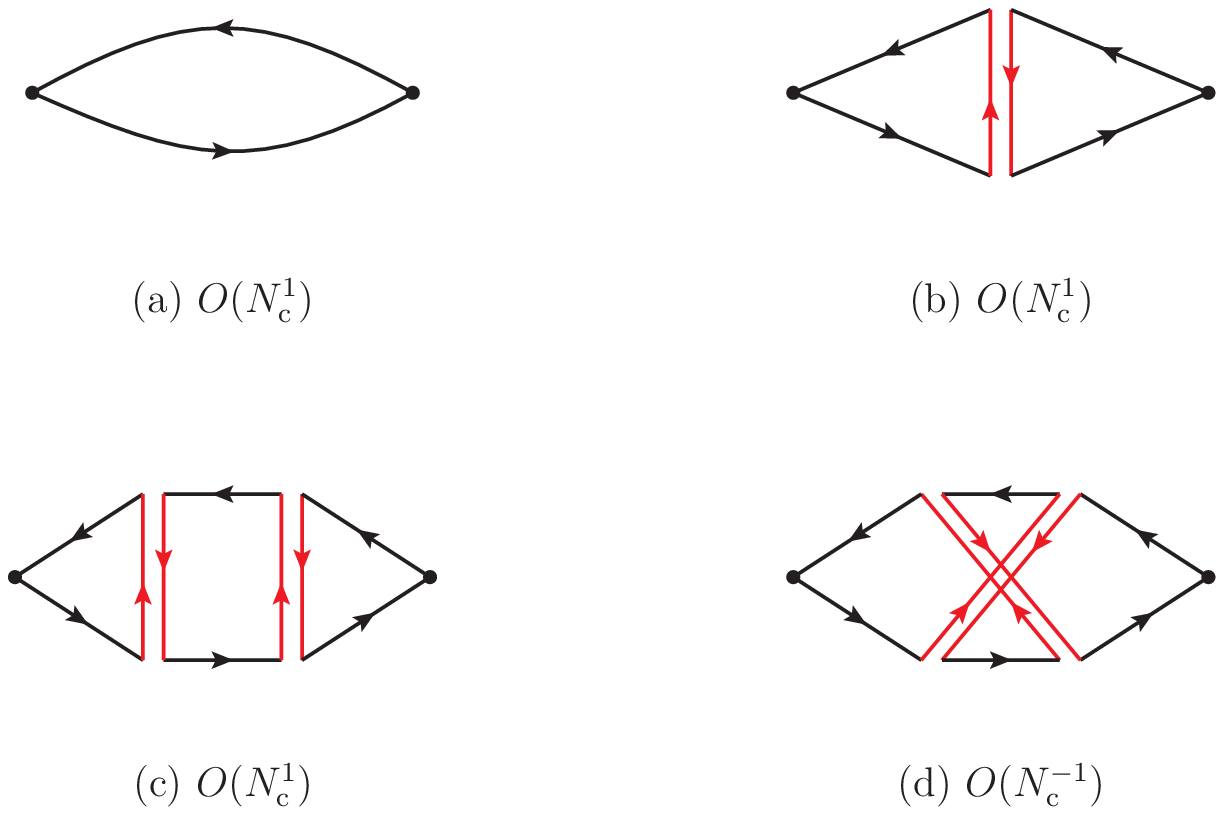,scale=0.8}
\caption{A few Feynman diagrams involved in the perturbative
expansion of the correlation function
$\langle j(x)j^{\dagger}(y)\rangle$.   
(a) A quark loop; (b) one-gluon exchange; (c) two-gluon ladder
diagram; (d) two-gluon crossed-ladder diagram.
The corresponding orders in large-$N_c$ behavior are also indicated.}
\lb{2f5}
\ec
\efg
\par
Nonplanar diagrams can be characterized by a specific topological
invariant, the number of \textit{handles}. Keeping in Fig.~\rf{2f5}d
one of the gluon propagators in space with respect to the projection
plane, one observes that it plays the role of a handle of a
two-dimensional surface. From the previous analysis, one deduces that
each handle introduces a factor $N_c^{-2}$ with respect to the planar
diagram contribution.
\par
The diagrams of Fig.~\rf{2f5} are examples of vacuum-to-vacuum
diagrams corresponding to the connected part of correlation functions
of gauge invariant local currents, each made of a quark and an
antiquark field.
Vacuum-to-vacuum diagrams can also be generated by gluon field
currents, made of bilinear functions of gluon field strengths.
An example is the current
\be \lb{2e9}
G_{[\mu\nu][\eta\sigma]}=F_{\mu\nu}^AF_{\eta\sigma}^A=
2F_{\ b,\mu\nu}^aF_{\ a,\eta\sigma}^b.
\ee
The leading-order behavior of the corresponding two-point function
is provided by  the planar diagrams, two of which are
represented in Fig.~\rf{2f6}.
\bfg
\bc
\epsfig{file=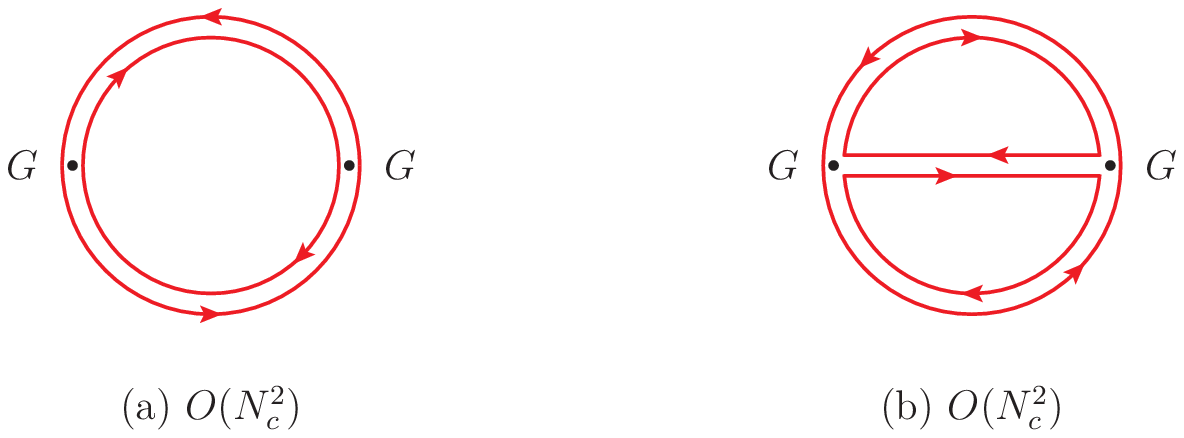,scale=0.7}
\caption{Vacuum-to-vacuum planar diagrams made of gluon lines,
corresponding to the two-point function of the gluonic current
$G$ [Eq.~(\rf{2e9})].}
\lb{2f6}
\ec
\efg
Their large-$N_c$ behavior is
$O(N_c^2)$. Comparing this with the leading-order behavior of the
quark bilinear case [Fig.~\rf{2f5}], we observe that the latter can
be deduced from the former by considering the external boundary
quark loop as a hole inside the planar diagram topology.
\par
We thus arrive at a general formula about the leading-order behavior
of a diagram, characterized by the number of two topological
invariants in color space, the hole, made of a color quark loop, and
the handle, made of a gluon propagator. Designating by $B$ the
number of holes and by $H$ the number of handles, the power of the
large-$N_c^{}$ behavior is
\be \lb{2e10}
N_c^{2-B-2H}\equiv N_c^{\chi},
\ee
where $\chi$ is called the Euler characteristic
(cf. Ref.~\cite{Coleman:1985rnk}, Appendix A).
\par
Planar diagrams without holes or with a number of holes determined
by the external boundary color quark loops will provide the
leading-order behavior at large $N_c^{}$. The inclusion of handles
and additional holes contributes to nonleading terms in $1/N_c^{}$.
Within this approach, $1/N_c^{}$ appears as the effective
perturbative expansion parameter of the theory. On the other hand,
planar diagrams are infinite in number; this means that, even with
the planarity approximation, one is not yet able to solve in a
simple way the theory. Nevertheless, one hopes that the $1/N_c^{}$
expansion will provide many qualitative simplifications and an
improvement in the understanding of the dynamics of the theory.
\par

\subsection{Mesons} \lb{s23}

Properties of physical states can be investigated by considering
correlation functions of gauge-invariant local currents, having
the same quantum numbers. For mesons, the natural candidates are
the quark bilinear currents, as defined in Eq.~(\rf{2e8}).
\par
Considering the two-point function of such a current (generally,
its connected part), typical
Feynman diagrams of its perturbative expansion in the coupling
constant have been presented in Fig.~\rf{2f5}, where the first
three, (a), (b) and (c), correspond to planar diagrams with
leading-order behaviors in $N_c^{}$. The most salient feature of
these, and of all planar diagrams, is the fact that they contain
only two quark lines (or propagators). A larger number of quark
lines can appear only in nonleading diagrams, containing an
additional number of holes. States characterized by a single
quark-antiquark pair, as a leading descriptive element, correspond
to the ordinary meson states. Therefore, the two-point function
of the current, saturated by a complete set of hadronic
intermediate states, reduces, at leading order in $N_c^{}$, to
a sum of meson poles:
\be \lb{2e11}
\int d^4x e^{ip.x}\langle j(x)j^{\dagger}(0)\rangle=
\sum_n \frac{iF_n^2}{p^2-M_n^2+i\epsilon}=O(N_c^1),
\ee
where $F_n^{}$ is defined as the matrix element of $j$ between
vacuum and the meson state $\vert n\rangle$:
\be \lb{2e12}
\langle 0|j|n\rangle = F_n^{}.
\ee
\par
The number of meson states must be infinite. This is dictated by
the asymptotic behavior of the left-hand side: because of asymptotic
freedom, its high-energy behavior is known and contains logarithmic
factors, which cannot be reproduced by a finite number of terms in
the sum (\rf{2e11}) \cite{Witten:1979kh}. This also entails a
generic behavior at large $N_c^{}$ for each term of the series in
connection with the behavior of the left-hand side ($O(N_c^{})$).
The most natural solution is that the meson masses (for finite $n$)
remain finite at large $N_c^{}$, while the couplings $F_n^{}$ increase
like $N_c^{1/2}$:
\be \lb{2e13}
M_n^{}=O(N_c^0),\ \ \ \ \ \ F_n^{}=O(N_c^{1/2}).
\ee
\par
From the complete decomposition of the two-point function into a
series of poles [Eq.~(\rf{2e11})], one also deduces that the
meson states are stable at large $N_c^{}$. If the mesons were
unstable, they would have finite widths, manifested as finite
imaginary parts in the pole terms, which, in turn, would imply,
through the unitarity property of the theory, the existence of
unitarity cuts and the appearance of many-particle states; these
would be manifested through the existence of more than two quark
lines in the leading-order diagrams, which is not the case.
\par 
In obtaining Eq.~(\rf{2e11}), we have assumed that all planar
diagrams containing two quark lines are perturbative representatives
of single meson states. Since planar diagrams contain, in general,
many gluon lines, the question arises as to whether such diagrams
may also contain independent glueball states, which might be formed
as gauge-invariant bound states of several gluon fields. This
question is best analyzed through the study of the singularities
and the imaginary part of the corresponding diagrams, by cutting
them with a vertical line. An example of this is presented
in Fig.~\rf{2f7}. 
\bfg
\bc
\epsfig{file=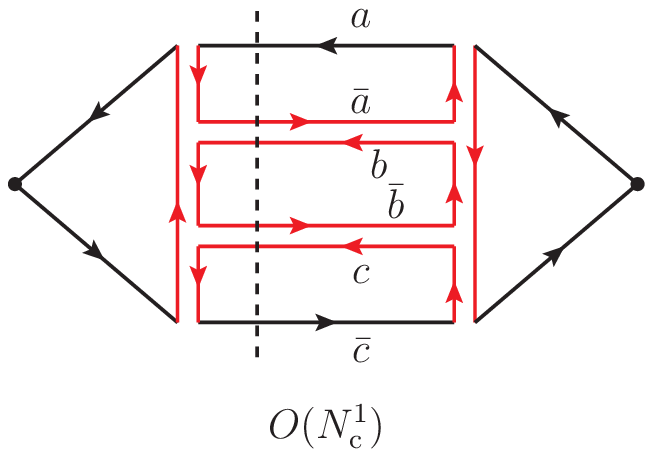,scale=0.8}
\caption{A planar diagram, contributing to the two-point function
of the current $j$, with two gluon propagators submitted, together
with the quark and antiquark propagators, to the vertical cut
(dashed). The color indices of the various lines on the right of the
vertical cut are explicitly indicated. Lower indices are distinguished
by bars.}
\lb{2f7}
\ec
\efg
One observes that each gluon propagator, cut
by the vertical line, is connected, with its color indices,
to the neighboring propagator, and nowhere a color-singlet
gluonic cluster emerges. The set of the above gluon propagators
belongs to the adjoint representation of the gauge group; on the
other hand, the set of the quark and antiquark propagators belongs
also to the adjoint representation. It is the connection of the
two sets that produces a color-singlet representation. Therefore,
the corresponding state is color-irreducible, in the sense that
it is not decomposable into the product of other color-singlet
representations. Hence, no independent glueball state may be
generated from such a diagram. This property is very general
for the two-point function and may be verified on more
complicated planar diagrams.
\par
Equation (\rf{2e11}) can be diagrammatically described by
representing the meson propagators by straight line segments
and displaying the large-$N_c^{}$ behavior of the couplings
[Eqs. (\rf{2e12}) and (\rf{2e13})]  (cf.~Fig.~\rf{2f8}, where
the connected part of the two-point function is schematically
represented in the form $\langle jj\rangle_c^{}$).
\bfg
\bc
\epsfig{file=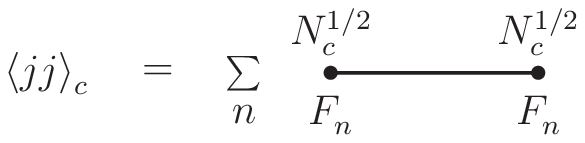,scale=0.8}
\caption{Equation (\rf{2e11}) in diagrammatic form.}
\lb{2f8}
\ec
\efg
\par
The study of the correlation functions of the currents $j$ can
be continued with the case of the three-point function
$\langle jjj\rangle$, where $j$ is a generic current, such that
connections between three $j$s are possible with quark lines.
The simplest planar diagram, for the connected part, is presented
in Fig.~\rf{2f9}.
\bfg
\bc
\epsfig{file=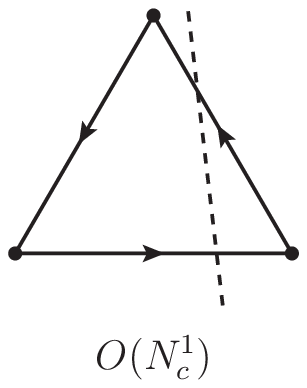,scale=0.8}
\caption{A planar diagram for the three-point function 
$\langle jjj\rangle_c^{}$; here, the arrowed lines indicate quarks,
whereas the dashed straight line indicates the cut.}
\lb{2f9}
\ec
\efg
\par
Cutting the diagram with a straight line in any direction and
position, one meets always a pair of quark-antiquark lines, which
means that the
singularities of the diagram are located at meson poles\footnote
{The choice of the form and geometry of the cuts allows a selection
of particular solutions of the Landau equations \cite{Landau:1959fi};
more generally, if $p$, $q$ and $r$ are the momenta associated with
the external currents, then the singularities occur, according to
the Landau equations, in the variables $p^2$, $q^2$ and $r^2$.}.
These may be three or two in number. The first category involves,
as a residue, the (amputated) vertex function of three meson sources,
providing the three-meson coupling constant. The second category
provides the coupling of a current $j$ to two mesons. The corresponding
equation is represented diagrammatically, together with the
relevant large-$N_c^{}$ behaviors, in Fig.~\rf{2f10}.
\bfg
\bc
\epsfig{file=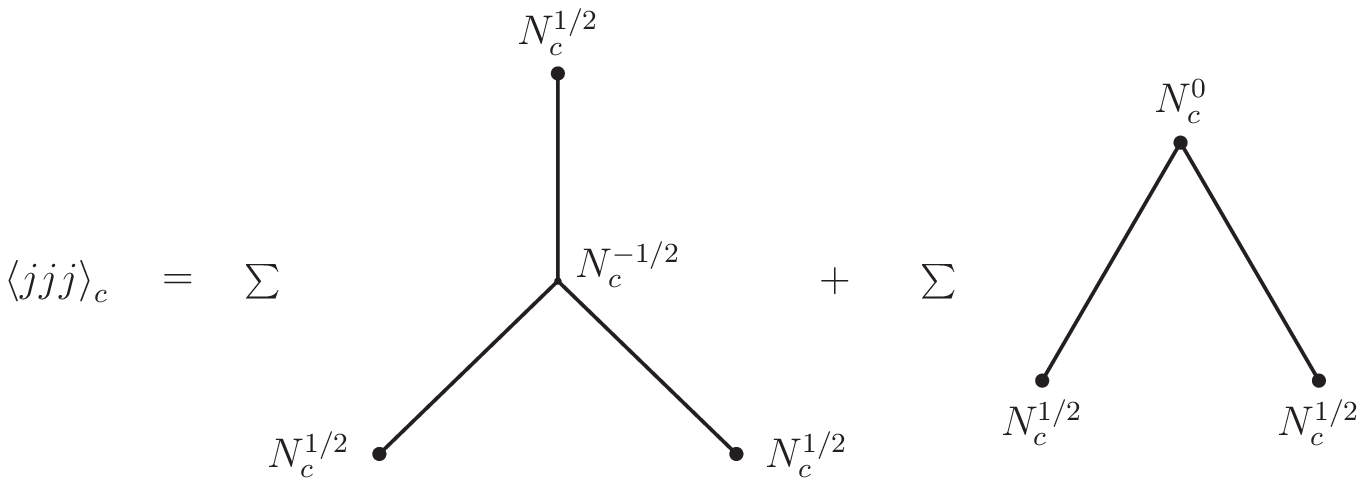,scale=0.7}
\caption{Diagrammatic representation of the three-point function
$\langle jjj\rangle_c^{}$ in terms of meson propagators and couplings,
at leading order of $N_c^{}$; here, the unarrowed lines indicate mesons.
The large-$N_c^{}$ behaviors of the various couplings are explicitly
indicated.}
\lb{2f10}
\ec
\efg
\par
One finds that the three-meson couplings behave, at large $N_c^{}$,
as $N_c^{-1/2}$. This also determines the decay amplitude of mesons
into two mesons, which vanishes at large $N_c^{}$; the mesons
are thus stable in this limit, a result that
confirms the stability property already deduced from the
decomposition of the two-point function [Eq.~(\rf{2e11})].
The second type of diagram in Fig.~\rf{2f10} determines the
coupling of the current $j$ to two mesons, or, equivalently, to two
pairs of quarks and antiquarks. Its behavior is $N_c^0$, a factor
$1/N_c^{1/2}$ less than the coupling $F_n$ [Eqs.~(\rf{2e12}) and
(\rf{2e13})].
If one interprets these couplings as probability amplitudes of
creating, by the current $j$, from the vacuum quark-antiquark
pairs, one deduces that at large $N_c^{}$ mesons are made of
one pair of quark-antiquark, while sea quarks, represented by
additional quark-antiquark pairs, occur only as nonleading
effects. This fact is a phenomeno\-lo\-gically confirmed
property, which is explained here in a natural way through the
$1/N_c^{}$ expansion method. This could also explain why meson-meson
type interpolators are generally needed in lattice-QCD calculations
in order to get the correct mass for states that couple strongly to
these two mesons \cite{Liu:2012zya,Lang:2014yfa}.
\par
We next consider four-point functions $\langle jjjj\rangle$ of
generic currents $j$, such that connections between
neighboring currents can be realized with quark lines.
The simplest planar diagram, for the connected part, is presented
in Fig.~\rf{2f11}.
\bfg
\bc
\epsfig{file=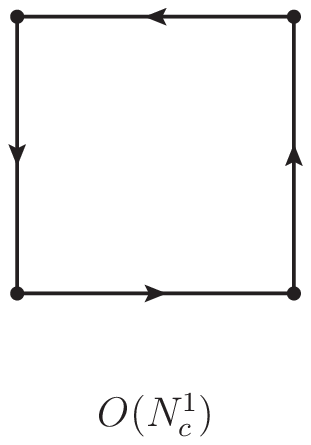,scale=0.8}
\caption{A planar diagram for the four-point function.}
\lb{2f11}
\ec
\efg
\par
The singularities are again located at meson poles.
The decomposition of the connected part of the four-point function
in terms of meson propagators and couplings is diagrammatically
presented, together with the relevant large-$N_c^{}$ behaviors of
the couplings, in Fig.~\rf{2f12}.
\begin{figure}[t]
\bc
\epsfig{file=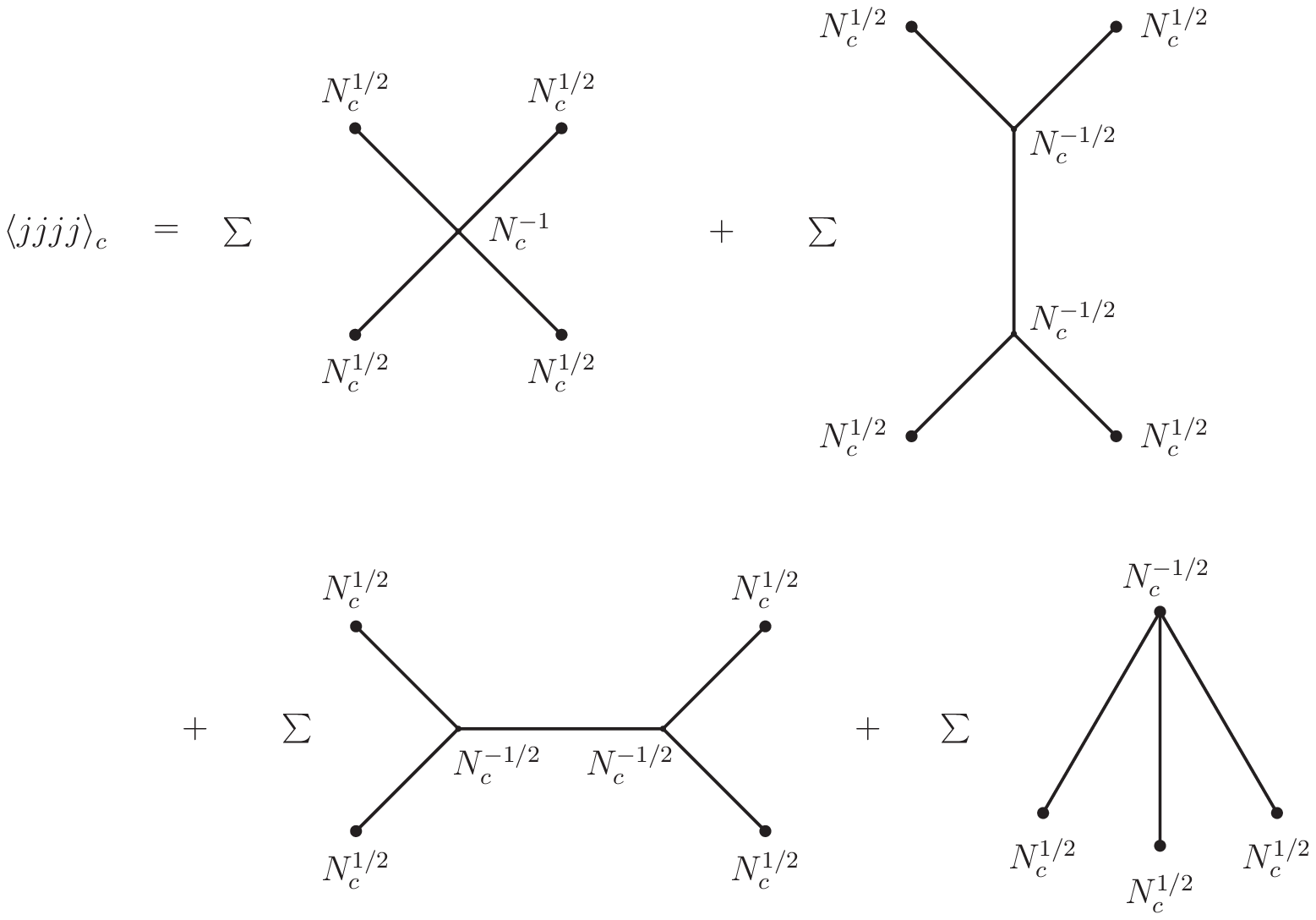,scale=0.7}
\caption{Diagrammatic representation of the four-point function
$\langle jjjj\rangle_c^{}$ in terms of meson propagators and couplings,
at leading order of $N_c^{}$. The large-$N_c^{}$ behaviors of the
various couplings are explicitly indicated.}
\lb{2f12}
\ec
\end{figure}
One finds that the four-meson couplings are of order $N_c^{-1}$,
smaller by a factor $1/N_c^{1/2}$ than the three-meson couplings.
This, in turn, entails that the decay amplitudes of mesons into
three mesons are also of order $N_c^{-1}$. The couplings of the
currents $j$ to three-meson states, or, equivalently, to three
pairs of quarks and antiquarks, are of order $N_c^{-1/2}$.
\par
Factoring out, in the above decomposition, the four meson propagators,
together with their couplings to the external currents, one obtains
the scattering amplitude of two mesons into two mesons, which is
of order $N_c^{-1}$:
\be \lb{2e14}
\mathcal{T}(MM\rightarrow MM)=O(N_c^{-1}).
\ee
It is worthwile noticing that, at that leading order, the scattering
amplitude is expressed as a series of tree diagrams involving the
infinite number of meson propagators and mutual couplings. This is
in qualitative accordance with Regge phenomenology, where the dominant
contributions come from tree diagrams of hadron exchanges and couplings
\cite{Veneziano:1968yb}, even though the Regge behavior itself cannot
be demonstrated by the sole large-$N_c^{}$ limit.
\par
The above procedure can be continued to higher numbers of currents.
One particular outcome, as was already evident, is the relative
decrease of the order in $N_c^{}$-behavior of the meson couplings
when the number of mesons increases. Thus, the $n$-meson couplings
behave as $N_c^{1-n/2}$. One consequence of this property is that
when a meson may decay into many mesons, it preferentially decays
first into two mesons (or, possibly, into three mesons if there is
a selection rule), which in turn decay into two or three mesons,
and so forth. This is also a phenome\-no\-lo\-gically confirmed
fact.
\par
Another phenomenon occurring in hadron physics is related to the
so-called OZI rule (after Okubo, Zweig and Iizuka)
\cite{Zweig:1964jf,Okubo:1963fa,Iizuka:1966fk}, which stipulates
that, in the case of three light quarks, mesons generally are members
of nonets of the flavor group U(3), rather than of separate octets
and singlets of the group SU(3) (best illustrated by the
$\varphi-\omega$ mixing)\footnote{The pseudoscalar mesons
are an exception, due to the chiral anomaly problem
\cite{Gasser:1984gg,'tHooft:1986nc}; however, the anomaly vanishes
at large $N_c^{}$.}.
A violation of the rule concerns processes where the quark lines
are completely disconnected between the initial and final states;
such processes should be negligible.
An example is illustrated by the two-point function 
$\langle j_{\bar kk}(x)j_{\bar \ell\ell}^{\dagger}(0)\rangle$, where $k$
and $\ell$ are different fixed flavor indices. The two currents
can mutually interact only through gluon exchange. A planar diagram
of the corresponding process is presented in Fig.~\rf{2f13}.
The diagram has two holes, due to the external color quark loop
boundaries; its behavior at large $N_c^{}$ is $O(N_c^{0})$, smaller
by a factor $1/N_c^{}$ with respect to the behavior of the planar
diagrams of Fig.~\rf{2f5}. Once again, the large-$N_c$ limit
provides a natural explanation of this qualitative effect, widely
verified by experimental data.
\bfg 
\bc
\epsfig{file=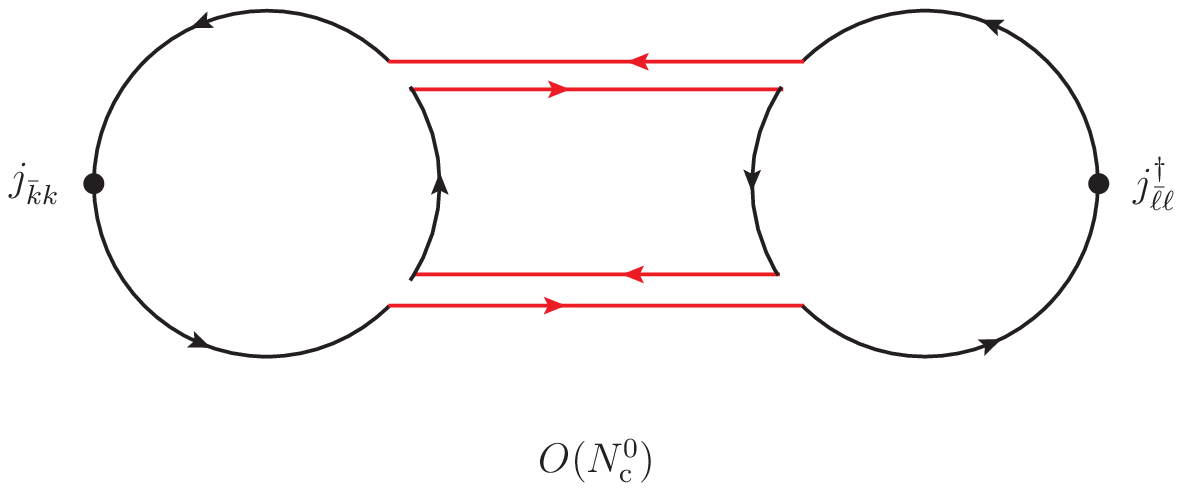,scale=0.6}
\caption{OZI-rule violating diagram: Two gluon propagators joining
two quark loops with completely different flavor content. The order
in large-$N_c^{}$ behavior is to be compared with the leading-order
behaviors of Fig.~\rf{2f5}.}  
\lb{2f13}
\ec
\efg 
\par
To summarize, the large-$N_c^{}$ limiting procedure leads, according
to the orders of $N_c^{}$, to a hierarchical classification of the
various processes that occur in QCD in its nonperturbative
regime, providing a qualitative understanding of many typical
phenomena that characterize the strong-interaction physics of hadrons.
As far as the meson sector is concerned, at leading order in
$N_c^{}$, QCD reduces to a theory of an infinite number of
free stable mesons, made of a quark-antiquark pair and of gluons,
whose masses squared are expected to lie along Regge trajectories.
At this level, all mesons are on equal footing; their differences
arise only from their specific quantum numbers. The interactions
among these mesons, which are responsible for their strong decays
and nontrivial scattering processes, appear at nonleading orders of
$N_c$. From this point of view, strong-interaction physics of
mesons corresponds to a weakly interacting effective field theory,
with an expansion parameter given by $1/N_c^{}$, dominated by tree
diagrams of meson exchanges and contact terms, as compared to the
underlying strongly interacting theory, which is responsible for
the confinement of quarks and gluons.  
\par
Studies of the influence of the large-$N_c^{}$ limit on meson
properties can be found in Refs.~\cite{Gasser:1984gg,Knecht:1997ts,
Peris:1998nj,Knecht:1999gb,Cirigliano:2003yq,Pich:2008jm,
Rosell:2009yb,Bali:2013kia}.

\subsection{Baryons} \lb{s24}

One would like to complete the large-$N_c^{}$ approach by extending
it to the physics of baryons. However, here, the method applied to
the case of mesons turns out to be inapplicable. 
\par
The main reason of that difficulty is related to the description
itself of baryonic
states at large $N_c^{}$. While for mesons, the change of the gauge
group from SU(3) to SU($N_c^{}$) did not need any change in their
description, characterized by their couplings to the local bilinear
currents (\rf{2e8}), baryons, and, more precisely, the currents to
which they may couple preferentially, require a change of description.
In SU(3), baryonic states are coupled to currents that are
trilinear in quark fields and completely antisymmetric in color
indices to ensure gauge invariance. A typical such current is:
\be \lb{2e15}
j_{B}^{(3)}(x)=\frac{1}{3!}\epsilon_{abc}\psi^a(x)\psi^b(x)\psi^c(x),
\ee
where $\epsilon$ is the Levi-Civita symbol (a completely
antisymmetric tensor) and, for simplicity, we have considered quarks
with the same flavor and omitted the spin indices. In passing to
SU($N_c^{}$), one has to generalize the above definition by using
the Levi-Civita tensor in $N_c^{}$ dimensions, involving $N_c^{}$
indices, which, in turn, requires the use of $N_c^{}$ quark fields.
The baryonic currrent then becomes:
\be \lb{2e16}
j_{B}^{(N_c^{})}=\frac{1}{N_c^{}!}\epsilon_{a_1^{}a_2^{}\cdots a_{N_c{}}^{}}
\psi^{a_1^{}}\psi^{a_2^{}}\cdots\psi^{a_{N_c^{}}^{}}.
\ee
Considering now the two-point function of this current, one can
try to evaluate the $N_c^{}$ dependence of the corresponding
Feynman diagrams, as was done in Fig.~\rf{2f5} for the mesonic
currents. Typical diagrams are presented in Fig.~\rf{2f14}. 
\bfg 
\bc
\epsfig{file=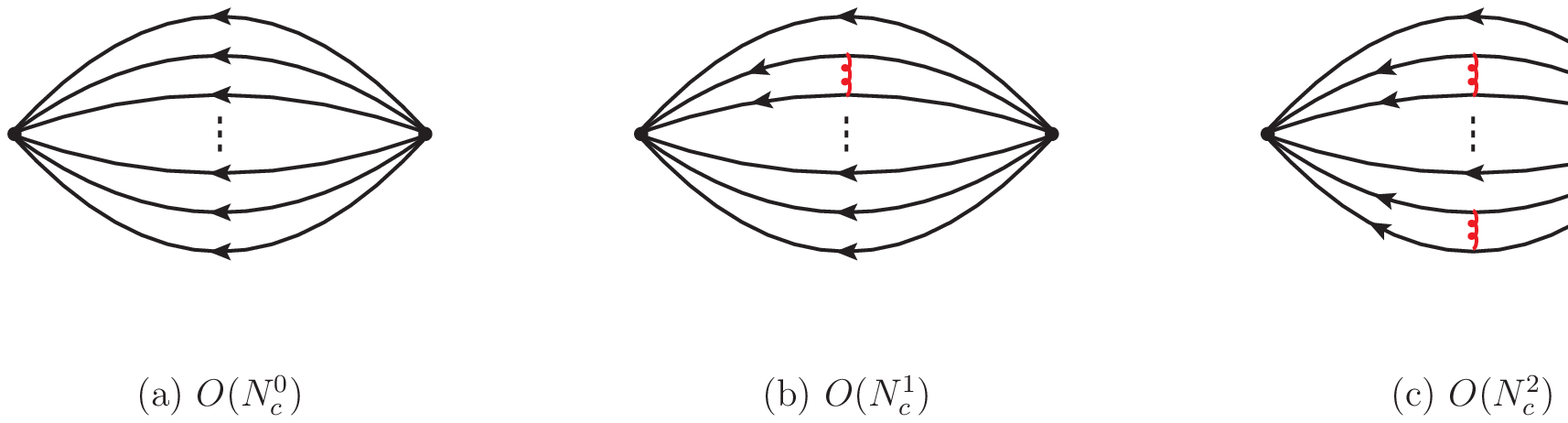,scale=0.7}
\caption{Typical Feynman diagrams of the two-point function
of the baryonic current (\rf{2e16}): (a) without gluon propagators;
(b) with one gluon propagator (sample); (c) with two gluon
propagators (sample). The order in large-$N_c^{}$ behavior of the
contributions of all diagrams of each category is also indicated.}
\lb{2f14}
\ec
\efg 
\par
Taking into account the normalization factor included in the
definition of the current (\rf{2e16}), the class of diagrams
not containing gluon propagators [Fig.~\rf{2f14}a] behaves as
$O(N_c^0)$, which fixes the normalization for the other types of
diagrams. A diagram containing one gluon propagator
[Fig.~\rf{2f14}b], joining two quark propagators, contains a
damping factor $1/N_c^{}$ coming from the coupling constant squared
[Eq.~(\rf{2e3})]. However, there are
$N_c^{}(N_c^{}-1)/2\sim N_c^2/2$ such diagrams; therefore the
total contribution of this category of diagrams is
$O(N_c^2/N_c^{})=O(N_c^1)$. A diagram containing two gluon propagators
[Fig.~\rf{2f14}c], joining different quark lines, contains the damping
factor $1/N_c^2$ coming from the coupling constants at the vertices;
this is to be multiplied by the total number of such diagrams, which
is of the order of $N_c^4$; the total contribution of this category of
diagrams is therefore $O(N_c^2)$. We observe that the perturbative
expansion of the two-point function introduces at each order of the
expansion a new factor $N_c^{}$, which makes the corresponding series
formally divergent at large $N_c^{}$. Contrary to the case of the
two-point functions of the mesonic currents, it is not possible here
to group, in a stable way, elements of the perturbative series into
topological classes having well-defined $N_c^{}$ behaviors.
\par
Another complication arises from the fact that, inside baryons, the
recognition of the color topological categories of diagrams is less
evident than for mesons; the reason for this is related to the fact
that, in baryons, all quark colors flow in the same direction,
while in mesons the color of the antiquark flows in the opposite
direction to that of the quark. We illustrate this phenomenon by
focusing on the first gluon-exchange diagrams between two quark
lines. They are presented, in the double-line representation,
in Fig.~\rf{2f15}.
\par
\bfg 
\bc
\epsfig{file=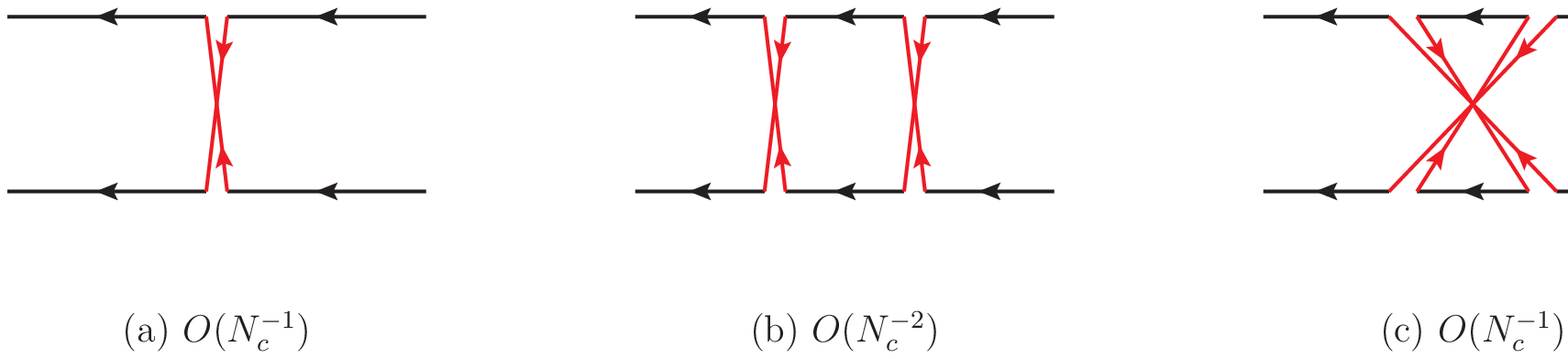,scale=0.7}
\caption{Gluon exchanges between two quark lines in the double-line
representation: (a) one-gluon exchange; (b) two-gluon ladder
exchange; (c) two-gluon crossed-ladder exchange. The order in
large-$N_c^{}$ behavior of each diagram, with external color indices
fixed, is also indicated.}
\lb{2f15}
\ec
\efg 
\par
Figure \rf{2f15}a corresponds to the one-gluon exchange diagram. We
observe that, because of the similar directions of the quark color
flows, the two color lines of the gluon propagator cross each other.
As it stands, this diagram cannot be factorized in a plane, as in the
planar-diagram case, into separate color flows without crossing.
One could, of course, unfold one quark line, by inversing its drawing
and flow, to make the diagram planar, however, this operation may,
in general, be forbidden by the rest of the bigger diagram into which
the above diagram is embedded. The behavior of the diagram, with
fixed external color indices, is $O(N_c^{-1})$. Figure \rf{2f15}b
corresponds to the ladder exchange of two gluon propagators. Here
also, we meet the previous phenomenon. Contrary to the meson case,
the diagram does not have any color loop and behaves at large
$N_c^{}$, with fixed external color indices, like $O(N_c^{-2})$, a
factor $1/N_c^{}$ less than in the case of mesons.
Figure \rf{2f15}c corresponds to the crossed-ladder exchange of
two gluon propagators. Here, the diagram contains an internal color
loop and behaves like $O(N_c^{-1})$. It thus appears as the partner
of the one-gluon exchange diagram for summation purposes. This
property remains true for $n$-gluon exchange diagrams where each
gluon line is crossed by the other $(n-1)$ gluon lines.
\par
The above results seem to prevent the consideration, in a simple way,
of the large-$N_c^{}$ limit in the sector of baryons, at least in a
way that is parallel to that of mesons.
\par
A way out of this difficulty was proposed by Witten
\cite{Witten:1979kh}. He noticed that whereas the perturbative
interaction between two quarks is small, of the order of $1/N_c^{}$,
it is the big number of quarks inside the baryons and the sum of
the mutual interactions that are at the origin of the divergence.
In such a case, diagrammatic considerations, which focus on
the mutual interactions of a few neighboring quarks, are of little
help. Every quark experiences a global strong force representing
an average form of the sum of the forces exerted by the other 
quarks. One is then in a situation where a self-consistent mean-field
approximation can be used. The problem simplifies further in
the case of heavy quarks, where nonrelativistic theory applies in
the form of the Hartree equations. Witten showed that these
equations can be consistently solved, yielding a coherent
description of the baryonic sector in the large-$N_c^{}$ limit.
\par 
The main property that characterizes these equations is that they
globally scale as $N_c^{}$ at large $N_c^{}$; this is due to
the fact that each of their components -- the total kinetic energy,
the total potential energy and the mass of the baryon -- has the
same $N_c^{}$-behavior. Therefore, $N_c^{}$ factors out of the
equations, leaving $N_c^{}$-independent equations, which ensure the
stability of the result under perturbations with respect to $1/N_c^{}$.  
In particular, the size and shape of the baryons turn out to be
independent of $N_c^{}$. The method is applied to the case of the
ground state, as well as to the excited states, of baryons.
\par
The above procedure is also applied to the study of processes like
baryon-baryon, baryon-antibaryon and baryon-meson scatterings.
All these processes have the property of leading to equations
that globally scale like $N_c^{}$, which is then factoured out.
\par
The properties and dynamics of the baryons are thus mainly described,
at large $N_c^{}$, by semi-classical equations, rather than by
microscopic quantum equations.
\par
How to interpret the dissymmetry that emerges, at large $N_c^{}$,
between the mesonic and baryonic sectors? In this respect, Witten
has made the following crucial observation \cite{Witten:1979kh}:
The mesonic sector of QCD is described by a weakly interacting
effective field theory, whose interaction scale is of the order of
$1/N_c^{}$.
On the other hand, weakly coupled field theories often develop
nonperturbative solutions, like solitons or monopoles
\cite{tHooft:1974kcl,Polyakov:1974ek,Polyakov:1974wq,Faddeev:1977rm},
whose mass scale is governed by the inverse of the weak coupling.
This is, in particular, the case for some electroweak theories with
spontaneously broken symmetry, characterized, say, by a coupling
constant squared $\alpha$, which possess a magnetic monopole type
solution, whose mass is of the order of $1/\alpha$
\cite{tHooft:1974kcl,Polyakov:1974ek,Polyakov:1974wq}.
The structure of the monopole is determined, for small
$\alpha$, by solving classical equations, from which $\alpha$
drops out, the size and shape of the monopole then becoming
independent of $\alpha$. Similarly, the mass of the baryons in QCD
is of the order of the inverse of the coupling $1/N_c^{}$, i.e.,
of the order of $1/(1/N_c^{})=N_c^{}$. Therefore, baryons can be
considered as the QCD analogs of the solitons or magnetic monopoles,
while mesons and glueballs are the analogs of ordinary particles.
\par
More detailed investigations in the baryonic sector can be found
in Refs. \cite{Manohar:1998xv,Dashen:1993jt,Dashen:1994qi,
Jenkins:1993zu,Dashen:1993as,Kaplan:1996rk,Pirjol:1997bp,
Pirjol:1997sr,Goity:2002pu,Cohen:2011cw,Cohen:2011hq,Matagne:2014lla}.

\subsection{Electric charges of quarks} \lb{s25}

We discuss, in this subsection, the generalization of the Standard
Model (SM) to the case when the color subgroup of the SM symmetry
group becomes $\mathrm{SU}(N_c)$ instead of SU(3).
\par
We first briefly recall the quantum numbers of the quark and lepton
fields in the SM (see, e.g., \cite{Abers:1973qs,Itzykson:1980rh,
Cheng:1985bj,Kaku:1993ym}).
The SM of elementary particles is a gauge theory based on the
spontaneously broken
$\mathrm{SU}_L(2)\times \mathrm{U}_Y(1)\times \mathrm{SU}(N_c)$
symmetry, where the
$\mathrm{SU}_L(2)\times \mathrm{U}_Y(1)$ sector describes electroweak
(EW) interactions
of the fundamental fermion fields of the SM, quarks and leptons.
\par
The SM contains three generations of fermion matter fields, quarks
and leptons. In each generation, left-handed fermions compose
doublets with respect to the $\mathrm{SU}_L(2)$ group, whereas the
right-handed matter fields are $\mathrm{SU}_L(2)$ singlets.
For instance, in the first generation, the SM contains two weak
doublets
\begin{eqnarray} \lb{2e17}
q_L=
\left(\begin{array}{c}
u_L\\
d_L
\end{array}\right),\ \ \ \ \ 
l_L=
\left(\begin{array}{c}
\nu_L\\
e_L
\end{array}\right),
\end{eqnarray}
and four right-handed $\mathrm{SU}_L(2)$ singlets (if one includes
the Dirac right-handed neutrino field in the set of the SM fields):
\begin{eqnarray} \lb{2e18}
u_R,\ d_R,\ e_R,\ \nu_R.
\end{eqnarray}
The quark and lepton doublets have the same $\mathrm{SU}_L(2)$ charges:
$+1/2$ for the upper components and $-1/2$ for its lower components.
The right-handed quark and lepton singlets have zero $\mathrm{SU}_L(2)$
charges.
The $\mathrm{U}_Y(1)$ quantum numbers -- the weak hypercharges $Y$ --
for each $\mathrm{SU}_L(2)$ multiplet are independent from each other
(the upper and the lower components of any doublet have the same $Y$).
In the SM one has
\begin{eqnarray} \lb{2e19}
&Y^l_{L}=-1,\ \ \ Y^e_{R}=-2, \ \ \ Y^\nu_{R}=0,\nonumber\\
&Y^q_{L}=\frac{1}{3},\ \ \ Y^u_{R}=\frac{4}{3}, \ \ \
Y^d_{R}=-\frac{2}{3}.
\end{eqnarray}
The electric charge is related to the $\mathrm{SU}_L(2)$ and $U_Y(1)$
quantum numbers by the Gell-Mann--Nishijima relation
\begin{eqnarray} \lb{2e20}
Q=I_3+Y/2,
\end{eqnarray}
where $I_3$ is the eigenvalue of the third component of the
\textit{weak isospin}.
For the left-handed doublets, $I_3=\pm 1/2$, for the right-handed
singlets, $I_3=0$.
\par
The SM is free from the chiral (axial) anomaly, since the quark-loop
contribution to the anomaly cancels against the lepton-loop
contribution. This happens since quark and lepton charges satisfy
the relation
\begin{eqnarray} \lb{2e21}
\sum_{\rm leptons}Q^l+\sum_{\rm quarks}Q^q=0.
\end{eqnarray}
Quark fields belong to the fundamental representation of the SU(3)
color group, so summation over quarks includes summation over color
indices running from 1 to 3. Notice that the left-handed fermion fields
and the right-handed fermion fields satisfy Eq.~(\rf{2e21}) separately.
\par
When one generalizes the color group SU(3) to $\mathrm{SU}(N_c)$,
the lepton quantum numbers remain unchanged, as leptons do not
interact with the gluons, but the EW, i.e.,
$\mathrm{SU}_L(2)\times \mathrm{U}(1)$, quantum
numbers of the quark fields should be changed.
To obtain the electric charge and weak hypercharge of the quarks for
arbitrary $N_c$, the following constraints are imposed:
(i) The left-handed quark fields
$\left(\begin{array}{c} u_L\\ d_L \end{array}\right)$,
$\left(\begin{array}{c} c_L\\ s_L \end{array}\right)$
and $\left(\begin{array}{c} t_L\\b_L \end{array}\right)$ remain
\textit{weak isospin doublets} with $I_3 = \pm 1/2$,
while the right-handed quark fields remain $\mathrm{SU}_L(2)$ singlets.
Electroweak quantum numbers of all quarks satisfy
the Gell-Mann--Nishijima relation (\rf{2e20}).
(ii) The $\mathrm{SU}_L(2)\times \mathrm{U}(1)\times \mathrm{SU}(N_c)$
Standard Model should be
free of axial anomalies. Quark fields are in the fundamental
representation of the $SU(N_c)$ color group, so there are $N_c$
different quark colors. The anomaly cancellation condition requires
that the sum of all electric charges (of leptons and quarks) vanishes.
\par
In calculating the sum of charges in (\rf{2e21}) for the left-handed
particles, the terms $\pm 1/2$ in each left-handed doublet
cancel out; the sum over quark colors gives a factor $N_c$ and thus
one comes to the following relation
\begin{eqnarray} \lb{2e22}
Y^l_L+N_c Y^q_L=0.
\end{eqnarray}
Taking into account that the hypercharge for the left-handed doublet
remains the same as in the SM, $Y^l_L=-1$, one obtains
$Y^q_L=\frac1{N_c}$. Proceeding in the same way for the right-handed
particles, one finds
\begin{eqnarray} \lb{2e23}
&Y_L^l=-1,\ \ \  Y_R^e=-2,\ \ \  Y_R^\nu=0, \\
&Y_L^q=\frac{1}{N_c},\ \ \ Y_R^u=1+\frac{1}{N_c},\ \ \ 
Y_R^d=-1+\frac{1}{N_c}.
\end{eqnarray}
The quark electric charges therefore become
\begin{eqnarray} \lb{2e24}
Q_{u,c,t} = \frac12 +\frac1{2N_c}, \ \ \ \  Q_{d,s,b} = -\frac12
+\frac1{2N_c}.
\end{eqnarray}
\par
The electric charges of mesons are not changed compared to the
case $N_c=3$, since they contain one quark and one antiquark,
whose $Y$-terms cancel each other. Masses of mesons, as
quark-antiquark composites, remain finite, $O(N_c^0)$, at large
$N_c$.
\par
Baryons are bound states of $N_c$ quarks. 
Their electric charges change compared to the SU(3) case and
generally increase with $N_c$: for odd $N_c$ their charges are
integers, while for even $N_c$ they are half-integers.
\par
For the classification of baryons, one first uses \textit{strong
isospin}; this approximate symmetry of strong interactions
is related to the smallness of the masses of the $u$ and $d$ quarks
compared to
$\Lambda_{\mathrm{QCD}}$ and to the values of the QCD vacuum condensates.
One may impose the conditions that, in the $\mathrm{SU}(N_c)$ theory,
$u$ and $d$
quarks still form a doublet with $I_3 = +1/2$ and $-1/2$, respectively,
while all other quarks are \textit{strong isosinglets},
and that all quarks satisfy the
Gell-Mann--Nishijima relation (\rf{2e20}) for \textit{strong
hypercharges} of the quarks. For the light quarks $u$, $d$ and
$s$, one finds \cite{Buchmann:2000wf}:
\begin{eqnarray} \lb{2e25}
Y_u = Y_d = \frac{1}{N_c}, \quad Y_s = -1+\frac{1}{N_c}.
\end{eqnarray}
Baryons in large $N_c$ have masses of $O(N_c)$. One then is entitled to
use spin-flavor symmetry \cite{Dashen:1993as} and to classify
baryons into spin-flavor representations at leading order in $1/N_c$.
In general, the latter are decomposed into distinct flavor multiplets
with increasing spins. It is then possible to assign the familiar
baryons into representations where they keep the same values of
spin, isospin, hypercharge and electric charge as in the SU(3) case.  
For instance, the proton is composed of $(N_c+1)/2$ $u$ quarks and
$(N_c-1)/2$ $d$ quarks, $N_c$ being now odd, and one may verify that
its electric charge is +1.
The reader may consult Ref. \cite{Buchmann:2000wf} for a more
detailed account of the classification scheme of baryons. 
\par

\subsection{Corrective effects} \lb{s26}

Let us briefly comment on the possible magnitude of the corrective
factors in the $1/N_c^{}$ expansion method in phenomelogical
applications, where one has to use the physical value 3 of $N_c^{}$.
\par
The qualitative successes that the $1/N_c^{}$ expansion method has
obtained in QCD in the understanding of the hierarchy of various
processes and phenomena brings an indirect or implicit
justification of the validity of the method, which hinges, like
other perturbative methods, on the smallness of the expansion
parameter $1/N_c^{}$. When the corrective factors to the leading
terms are of order $1/N_c^2$, the latter become, for $N_c^{}=3$,
of the order of 1/10, which is indeed a small quantity. For
corrections of order $1/N_c^{}$ (not to be confounded with
leading terms of order $1/N_c^{}$), one has a quantity of the
order of 1/3, which is not fairly small. Generally, corrections
of order $1/N_c^{}$ come from internal quark loops, which,
furthermore, are proportional to the flavor number $N_f^{}$.
Considering only the case of three light quarks ($u,\ d,\ s$),
$N_f=3$, one would have $N_f^{}/N_c^{}=1$, which is not a
perturbative parameter. Actually, it becomes important here to
know about the size of the coefficient that accompanies the factor
$N_f^{}/N_c^{}$. A hint about
the latter coefficient is provided by the expression of the
beta function, Eq.~(\rf{2e2}). Comparing both terms of the
right-hand side of Eq.~(\rf{2e2}), one deduces that, for $N_f^{}=3$
and $N_c^{}=3$, the corrective term coming from the quark loops
with respect to the leading term, is approximately equal to 0.18,
which is a small quantity that might represent an acceptable
value for an expansion procedure. In reality, on phenomenological
grounds, the corresponding corrective factors, examples of which
are provided by the quark-sea contributions inside hadrons
(in the nonperturbative regime) and by the OZI-rule violations,
are even much smaller, being of the order of a few percent.
\par
Therefore, one may consider the perturbative expansion, in terms
of $1/N_c^{}$, at least for the first corrective terms, as a
phenomenologically well-established procedure.
\par
In conclusion, the large-$N_c^{}$ limit of QCD, proposed by
't Hooft and completed by Witten's proposal about baryons,
leads to a consistent and simplified picture of the hadronic world
and of its strong interaction dynamics.
\par

\subsection{The AdS/CFT correspondence} \lb{s27}

In this subsection, we briefly outline the role of the large-$N_c^{}$
limit in the correspondence established between string theories
and quantum field theories, mostly known as the AdS/CFT
correspondence \cite{Maldacena:1997re,Maldacena:1998im,Witten:1998qj,
Witten:1998zw,Gubser:1998bc,Aharony:1999ti,Maldacena:1999fi}.
This subject being out
of the scope of the present review, we only focus here on the philosophy
that has guided the related investigations. The interested reader is
invited to consult the quoted references.
\par
String theory has been present in hadronic physics from the early
days of the discoveries of hadron resonances. The existence of a
large number of hadron resonances, lying along Regge trajectories,
was very suggestive of string theory spectra. Later, the advent
of the quark model and of QCD introduced the concept of
confinement of quarks and gluons, a property that is also shared
by strings, at the endpoints of which are attached quarks.
(For a review, cf., e.g., Ref. \cite{Scherk:1974jj}.) 
\par
It was noticed by 't Hooft, on the basis of the diagrammatic
expansion in the large-$N_c^{}$ limit, that the resemblance between
QCD theory in its nonperturbative regime and string theory is
much enforced \cite{'tHooft:1973jz} in that limit. 
The spectrum of mesons is then very similar
to that of free strings, whose coupling constant would be of the
order of $1/N_c^{}$. The string picture is expected to be induced
in QCD by the chromo\-elec\-tric flux tubes and by the Wilson lines
that ensure gauge invariance of multilocal operators.
This suggests a correspondence between the two theories. 
However, flat-space string theories are consistently formulated in
ten dimensions and the scales that they involve are of the order of
the Planck scale, rather than of the hadronic scale
\cite{Green:1987sp,Green:1987mn,Polchinski:1998rq,Polchinski:1998rr}.
The stringy behavior of QCD at large distances would then be
understood as resulting from an effective field theory formulation
\cite{Luscher:1980ac,Polchinski:1991ax,Makeenko:2012ug}. 
\par
On the other hand, the analysis applied to the large-$N_c^{}$ limit
of QCD is rather general and could be applied to other gauge
theories with different types of gauge symmetries. 
Duality relations could thus be searched for in a wider area of
theories, general gauge field theories, on
the one hand, and general string theories, on the other. The basic
idea is that some physical theories might have two
equivalent descriptions, each with different variables, such that
the strong-coupling regime of one of them corresponds to the
weak-coupling regime of the other.
The corresponding investigations have been based on searches for
theories having common global-symmetry properties, which survive
the changes of description.
\par
From the latter point of view, conformal invariance is the
simplest symmetry that solves the problem. Conformal invariance
requires the absence of any mass or energy scale in the theory.
In four dimensions, where the gauge coupling constant is
dimensionless, this amounts to the requirement that the latter
be scale independent, which means that the $\beta$-function is
null. It is only in supersymmetric theories that simple
realizations of conformal invariance can be found. The most
interesting example is provided by $\mathcal{N}=4$ SU($N_c$)
super-Yang-Mills (SYM) theory, which involves four spinor
supercharges. Another global symmetry, the $R$-symmetry, with the
symmetry group SU(4), transforms the four supercharges into each
other. The supersymmetry field content is then the following: the
gluon fields belong to the singlet representation of the $R$-symmetry
group, the fermion fields to the fundamental ($\mathbf{4}$)
representation and the scalar fields to the two-index antisymmetric
($\mathbf{6}$) representation, all of them belonging to the adjoint
representation of the color-gauge group.
\par
If the above theory has a dual string theory, then its global
invariance properties should be reflected by the latter. The
conformal group is SO(2,4); locally, there is only one space with
SO(2,4) isometries: five-dimensional anti-de Sitter space, or AdS$_5$.
The SU(4) group of $R$-symmetry is equivalent to SO(6) and the
latter is the symmetry group of a 5-sphere $S^5$. One therefore
expects that the dual string theory should be formulated in a
ten-dimensional space containing, at least in some regions, the
product of the two previous spaces, AdS$_5\times S^5$; notice that
the total spacetime dimension precisely corresponds to the dimension
in which flat-space string theory can be consistently formulated;
in the present case, one should have a five-dimensional
compactification into the 5-sphere and the introduction of a
curvature through the anti-de Sitter space.
\par
The link between string theory and gauge theory is realized
by means of $D$-branes (``$D$'' from Dirichlet) \cite{Polchinski:1995mt}.
These are solitonic solutions of type-II strings and come out in
various dimensionalities; $Dp$-branes have $p$ spatial dimensions.
In string perturbation theory, $D$-branes are defined as surfaces where
open strings can end. The latter have massless modes representing gauge
fields and their fermionic partners. If one has $N_c$ coincident 
$D$-branes, open strings can start and end on two different branes;
this implies labeling of the gauge fields and their fermionic
partners with two different color indices, typical of the adjoint
representation of an SU($N_c$) gauge theory, which would then describe
the low-energy dynamics of the theory.
\par
The presence of $D$-branes changes the original flat-space metric to
a curved one. $D$-branes, like black holes, contain event horizons.
The low-energy limit of the new theory is represented by
supergravity, itself containing classical brane-type solutions.
Considering the case of $N_c$ coincident $D3$-branes,
appearing in type-IIB string theory ($p=3$ being odd),
one finds that the near horizon geometry is described by
AdS$_5\times S^5$. On the other hand, the low-energy limit of the
dynamics in the worldvolume of the branes is described by  
$\mathcal{N}=4$ SU($N_c$) SYM theory. One thus obtains two different
descriptions of the $D3$-branes at low energies, the first as
a gravity theory in AdS$_5\times S^5$ space, the second as an
$\mathcal{N}=4$ SU($N_c$) SYM theory in four-dimensional Minkowski
space. Designating the string theory coupling constant by $g_s^{}$,
its connection with 't Hooft's coupling constant $\lambda$
[Eq. (\rf{2e3})] is $4\pi g_s^{}=\lambda/N_c$. In the large-$N_c$
limit, with $\lambda$ fixed, $g_s^{}$ is generally small and the
string theory is reduced, at leading order, to its classical limit.
The precise outcome depends, however, on the domain of values
of $\lambda$. When $\lambda$ is sufficiently large, but finite,
corresponding to the strong-coupling regime of the gauge theory,
$g_s^{}$ is relatively large, but, however, the curvature radii of the
AdS space and of the 5-sphere are also large, being proportional
to $\lambda^{1/4}$.
Then the dynamics is described by the AdS$_5\times S^5$ near-horizon
geometry, with small curvature, dominated by classical gravity.
For small values of $\lambda$, it is the perturbative regime of
the gauge theory that provides the simplest description.
It was conjectured, on the basis of the property of the gauge theory
being a unitary theory, that the gauge/gravity correspondence,
demonstrated
in the supergravity approximation at large $N_c$, might go beyond
that approximation and could be valid for the full string theory
\cite{Maldacena:1997re}. (Cf. also \cite{Aharony:1999ti} for a
detailed discussion of various aspects of this conjecture.)
\par
Can one claim at this stage equivalence of gravitational theory and
gauge theory? The question is pertinent, since the two theories
live in different dimensions of spacetime. The gravitational theory
is defined in ten-dimensional curved space, while the gauge theory
is defined in four dimensions;
the mapping between the degrees of freedom of the two theories does
not seem trivial. A possible solution to this issue had been
proposed by several authors, who had
observed that the information carried by gravitational theories,
defined in $(d+1)$ dimensions, might be stored, according to
a holographic principle, in a $d$-dimensional boundary region
\cite{tHooft:1993dmi,Thorn:1991fv,Susskind:1994vu,Brown:1986nw}.
It is such a principle that is conjectured in the presently considered
AdS/CFT correspondence, allowing the extension of the duality
property away from the $D$-brane horizon region in
AdS$\times S^5$ space \cite{Aharony:1999ti}.
\par
In spite of this theoretical progress, the above duality relation
does not directly apply to QCD theory, for several reasons.
First, the latter theory is not conformally invariant; scale symmetry
is broken by quantization and the bound-state spectrum displays there
a mass gap with towers of discrete masses.
Second, QCD is not supersymmetric.
Third, quarks belong to the fundamental representation of the color
group, while in $\mathcal{N}=4$ SYM fermions belong, in four
(identical) copies, to the (sole) adjoint representation with the
six-fold presence of scalar partners. For these
reasons, the treatment of QCD needs more elaborate pathways to
establish the bridge to string theory \cite{Witten:1998zw}.
The AdS/QCD correspondence remains, for the moment, at the level of
phenomenological approaches or of model building
\cite{Karch:2006pv,Andreev:2006ct,Erdmenger:2007cm,Brodsky:2008pg,
dePaula:2008fp,Colangelo:2008us}.
\par

\section{How to describe multiquark states?} \lb{s3}

Properties of physical states are usually probed in quantum field
theory by the study of Green's functions or correlation
functions, using interpolating currents having nonvanishing
couplings to them. This was the case for mesons [Eq.~(\rf{2e8})]
and baryons [Eq.~(\rf{2e15})]. In the case of bound states, more
detailed informations are obtained from the solution of
bound-state equations, which generally require the use of multilocal
operators as interpolating probes.
\par
The problem is similar, in principle, in the case of exotic
hadrons or multiquark states. One has to find appropriate
interpolating currents to extract from correlation functions
their specific properties. Here, however, additional complications
arise. First, because of the increasing number of quarks 
in multiquark states, the number of the corresponding
interpolating currents also increases and several combinations of
them may be as good candidates as the individual ones. An optimal
choice, for practical purposes, would be the one that would provide
the strongest coupling to the state under study. However, the
physical properties of the states are independent of the initial
choice of interpolating currents, provided the latter have
nonvanishing couplings to them. Second, in varying $N_c^{}$, the
definition of the multiquark state itself may change. An example
of this phenomenon has been met with the ordinary baryons, for
which the interpolating current had to be modified [Eq.~(\rf{2e16})].
This phenomenon is rather general for multiquark states, in
which case the large-$N_c^{}$ generalization of the interpolating
currents is no longer unique: one has to deal with different
schemes of well-known multiquark states of the case $N_c^{}=3$,
such as tetraquarks, pentaquarks and hexaquarks.
\par
We shall review, in this section, the various possibilities that one
meets for the choice of interpolating currents and operators for the
study of the properties of multiquark states. We shall first
consider the case of the gauge group SU(3) and then its
generalization to SU($N_c^{}$). For simplicity, we shall ignore
spin/Dirac indices and concentrate on color and flavor indices.
The inclusion of spin/Dirac indices can be done with the incorporation
of appropriate Dirac matrices, taking into account the total spin
and parity of the states.
\par

\subsection{SU(3)} \lb{s31}

We first consider the case of tetraquarks, which are mesons expected
to be represented by two pairs of valence quarks and
antiquarks\footnote{For brevity, we shall often refer to them as
four-quark states.}. To avoid mixing problems with ordinary meson
states, we shall consider four different quark flavors, referred to
by indices $i,j,k,\ell$; color indices will be designated by
$a,b,c,\ldots$ .
\par
Since the tetraquark is a color-singlet state, one has to find
interpolating currents that are globally color singlets.
As mentioned in the Introduction, an evident choice is the product
of two mesonic color-singlet currents of the type of Eq.~(\rf{2e8}).
Designating by $T^{(1,1)}$ the tetraquark current, one has two
different choices:
\bea
\lb{3e1}
& &T_{\bar ij,\bar k\ell}^{(1,1)}(x)=
j_{\bar ij}^{}(x)j_{\bar k\ell}^{}(x)
=\Big(\overline\psi_{a,i}^{}\psi_{j}^{a}\Big)(x)
\Big(\overline\psi_{b,k}^{}\psi_{\ell}^{b}\Big)(x),\\
\lb{3e2}
& &T_{\bar i\ell,\bar kj}^{(1,1)}(x)=
j_{\bar i\ell}^{}(x)j_{\bar kj}^{}(x)
=\Big(\overline\psi_{a,i}^{}\psi_{\ell}^{a}\Big)(x)
\Big(\overline\psi_{b,k}^{}\psi_{j}^{b}\Big)(x).
\eea
\par
Another choice corresponds to the ``diquark'' combinations,
by grouping the two quarks
and the two antiquarks into antisymmetric or symmetric
representations. In the first case, one obtains with the two
quarks the antitriplet representation, $\mathbf{\bar 3}$, and
with the two antiquarks the triplet representation, $\mathbf{3}$;
the two may then be combined into the singlet representation.
In the second case, the two quarks are in the sextet representation,
$\mathbf{6}$, and the two antiquarks in the antisextet representation,
$\mathbf{\bar 6}$, which also can be combined to yield the
singlet representation. Designating by $T^{(\mp,\mp)}$ the
corresponding currents, one has
\bea
\lb{3e3}
& &T_{\bar i\bar k,j\ell}^{(-,-)}(x)=\frac{1}{2}\epsilon^{abc}
\Big(\overline\psi_{a,i}^{}\overline\psi_{b,k}^{}\Big)(x)
\epsilon_{dec}^{}\Big(\psi_{j}^{d}\psi_{\ell}^{e}\Big)(x),\\
\lb{3e4}
& &T_{\bar i\bar k,j\ell}^{(+,+)}(x)=\frac{1}{4}
\Big(\overline\psi_{a,i}^{}\overline\psi_{b,k}^{}+
\overline\psi_{b,i}^{}\overline\psi_{a,k}^{}\Big)(x)
\Big(\psi_{j}^{a}\psi_{\ell}^{b}+\psi_{j}^{b}\psi_{\ell}^{a}
\Big)(x),
\eea
where $\epsilon$ is the Levi-Civita tensor, already introduced in
Eq.~(\rf{2e15}). The two currents $T^{(\mp,\mp)}$ are not independent
of the two former currents $T^{(1,1)}$. By using in Eq. (\rf{3e3})
the relation
\be \lb{3e5}
\epsilon^{abc}\epsilon_{dec}^{}=\delta_{\ d}^a\delta_{\ e}^b
-\delta_{\ e}^a\delta_{\ d}^b,
\ee
and grouping in Eq.~(\rf{3e4}) the quark fields in bilinear
current forms, one finds
\be \lb{3e6}
T_{\bar i\bar k,j\ell}^{(-,-)}=
-\frac{1}{2}(T_{\bar ij,\bar k\ell}^{(1,1)}+
T_{\bar i\ell,\bar kj}^{(1,1)}),\ \ \ \ \ \ 
T_{\bar i\bar k,j\ell}^{(+,+)}=
-\frac{1}{2}(T_{\bar ij,\bar k\ell}^{(1,1)}-
T_{\bar i\ell,\bar kj}^{(1,1)}).
\ee
Clearly, one can also reexpress the currents $T^{(1,1)}$ as
combinations of the currents $T^{(\mp,\mp)}$.
\par
Finally, one can also choose tetraquark currents made of products
of bilinear currents in the octet representation, $\mathbf{8}$:
\bea
\lb{3e7}
& &T_{\bar ij,\bar k\ell}^{(8,8)}(x)=
\Big(\overline\psi_{a,i}^{}(T^A)_{\ b}^a\psi_{j}^{b}\Big)(x)
\Big(\overline\psi_{c,k}^{}(T^A)_{\ d}^c\psi_{\ell}^{d}\Big)(x),\\
\lb{3e8}
& &T_{\bar i\ell,\bar kj}^{(8,8)}(x)=
\Big(\overline\psi_{a,i}^{}(T^A)_{\ b}^a\psi_{\ell}^{b}\Big)(x)
\Big(\overline\psi_{c,k}^{}(T^A)_{\ d}^c\psi_{j}^{d}\Big)(x),
\eea
where the $T^A$s are the generators of SU(3) in the fundamental
representation. By using the relation
\be \lb{3e9}
(T^A)_{\ b}^a(T^A)_{\ d}^c=
\frac{1}{2}\Big(\delta_{\ d}^a\delta_{\ b}^c
-\frac{1}{N_c^{}}\delta_{\ b}^a\delta_{\ d}^c\Big),
\ee
with $N_c^{}=3$, one can reexpress these currents in terms of the
currents $T^{(1,1)}$:
\be \lb{3e10}
T_{\bar ij,\bar k\ell}^{(8,8)}=
-\frac{1}{2}\Big(\frac{1}{3}T_{\bar ij,\bar k\ell}^{(1,1)}
+T_{\bar i\ell,\bar kj}^{(1,1)}\Big),\ \ \ \ \ \
T_{\bar i\ell,\bar kj}^{(8,8)}=
-\frac{1}{2}\Big(T_{\bar ij,\bar k\ell}^{(1,1)}
+\frac{1}{3}T_{\bar i\ell,\bar kj}^{(1,1)}\Big).
\ee
\par
Therefore, only two currents are independent for the probe of
tetraquarks with four different quark flavors. Their specific
choice is a matter of taste or practical usefulness and does
not prejudge in any way the physical structure of the tetraquark.
It is the calculation of their couplings to the latter which
ultimately may provide the physical information.
\par
The above procedure of construction of currents can readily be
generalized to other multiquark states. We briefly sketch some
of them.
\par
Pentaquarks are expected to be dominated by four valence quarks
and one valence antiquark. We consider the case of four different
flavors for the quarks, the antiquark having one of these flavors.
A pentaquark current is most easily constructed as a product of
a bilinear mesonic current [Eq.~(\rf{2e8})] and of a trilinear
baryonic current [Eq.~(\rf{2e15})], an example of which is
\be \lb{3e11}
P_{\bar ij,ik\ell}^{(1,1)}(x)=j_{\bar ij}^{}(x)j_{B,ik\ell}^{}(x)
=\Big(\overline\psi_{a,i}^{}\psi_{j}^{a}\Big)(x)
\frac{1}{3!}\epsilon_{bcd}\Big(\psi_{i}^{b}\psi_{k}^{c}
\psi_{\ell}^d\Big)(x).
\ee
\par
Other currents commonly used are based on the diquark
antisymmetric representation:
\be \lb{3e12}
P_{\bar i,ij,k\ell}^{(\bar{3},-,-)}(x)=
\frac{1}{4}\epsilon^{abc}\overline\psi_{a,i}^{}(x)
\epsilon_{bde}\Big(\psi_{i}^{d}\psi_{j}^{e}\Big)(x)
\epsilon_{cd'e'}\Big(\psi_{k}^{d'}\psi_{\ell}^{e'}\Big)(x).
\ee
\par
Hexaquarks are dominated by six valence quarks (or by three quarks
and three antiquarks, a case that we omit below). Their currents
can commonly be represented as products of two baryonic currents
or products of three antisymmetric diquark currents (here considered
with four different quark flavors):
\bea \lb{3e13}
H_{ijk,ij\ell}^{(1,1)}(x)&=&j_{B,ijk}^{}(x)j_{B,ij\ell}^{}(x)
\nonumber \\
&=&\frac{1}{3!}\epsilon_{abc}\Big(\psi_{i}^{a}\psi_{j}^{b}
\psi_{k}^c\Big)(x)
\frac{1}{3!}\epsilon_{a'b'c'}\Big(\psi_{i}^{a'}\psi_{j}^{b'}
\psi_{\ell}^{c'}\Big)(x),
\eea
\bea \lb{3e14}
& &H_{ij,ik,j\ell}^{(-,-,-)}(x)=\frac{1}{8}\epsilon^{abc}
\epsilon_{aa_1a_2}\Big(\psi_{i}^{a_1}\psi_{j}^{a_2}\Big)(x)
\epsilon_{bb_1b_2}\Big(\psi_{i}^{b_1}\psi_{k}^{b_2}\Big)(x)
\epsilon_{cc_1c_2}\Big(\psi_{j}^{c_1}\psi_{\ell}^{c_2}\Big)(x)
\nonumber\\
& &
\eea
(no summation over repeated flavor indices).
\par

\subsection{SU($N_c^{}$)} \lb{s32}

In passing to SU($N_c^{}$), nontrivial modifications occur in
the definitions of the interpolating currents that we met in
the case of SU(3). This is related to the fact that these
currents are generally defined as products of irreducible tensors
and with the increase of $N_c^{}$ the number of such tensors
increases in turn. Adopting the notation of irreducible
representations based on Young tableaux (see, e.g., 
Ref.~\cite{Georgi:1982jb}), $[\ell_1^{},\ell_2^{},\cdots]$, where
the nonincreasing integers $\ell_i^{}$ ($i=1,2,\ldots$) denote the
number of boxes in each column, the defining fundamental
representation to which belongs the quark field is simply $[1]$,
while the antifundamental representation to which belongs the
antiquark field is $[N_c^{}-1]$. When $N_c^{}=3$, the latter
reduces to $[2]$, which implies that a two-index
antisymmetric tensor of quark fields belongs to the antitriplet
representation $\mathbf{\bar 3}$. When $N_c^{}>3$, one has
$(N_c-2)$ distinct antisymmetric irreducible representations,
which generalize the case of $N_c=3$. The two-index antisymmetric
representation is then $[2]$, while the two-index symmetric
representation is $[1,1]$.
\par
To construct the generalized antisymmetric multiquark
representations, we introduce (generally colored) tensor currents,
made of $(N_c-J)$ quark fields, or $(N_c-J)$ antiquark fields, where
$J=0,1,2,\ldots,(N_c-1)$, combined together with the antisymmetric
$\epsilon$ tensor, which now contains $N_c$ indices:
\bea
\lb{3e15}
& &j_{a_1a_2\cdots a_{J}^{}}^{(J,-)}=
\epsilon_{a_1a_2\cdots a_{J}^{}b_1b_2\cdots b_{N_c^{}-J}^{}}
\psi^{b_1}\psi^{b_2}\cdots\psi^{b_{N_c^{}-J}^{}},\\
\lb{3e16}
& &\overline j_{(J,-)}^{\ a_1a_2\cdots a_{J}^{}}=
\epsilon^{a_1a_2\cdots a_{J}^{}b_1b_2\cdots b_{N_c^{}-J}^{}}
\overline\psi_{b_1}^{}\overline\psi_{b_2}^{}
\cdots\overline\psi_{b_{N_c^{}-J}^{}}^{},
\eea
flavor indices being ignored. The case $J=0$ reproduces, up
to a multiplicative constant, the baryonic color singlet current
(\rf{2e16}).
\par
The equivalent forms of representation (\rf{3e3}) are then 
\be \lb{3e17}
T^{(J,-)}=\frac{1}{(N_c^{}-J)!}\
\overline j_{(J,-)}^{\ a_1a_2\cdots a_{J}^{}}
j_{a_1a_2\cdots a_{J}^{}}^{(J,-)},\ \ \ \ \ \ J=1,2,\ldots,(N_c-2).
\ee
The choice $J=(N_c^{}-2)$ reproduces the antisymmetric
representation of Eq.~(\rf{3e6}).
The choice $J=1$ corresponds to the grouping of the quark
fields into the antifundamental representation.
\par
The above currents do not exhaust all the possibilities of
constructing interpolating currents. One still has the
possibility of incorporating symmetric representations, as in
Eq.~(\rf{3e4}), which we omit here for simplicity.
\par
Other types of representation are the products of ordinary
mesonic currents, like in Eqs.~(\rf{3e1}) and (\rf{3e2}).
Using in Eqs.~(\rf{3e17}) contractions of the two $\epsilon$
tensors, as in Eq.~(\rf{3e5}), which are present in $j$ and
$\overline j$, one can reexpress the antisymmetric-type
tensor currents as combinations of products of such mesonic
bilinear currents, as in Eq.~(\rf{3e6}).
(In the case of one quark flavor, these reduce to powers of a
single current.) The same result is also obtained with
the symmetric representations. We notice, in particular, that
the simplest quadrilinear currents of the types of
Eqs.~(\rf{3e1}) and (\rf{3e2}), met in the SU(3) case, may continue
playing the role of interpolating currents in the SU$(N_c)$ case,
corresponding to the choice $J=N_c^{}-2$ in the antisymmetric
representation (\rf{3e17}) and in the corresponding symmetric one.
\par
It is the calculation of the couplings of the above currents to
the hypothetical tetraquark state that may provide an indication
about its internal structure. In the case of dominance of the
currents of the types (\rf{3e1})--(\rf{3e4}), the tetraquark will
have a very similar structure as in the SU(3) case. In the case of
dominance of the currents of the other extreme cases, like in
(\rf{3e17}) with $J$ close to $1$, the tetraquark will have a
many-body structure, as in the case of
baryons\footnote{For simplicity of language, we shall continue
using for the exotic states the same names as in SU(3),
independently of their internal structure.}.
\par
The fact that all tetraquark currents can be reexpressed as
combinations of products of ordinary mesonic currents is an
indication that they are color-reducible, unlike the currents
of ordinary mesons and baryons. This has the consequence, that, at
large $N_c$, the leading behavior of their correlation functions
is given by that of products of correlation functions of
ordinary mesonic currents, representing disconnected propagation of 
free mesons and not of tetraquarks
\cite{Witten:1979kh,Coleman:1985rnk,Jaffe:2008zz}. We consider here,
as an example, the case of the two-point function of the current
(\rf{3e1}). At large $N_c$, it behaves at leading order as
\be \lb{3e18}
\langle T_{\bar ij,\bar k\ell}^{(1,1)}(x)
T_{\bar ij,\bar k\ell}^{(1,1)\dagger}(0)\rangle
_{\stackrel{{\displaystyle=}}{N_c^{}\rightarrow\infty}}
\langle j_{\bar ij}^{}(x)j_{\bar ij}^{\dagger}(0)\rangle
\langle j_{\bar k\ell}^{}(x)j_{\bar k\ell}^{\dagger}(0)\rangle
=O(N_c^2),
\ee
where the $N_c$-behavior is obtained from Fig.~\rf{2f5}a.
This means that the search for tetraquark states in correlation
functions has to go beyond the leading order \cite{Weinberg:2013cfa}.
\par
For pentaquarks, the currents constructed as products of a mesonic
and a baryonic current are still valid, provided one uses for the
latter its expression of SU($N_c$) [Eqs.~(\rf{2e16}) and (\rf{3e15})]:
\be \lb{3e19}
P^{(1,1)}(x)=j(x)j^{(0,-)}(x).
\ee
\par
Generalizations of the antisymmetric tensor currents (\rf{3e12}) are:
\bea \lb{3e20}
P^{(J,K,-)}&=&\overline j_{(J+K,-)}^{\ a_1\cdots a_{J}^{},
b_1\cdots b_{K}^{}}\
j_{a_1a_2\cdots a_{J}^{}}^{(J,-)}\ j_{b_1b_2\cdots b_{K}^{}}^{(K,-)},
\nonumber \\
& &\ \ \ \ \ \ J,K=1,2,\ldots,(N_c-2),\ \ \ \ (J+K)\le (N_c-1).
\eea
\par
For hexaquarks, the analogs of representations (\rf{3e13}) are
\be \lb{3e21}
H^{(1,1)}(x)=j^{(0,-)}(x)j^{(0,-)}(x).
\ee
For the generalizations of the antisymmetric representations
(\rf{3e14}), we display here only the two extreme cases of interest:
\be \lb{3e22}
H^{(1,1,...,1,-)}=\epsilon^{a_1a_2\cdots a_{N_c^{}}}
j_{a_1}^{(1,-)}j_{a_2}^{(1,-)}\cdots j_{a_{N_c^{}}}^{(1,-)},
\ee
\be \lb{3e23}
H^{(1,1,N_c^{}-2,-)}=\epsilon^{a_1a_2\cdots a_{N_c^{}}}
j_{a_1}^{(1,-)}j_{a_2}^{(1,-)}j_{a_3\cdots a_{N_c^{}}}^{(N_c^{}-2,-)}.
\ee
\par
Like the tetraquark currents, pentaquark and hexaquark currents are
decomposable along combinations of products of ordinary hadronic
currents; their two-point functions satisfy properties similar to
that of Eq.~(\rf{3e18}).
\par
Graphical representations of the currents introduced in the present
subsection will be presented in Sec.~\rf{s33}, in the 
more general case of multilocal operators.
\par
Multiquark-state currents of the types introduced above have been
considered and studied in Refs.~\cite{Cohen:2014vta,Maiani:2018tfe}.
\par

\subsection{Multilocal operators} \lb{s33}

The description of multiquark states may also necessitate in some
instances the use of more general probes than local currents.
Bound-state equations require the use of multilocal fields.
Lattice gauge theory, which works in a discretized spacetime,
is another instance where the theory is formulated from the start
by means of such operators. It is therefore necessary to find the
corresponding generalizations of the various currents that we met
in our previous study. We shall focus our attention on representations
that preserve the gauge invariance of the theory.
\par
Gauge-invariant operators are constructed by using path-ordered
gluon-field phase factors \cite{Mandelstam:1962mi,
BialynickiBirula:1963,Mandelstam:1968hz,Nambu:1978bd}, also called
gauge links or Wilson lines, having the form
\be \lb{3e24}
U_{\ b}^a(C_{yx}^{})=\Big(Pe^{{\displaystyle ig\int_{C_{yx}^{}}
dz^{\mu}T^BA_{\mu}^B(z)}}\Big)_{\ b}^a,
\ee
where $C_{yx^{}}$ is an oriented curve going from $x$ to $y$
and $P$ represents the path-ordering of the gluon fields
according to their position on the line $C_{yx^{}}$; the
integration goes from $x$ to $y$ along that line. The phase
factors $U(C_{yx}^{})$ are the color parallel transporters along
the lines $C_{yx}^{}$ \cite{Corrigan:1978zg}. 
\par
The color-trace operation on a phase factor taken along a closed
contour $C_{xx}$ defines a gauge-invariant operator, called the
Wilson loop. Its vacuum expectation value plays an important
role in defining gauge-invariant static potential energies
\cite{Wilson:1974sk,Brown:1979ya,Kogut:1982ds}.
\par
Gauge-invariant operators coupling to mesons and baryons (here
for the group SU(3)) are\footnote{In this and the following subsections,
for ease of pictorial representations, the quark fields are
designated by the notation $q$, rather than $\psi$.}  
\be \lb{3e25}
M=\overline q_a^{}(y)U_{\ b}^a(C_{yx}^{})q^b(x),
\ee
\be \lb{3e26}
B=\frac{1}{3!}\epsilon_{abc}\ U_{\ d}^a(C_{xy}^{})q^d(y)\
U_{\ e}^b(C_{xt}^{})q^e(t)\ U_{\ f}^c(C_{xz}^{})q^f(z),
\ee
where flavor and spin indices have been omitted. 
A pictorial representation of them, with phase factor lines
chosen along straight line segments, is given in Fig.~\rf{3f1}.
Meson and baryon local currents [Eqs.~(\rf{2e8}) and (\rf{2e15}),
respectively] are obtained (up to the defining multiplicative
constants) by concentrating the quark and antiquark coordinates at
single points and by shrinking in the latter expressions the phase
factors to 1. 
\bfg
\bc
\epsfig{file=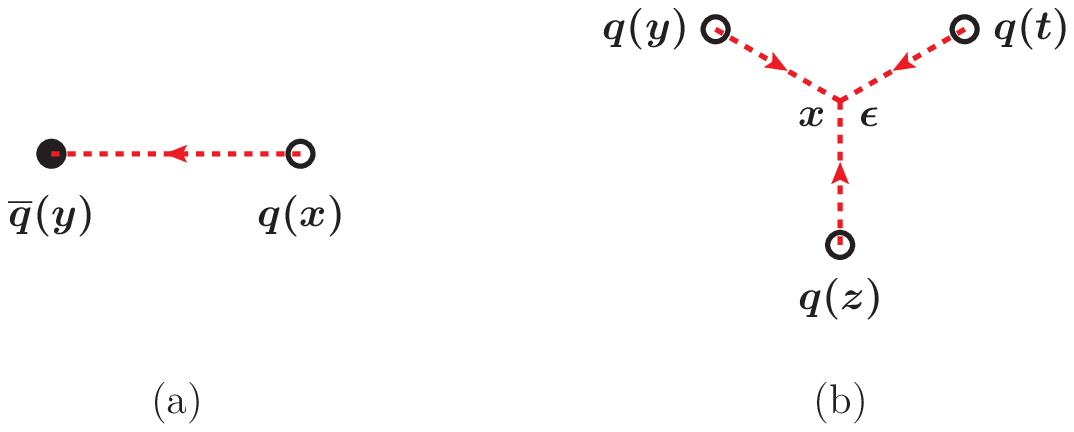,scale=0.8}
\caption{Pictorial representation of the gauge-invariant meson
(a) and baryon (b) operators; in the baryon case, $\epsilon$ is the
completely antisymmetric Levi-Civita tensor, indicating the
antisymmetric property of the three-line vertex.}
\lb{3f1}
\ec
\efg
\par
As a general remark, let us emphasize that physical properties of
states should be independent of the shape of the lines $C_{yx^{}}$,
provided they are continuous and smoothly varied upon
deformations from straight lines. The latter are generally chosen
for their simplicity and for their adequacy in lattice calculations
\cite{Wilson:1974sk}. The above property can be verified in the
case of bound-state energies, which are obtained from the behavior
of Wilson-loop vacuum averages at large time separations. In QCD, 
the Wilson-loop vacuum average is expected to satisfy in that limit
the area law and, more generally, the minimal surface property
\cite{Wilson:1974sk,Makeenko:1980wr,Jugeau:2003df}. Smooth
deformations of the phase-factor lines inside the bound-state
definition do not change its energy, but only affect
the expression of the wave function (cf.~Sec.~\rf{s72} below and
Ref.~\cite{Jugeau:2003df}, Appendix A).
\par
Similar constructions can also be applied to the multiquark states.
They were promoted in the past by Rossi and Veneziano
\cite{Rossi:1977cy,Rossi:2016szw} and are called ``string-junction''
or ``$Y$-shaped-junction'' type representations. They are mainly
considered for the antisymmetric representations, typical of the
diquark picture \cite{Jaffe:2003sg,Shuryak:2003zi,Maiani:2004vq},
in which the interquark forces are expected to be attractive,
leading to the emergence of multiquark bound states. Pictorial
representations of these are given in Fig.~\rf{3f2}.
The local multiquark antisymmetric representation currents
[Eqs.~(\rf{3e3}), (\rf{3e12}) and (\rf{3e14})] are obtained
(up to the defining multiplicative constants) 
by concentrating quark and antiquark coordinates at single
points and by shrinking the phase factors to 1.
\begin{figure}[t]
\bc
\epsfig{file=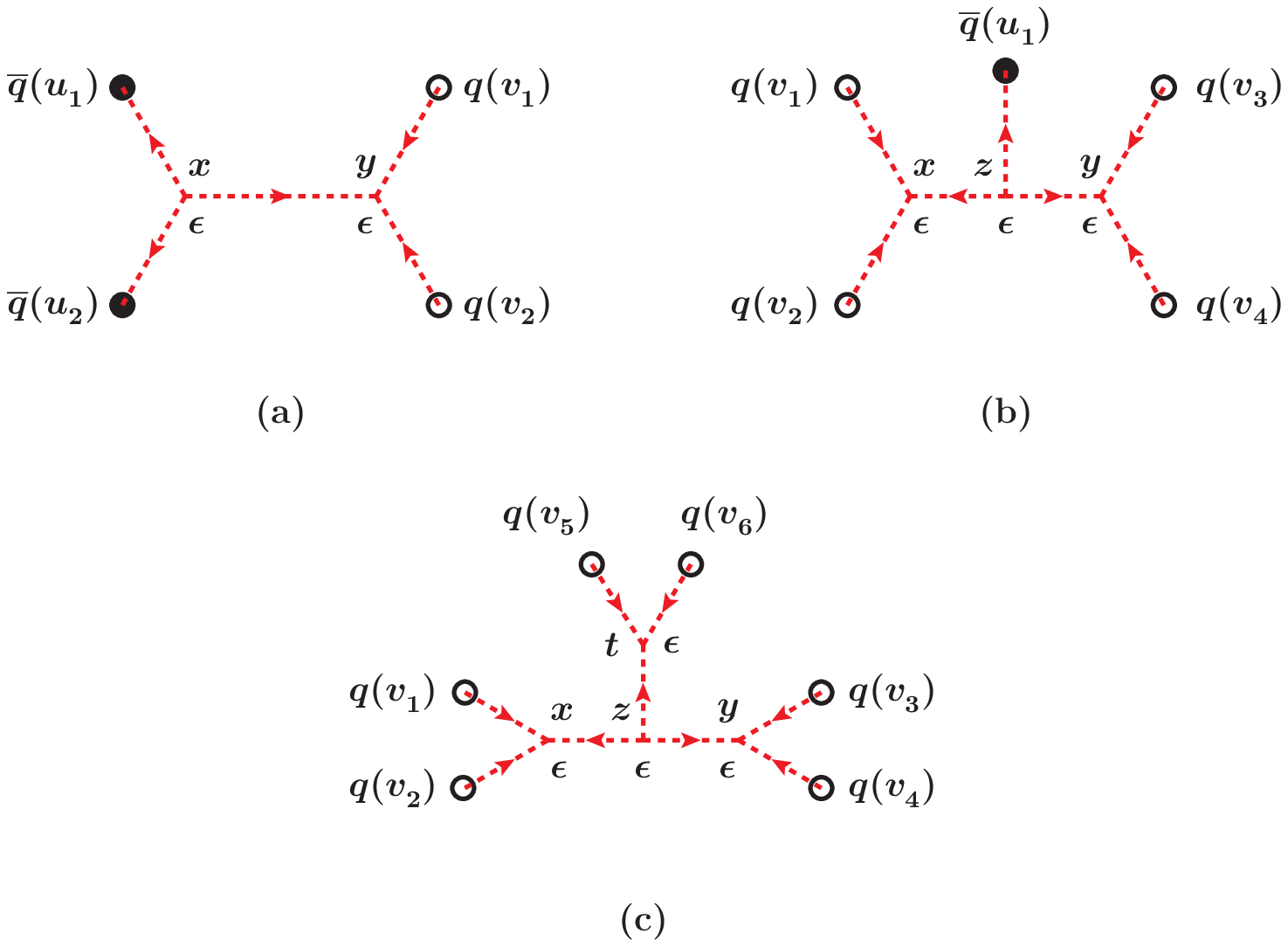,scale=0.7}
\caption{Pictorial representation of the gauge-invariant
(a) tetraquark, (b) pentaquark, and (c) hexaquark operators.}
\lb{3f2}
\ec
\end{figure}
\par
Generalizations of the previous operators to the SU($N_c^{}$)
case are straightforward, following the constructions of the
corresponding local currents [Eqs. (\rf{2e8}), (\rf{2e16}),
(\rf{3e17}), (\rf{3e20}), (\rf{3e22}) and (\rf{3e23})].
The case of mesons and baryons is displayed in Fig.~\rf{3f3}.
\bfg
\bc
\epsfig{file=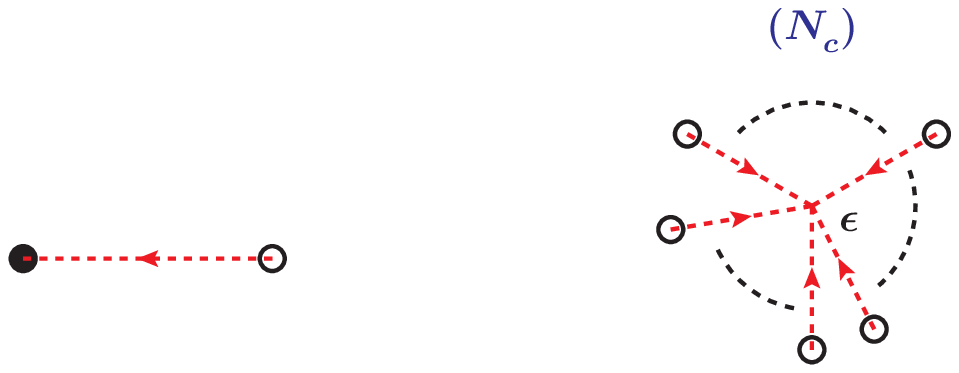,scale=0.7}
\caption{Meson and baryon operators in the SU($N_c^{}$) case.}
\lb{3f3}
\ec
\efg
Tetraquark, pentaquark and hexaquark operators are graphically
represented in Figs.~\rf{3f4}, \rf{3f5} and \rf{3f6}.
\bfg
\bc
\epsfig{file=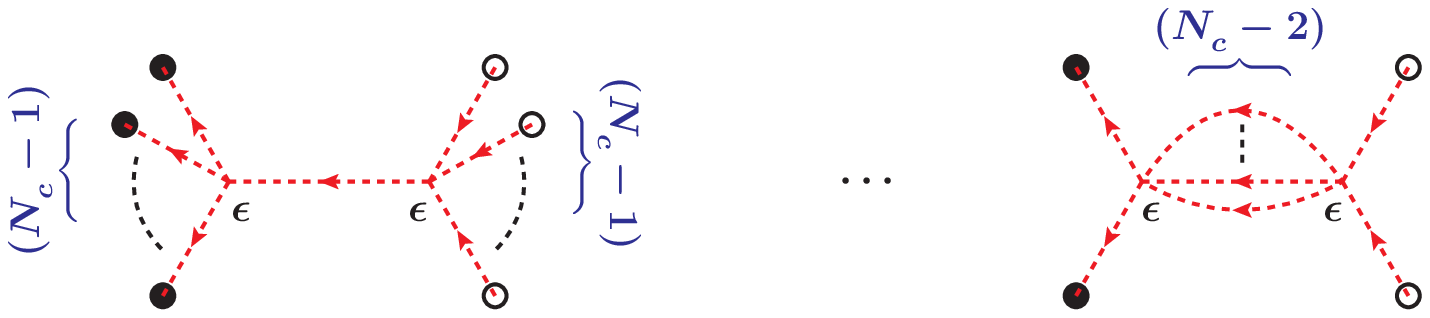,scale=0.8}
\caption{Tetraquark operators in the SU($N_c^{}$) case. The two
extreme cases, with $J=1$ and $J=(N_c^{}-2)$, corresponding to Eq.
(\rf{3e17}), are displayed. The first diagram contains $(N_c^{}-1)$
quarks and $(N_c^{}-1)$ antiquarks, with a single link between the
two string junctions. The last diagram contains two quarks and two
antiquarks, with $(N_c^{}-2)$ links between the string junctions.}
\lb{3f4}
\ec
\efg
\bfg
\bc
\epsfig{file=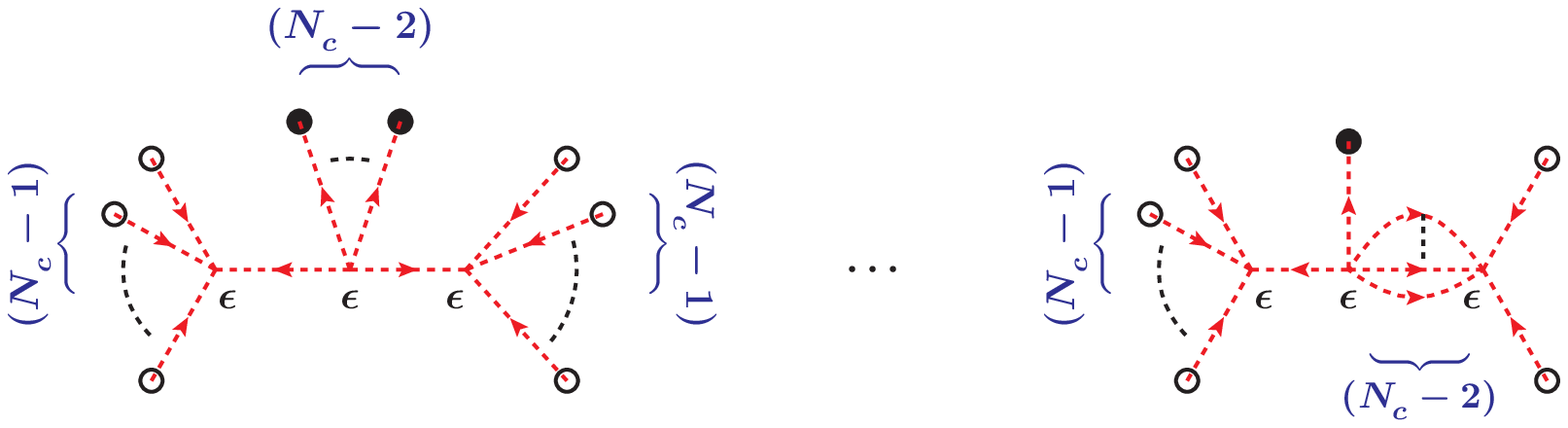,scale=0.7}
\caption{Pentaquark operators in the SU($N_c^{}$) case. Two
extreme cases, with $J=1$, $K=1$ and $J=1$, $K=(N_c^{}-2)$,
corresponding to Eq.~(\rf{3e20}), are displayed. The first diagram
contains $2(N_c^{}-1)$ quarks and $(N_c^{}-2)$ antiquarks.
The last diagram contains $(N_c^{}-1)+2$ quarks and one antiquark.}
\lb{3f5}
\ec
\efg
\bfg
\bc
\epsfig{file=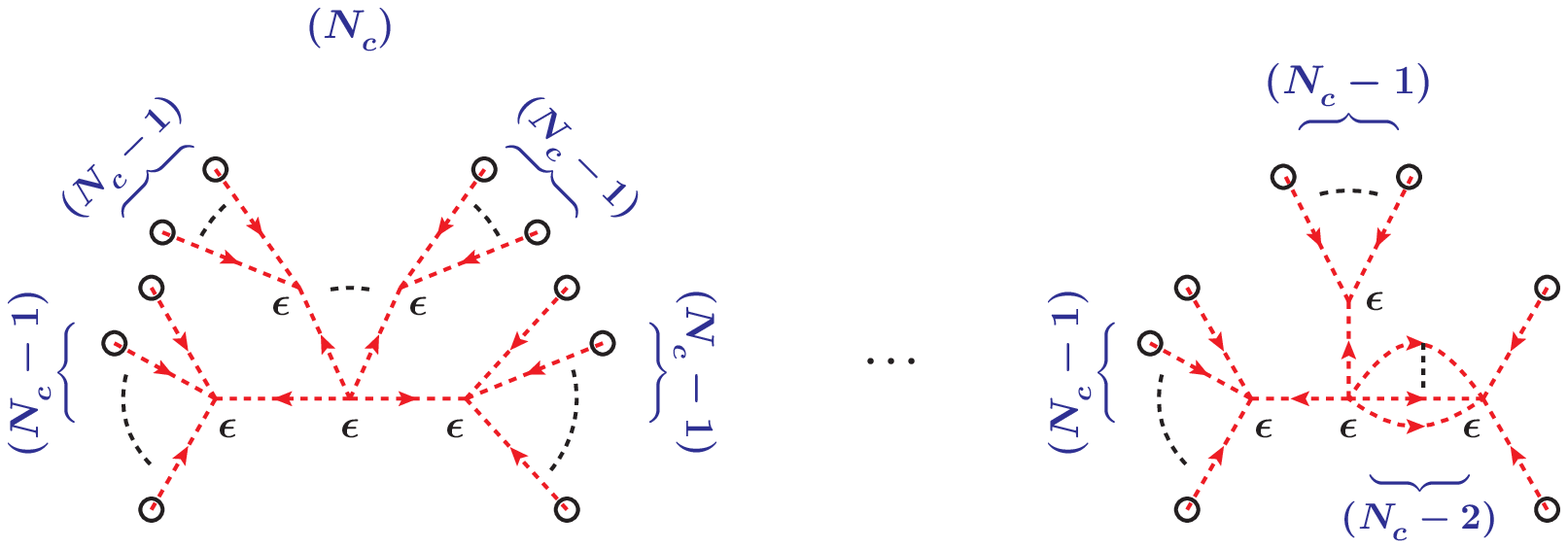,scale=0.7}
\caption{Hexaquark operators in the SU($N_c^{}$) case. Two
extreme cases, corresponding to Eqs.~(\rf{3e22}) and (\rf{3e23}),
are displayed. The first diagram contains $N_c(N_c^{}-1)$ quarks.
The last diagram contains $2(N_c^{}-1)+2$ quarks.}
\lb{3f6}
\ec
\efg
\par
In summary, multiquark states can generally be described, or
theoretically probed, by several operators, each highlighting
a particular aspect of the state under study. In passing to large
$N_c$, the number of these operators increases and the structure
of the multiquark states may become more complicated. One however
hopes that only few of them will represent the dominant
representative scheme, which might correspond to the outcome of
more dynamical investigations.
\par


\section{Singularities of Feynman diagrams connected with 
multiquark states} \lb{s4}

While the large-$N_c$ limit approach is a method aiming to explore
the properties of the theory in its nonperturbative regime,
here, for QCD, in its confining regime, it still hinges, as
we have seen in Sec.~\rf{s2}, on the analysis of Feynman diagrams,
which are representative of the perturbative regime of the
theory. Although this might seem contradictory, it should be
emphasized that one is not considering a single or a finite number
of Feynman diagrams, but rather classes of Feynman diagrams which
are distinguished by their topological properties in color space.
Thus, in Fig.~\rf{2f5}, the diagrams (a), (b) and (c) are parts of
the same class of planar diagrams, depicting the two-point
correlation function of meson currents, having the same large-$N_c$
behavior. This class contains an infinite number of other diagrams,
involving many-gluon exchanges, but the same number (viz., two) of
quark lines. The large-$N_c$ approach assumes that the infinite sum
of diagrams contained in this topological class produces the
bound states of mesons and maintains the confining property of
the theory \cite{'tHooft:1973jz,Witten:1979kh,Coleman:1985rnk}.
\par
Each Feynman diagram participating in the above summation 
process, though not explicitly displaying confining properties
or bound-state attri\-butes, should carry a minimum amount of
common qualitative features with the other diagrams in order to
produce at the end the desired nonperturbative effects. In the
example of the two-point function of the meson currents given
above, it is the number of quark lines which is common to all
the summed diagrams. It is this number that allows the introduction
of the notion of valence quarks. This is then manifested in
each Feynman diagram through the singularity structure in momentum
space, represented by a discontinuity in the total invariant mass
squared, the so-called Mandelstam $s$-variable, starting from the
two-quark threshold and going to infinity. The summation of diagrams
transforms this singularity into a series of meson poles.
\par
The same procedure also applies to the two-point functions of the
baryonic currents. For $N_c=3$, it is the Feynman diagrams with
three quark lines which should be representative of the leading
valence-quark structure. When the large-$N_c$ limit is taken, the
number of valence quarks for baryons increases with $N_c$, however,
the singularity content of each diagram should remain, in that
a threshold singularity in the $s$-variable should be present and
should lead after summation to a series of baryon poles.
\par
In passing to the case of exotic (multiquark) states, one expects
a generalization of the above phenomenon. An exotic state with
a number $A$ of valence quarks should be represented, at
leading order in $N_c$, by diagrams containing a number $A$ of
quark lines. Here, however, two kinds of difficulties emerge,
which were not present in the case of ordinary hadrons. 
First, Feynman diagrams with $A$ quark lines may contain
color-singlet disconnected pieces; this is a consequence of the
fact that the multiquark currents are generally expressible as
combinations of products of ordinary currents (cf. Sec.~\rf{s3},
Eqs.~(\rf{3e1}), (\rf{3e2}), (\rf{3e11}), (\rf{3e13}) and (\rf{3e18})).
Such diagrams, which represent propagation of free particles, cannot
participate in the formation of bound states and hence should not be
taken into account. Second, there are still connected Feynman diagrams,
having $A$ quark lines, which do not possess singularities in the
$s$-variable. Their singularities concern the $u$- or $t$-variables
and therefore cannot not participate in the multiquark pole
production process and should not be considered.
\par
In summary, the counting of quark lines in a given Feynman diagram
is no longer sufficient for its consideration in the formation
process of the multiquark bound state. A more precise criterion,
based on the analysis of the singularity structure in the
$s$-variable, is necessary. This criterion is provided by the
Landau equations \cite{Landau:1959fi,Itzykson:1980rh}, which
allow one to analyze in more detail the singularity properties
of Feynman diagrams. We shall briefly sketch below the Landau
equations and shall consider a few typical examples which will
be helpful in the analyses of multiquark-state properties.
\par
According to its quark-flavor structure (if one quark and one
antiquark are of the same flavor), a multiquark state may
have a mixing with an ordinary meson or baryon state. It is
understood that to receive the multiquark label, the multiquark
component having $A$ valence quarks is part of the $N_c$-leading
components of the total state. Otherwise, any ordinary hadron state
has multiquark-type components, due to the sea quarks, which are
parts of the $N_c$-subleading components. Therefore, independently
of possibly existing mixings with ordinary hadron states, it is
the structure of the multiquark component, assumed to provide, at
least partly, the $N_c$-leading behavior of the state, which is
the key ingredient of the present analysis.
\par
A generic expression of a Feynman diagram is
\be \lb{4e1}
I(p)=\int\prod_{\ell=1}^{L}\frac{d^4k_{\ell}^{}}{(2\pi)^4}
\prod_{i=1}^{I}\frac{1}{(q_i^2-m_i^2+i\epsilon)},
\ee
where $p$ represents a collection of external momenta and $q_i$
($I$ in number) are linear functions of the $p$s and of the loop
variables $k_{\ell}^{}$ ($L$ in number).
\par
The Landau equations are
\bea
\lb{4e2}
& &\lambda_i^{}(q_i^2-m_i^2)=0,\ \ \ \ \ \ \ i=1,\ldots,I,\\
\lb{4e3}
& &\sum_{i=1}^I\lambda_i q_i^{}\cdot\frac{\partial q_i^{}}
{\partial k_{\ell}^{}}=0,\ \ \ \ \ \ \ \ \ell=1,\ldots,L,
\eea  
where the $\lambda$s are Lagrange multipliers to be
determined. Some of the parameters $\lambda$ may vanish or may be
compatible with vanishing values. 
\par
We are mainly interested in the location of the singularities
produced by the quark propagators. Gluons being massless, the
singularities of their propagators generally start at the same
positions as those produced by the quark propagators. We therefore
shall not consider, in general, gluon propagators in the Landau
equations as independent sources of singularities; this is
realized by putting, from the start, the corresponding $\lambda$s
equal to zero. However, gluon lines may participate in the production
of quark singularities through the momentum they carry.
\par
Since multiquark states are expected to decay into ordinary
hadrons, or to have couplings with them in the case of bound
states, it is easier to study their properties through the
scattering amplitudes of ordinary hadrons and the corresponding
Feynman diagrams.
\par
We consider, for definiteness, the case of tetraquarks made of four
quarks with different flavors; the quarks will be designated by
indices 1 and 3 and the antiquarks by indices $\overline 2$
and $\overline 4$. As we have seen in Sec.~\rf{s32}, such a
description still remains valid for general $N_c$, although in the
latter case other representations also emerge (cf.~Fig.~\rf{3f4}).
Within the present representation, the tetraquark may couple to
two mesons and therefore may be probed in two-meson scattering
processes. Using the bilinear currents defined in Eq.~(\rf{2e8}),
one may consider, in momentum space, Fourier transforms of the
correlation functions 
\bea
\lb{4e4}
\lefteqn{\hspace{-0.3 cm}\Gamma_{D1}^{}\equiv
\langle j_{\overline 21}^{}(x)j_{\overline 43}^{}(y)
j_{\overline 43}^{\dagger}(z)j_{\overline 21}^{\dagger}(0)\rangle,
\ \ \ \Gamma_{D2}^{}\equiv
\langle j_{\overline 41}^{}(x)j_{\overline 23}^{}(y)
j_{\overline 23}^{\dagger}(z)j_{\overline 41}^{\dagger}(0)\rangle,}\\
\lb{4e5}
& &{\hspace{-1 cm}\Gamma_{R1}^{}\equiv
\langle j_{\overline 21}^{}(x)j_{\overline 43}^{}(y)
j_{\overline 41}^{\dagger}(z)j_{\overline 23}^{\dagger}(0)\rangle,
\ \ \ \Gamma_{R2}^{}\equiv
\langle j_{\overline 41}^{}(x)j_{\overline 23}^{}(y)
j_{\overline 21}^{\dagger}(z)j_{\overline 43}^{\dagger}(0)\rangle,}
\eea
where the subscript $D$ refers to direct-channel processes and the
subscript $R$ to quark recombination or rearrangement channel processes.
\par
We first consider two typical diagrams of the direct channel 1 process
-- a disconnected diagram and a connected one -- represented in
Figs.~\rf{4f1}a and b, respectively. (The diagram with one-gluon
exchange between the two disconnected color-singlet diagrams is zero.)
The total conserved momentum of the two-meson processes is designated
by $P$, with the usual definition $s=P^2$.
\par
\bfg
\bc
\epsfig{file=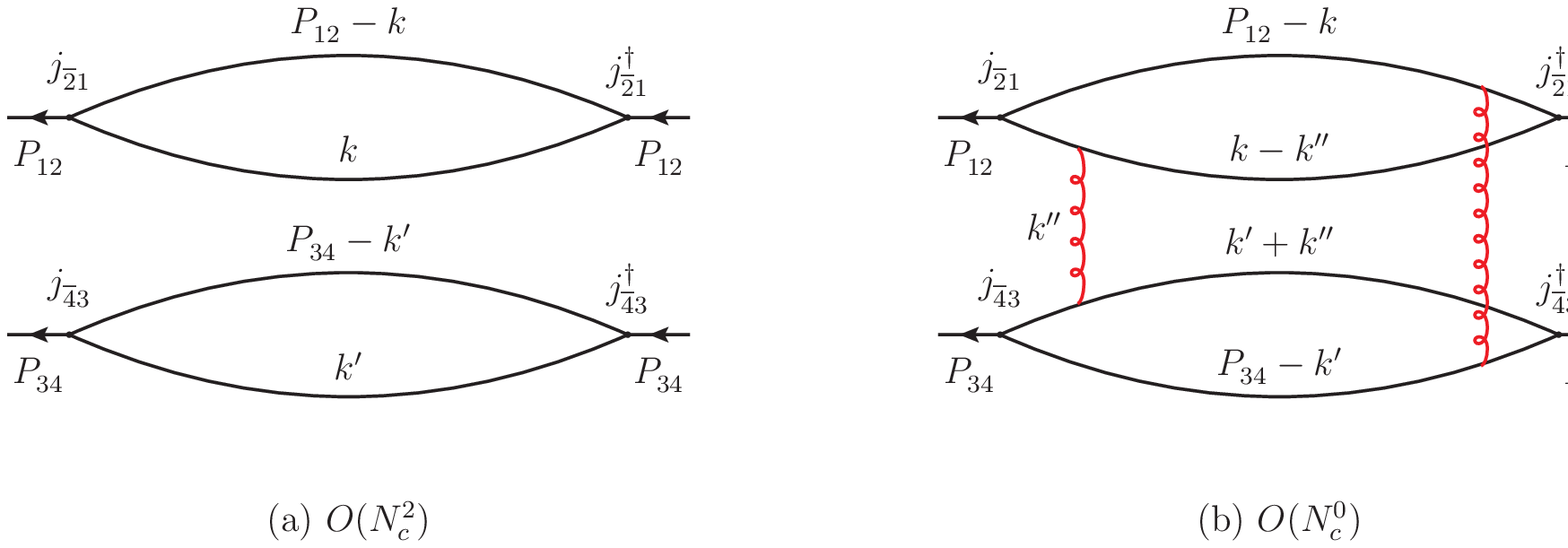,scale=0.65}
\caption{Disconnected (a) and connected (b) diagrams in the direct
channel $M_{\overline 21}^{}M_{\overline 43}^{}\rightarrow
M_{\overline 21}^{}M_{\overline 43}^{}$ of the meson-meson scattering
amplitude.}
\lb{4f1}
\ec
\efg
\par
The Landau equations of the disconnected diagram are themselves
separable into two independent subsets, leading to the physical
singularities $P_{12}^2=(m_1^{}+m_2^{})^2$ and
$P_{34}^2=(m_3^{}+m_4^{})^2$, which do not involve $s$. These
singularities refer to the internal structure of each meson, which
propagates freely and independently from the other. It is evident
that this diagram cannot be involved in the formation of a bound
state or of a resonance. One can also consider all other planar
diagrams which include the gluon exchanges inside each quark loop,
associated with the above disconnected diagram, still finding the
same singularities as above, confirming the fact that gluon
propagators generally do not modify the location of singularities
found with the sole quark propagators.
\par
For the connected diagram of Fig.~\rf{4f1}, it is sufficient to
consider a vertical cut passing between the two gluon lines. The
Landau equations are then:
\bea \lb{4e6}
& &\lambda_1^{}((P_{12}^{}-k)^2-m_1^2)=0,\ \ \ \ \ \ 
\lambda_2^{}((k-k'')^2-m_2^2)=0,\nonumber \\
& &\lambda_3^{}((k'+k'')^2-m_3^2)=0,\ \ \ \ \ \ 
\lambda_4^{}((P_{34}^{}-k')^2-m_4^2)=0,
\eea
\bea \lb{4e7}
& &-\lambda_1^{}(P_{12}^{}-k)+\lambda_2^{}(k-k'')=0,\ \ \ \ 
-\lambda_2^{}(k-k'')+\lambda_3^{}(k'+k'')=0,\nonumber \\
& &\lambda_3^{}(k'+k'')-\lambda_4^{}(P_{34}^{}-k')=0.
\eea
The following definitions hold:
\bea \lb{4e8}
& &P=P_{12}^{}+P_{34}^{}=P_{12}^{\prime}+P_{34}^{\prime},\ \ \ \ 
s=P^2,\nonumber \\
& &t=(P_{12}^{}-P_{12}^{\prime})^2,\ \ \ \
u=(P_{12}^{}-P_{34}^{\prime})^2.
\eea
The system of equations (\rf{4e6}) and (\rf{4e7}) can be solved,
leading to the physical singularity at $s=(\sum_{i=1}^4m_i^{})^2$.
The fact that the four quark masses are present means that we
have four-quark intermediate states, which, together with the
contributions of other diagrams involving more gluon lines,
will generate two-interacting-meson states and possibly tetraquark
states.
\par
We next consider, in Fig.~\rf{4f2}, two diagrams of the recombination
channel 1. Here, the following definitions hold: 
\bea \lb{4e9}
& &P=(P_{12}^{}+P_{34}^{})=(P_{14}^{\prime}+P_{23}^{\prime}),\ \ \ \ 
s=P^2,\nonumber \\
& &t=(P_{12}^{}-P_{14}^{\prime})^2,\ \ \ \
u=(P_{12}^{}-P_{23}^{\prime})^2.
\eea
\bfg
\bc
\epsfig{file=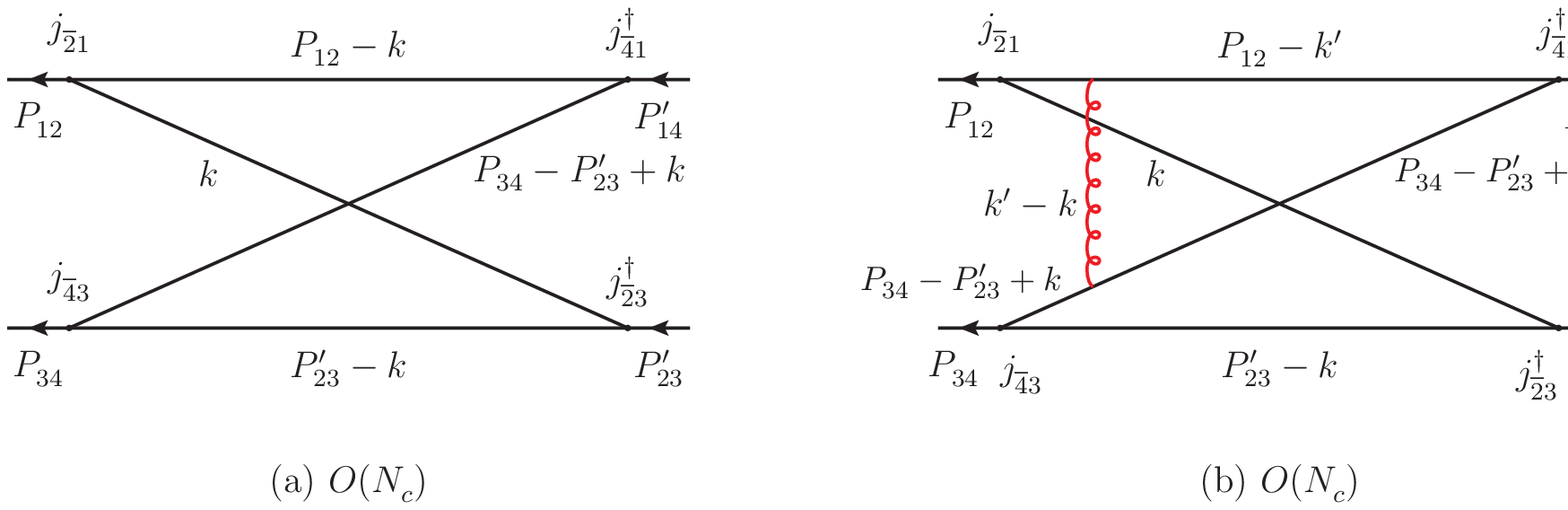,scale=0.65}
\caption{Diagrams in the quark-rearrangement channel
$M_{\overline 41}^{}M_{\overline 23}^{}\rightarrow
M_{\overline 21}^{}M_{\overline 43}^{}$ of the meson-meson scattering
amplitude, not concerned, at large $N_c$, with tetraquark states.}
\lb{4f2}
\ec
\efg
\par
The Landau equations of diagram (a) have several
subsets of physical solutions: $t=(m_2^{}+m_4^{})^2$,
$u=(m_1^{}+m_3^{})^2$, $P_{12}^2=(m_1^{}+m_2^{})^2$, etc., but
no singularities in $s$ are found. The $t$- and $u$-channel
singularities will be saturated, with similar diagrams involving
gluon lines, by one-meson states. The singularities in the
external momenta squared are those that are present in the
external meson propagators. The Landau equations of diagram (b) 
also lead to the same sets of singularities as diagram (a).
Therefore, the diagrams of Fig.~\rf{4f2}, which
apparently display four quark lines, do not participate in the
formation of two-meson interacting systems, nor to possibly
existing tetraquark states. Nevertheless, they produce, in the
mesonic world, contact-type interactions, as well as one-meson
exchange diagrams.
\par
One may have another view of the preceding results, by referring
to the topological properties of the diagrams in color space.
The latter are planar diagrams and can be unfolded, as suggested
in Ref.~\cite{Cohen:2014tga}, to make explicit the color flow.
The unfolded plane corresponds now to the $(u,t)$ plane (Fig.
\rf{4f3}). The $t$- and $u$-channel singularities are obtained
by cutting the box diagrams by horizontal and vertical lines,
respectively. It is evident, here, that the corresponding
singularities are two-quark singularities, typical of one-meson
states. The $s$-channel singularities are obtained by cutting
the box diagrams by oblique and curved lines passing through the
four quark propagators. However, when the diagram is color-planar,
as is the case here, the cuts produce disconnected singularities,
concentrated at opposite corners and corresponding to radiative
corrections of external-meson propagators or of current vertices.
$s$-channel singularities may arise only when the diagram is
color-nonplanar.
\par
\bfg
\bc
\epsfig{file=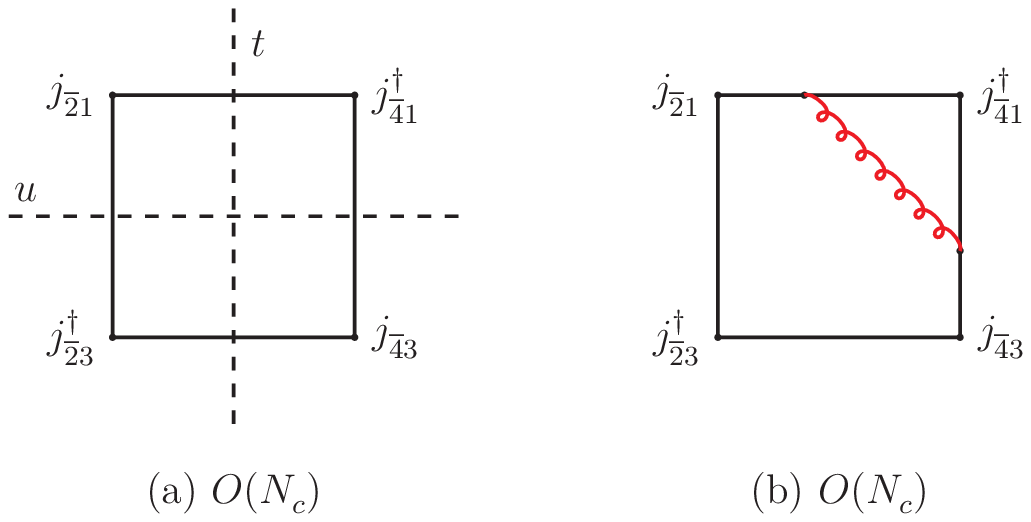,scale=0.7}
\caption{Diagrams of Fig. \rf{4f2} in unfolded form.}
\lb{4f3}
\ec
\efg
\par
A typical color-nonplanar diagram, which contributes to the
$s$-channel singularities, is presented in Fig.~\rf{4f4},
together with its unfolded form, where the nonplanarity is manifest.
\bfg
\bc
\epsfig{file=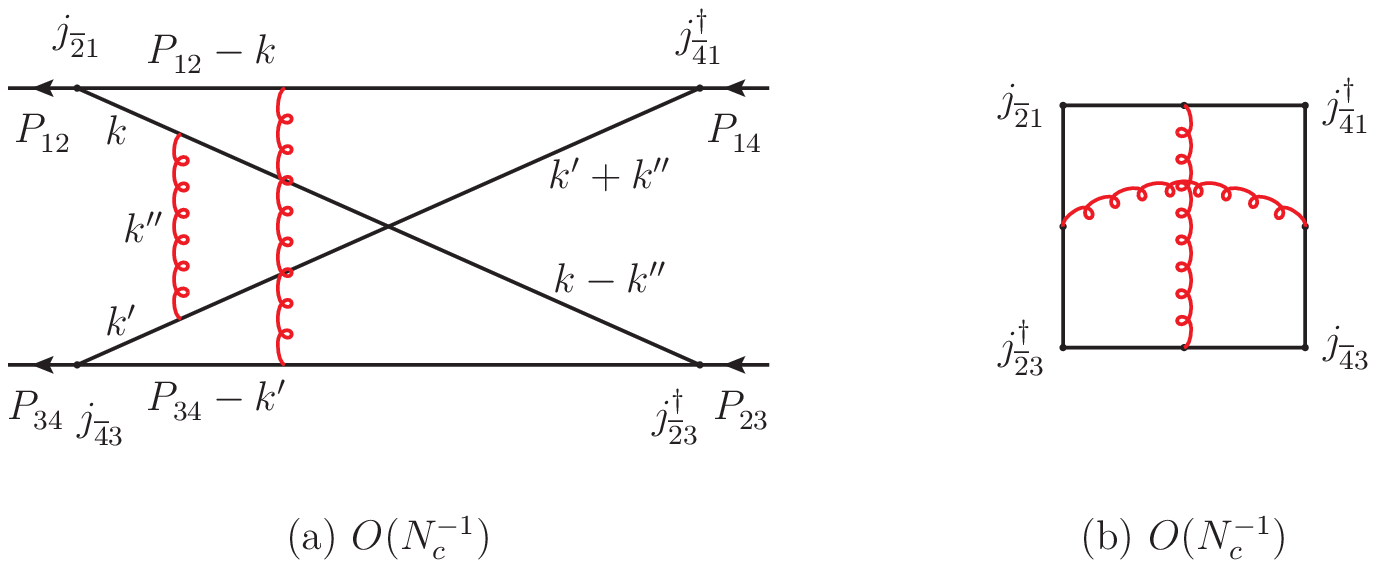,scale=0.7}
\caption{(a) Diagram in the quark-rearrangement channel
$M_{\overline 41}^{}M_{\overline 23}^{}\rightarrow
M_{\overline 21}^{}M_{\overline 43}^{}$ of the meson-meson scattering
amplitude, participating, at large $N_c$, in the formation of
possible tetraquark states. (b) The same diagram in the unfolded
$(u,t)$ plane, displaying its color-nonplanar character.}
\lb{4f4}
\ec
\efg
\par
The efficient Landau equations are obtained by cutting the
diagram (a) of Fig.~\rf{4f4} by a vertical line passing between
the two gluon lines. One now finds a physical singularity in the
$s$-channel at the position $s=(\sum_{i=1}^4m_i^{})^2$, as in
the case of Fig. \rf{4f1}b. Therefore, this diagram, together
with other diagrams of the same color-topological class, will
contribute to the formation of two-meson interacting states and
eventually to that of tetraquark states.
\par
It is worthwhile noticing the difference of behavior in $N_c$
of the diagram (b) of Fig.~\rf{4f1} and the diagram (a) of
Fig.~\rf{4f4} -- $O(N_c^0)$ for the first and $O(N_c^{-1})$ for the
second -- contributing to the formation of tetraquarks in the direct
and recombination channels, respectively. This outlines the
color-topological difference that exists between them: the former is
planar, while the latter is nonplanar, a feature that may have
consequences for the various couplings of tetraquarks to two-meson
states.
\par
Other examples or details of Landau equations can be found in
Ref.~\cite{Lucha:2017gqq}.
\par

\section{Tetraquarks at large $N_c$} \lb{s5}

This section is devoted to a detailed study of some of the properties
of tetraquarks at large $N_c$. We have seen, in Secs.~\rf{s32} and
\rf{s33}, that tetraquarks at large $N_c$ may be described by
$(N_c-2)$ inequivalent classes of operators having different numbers
of valence quarks [Eq.~(\rf{3e17})], generically each having $(N_c-J)$
valence quarks and $(N_c-J)$ valence antiquarks, where $J$ takes
values from 1 to $(N_c-2)$. In multilocal form, they are pictorially
represented, for the antisymmetric representation, in Fig. \rf{3f4}.
For $J=(N_c-2)$, the representation is similar, in local form, to that
of SU(3): the tetraquark continues to be described by two pairs of
valence quarks and antiquarks. In particular, the local currents
(\rf{3e1}), (\rf{3e2}) and (\rf{3e6}) continue representing sources of
tetraquarks. This situation is very similar to that of ordinary
mesons, for which the extension of the gauge group to SU($N_c$) does
not change their description. For the representations with
$1\le J\le (N_c-3)$, one has new descriptions and, for values of
$J$ close to 1, the latter become similar to the description found for
baryons: the number of valence quarks and antiquarks grows with $N_c$
and presumably also the mass of the tetraquarks. Here, the tetraquark
becomes rather a many-body object, requiring a different type of
treatment. It is not clear, for the time being, which representation
provides the most faithful description in the limit $N_c=3$.
The most general situation would correspond to a mixture of all the
above representations.
Since, however, the representation with $J=(N_c-2)$ does not require
any modification of treatment, it remains the most practical one from
the mathematical and phenomenological viewpoints. This is why we shall
concentrate, in the present section and review, on this representation.
The reader may consult Refs.~\cite{Cohen:2014vta,Maiani:2018tfe} for
a detailed account of the properties of tetraquarks in higher
representations.
\par
One particular feature of tetraquark and multiquark currents, met in
Sec.~\rf{s3}, is their decomposition property into combinations of
products of ordinary mesonic or baryonic currents, reflecting their
color reducibility. At large $N_c$, two-point functions of tetraquark
currents are dominated by the contributions of their disconnected
parts [Eq.~(\rf{3e18})]. This fact has led Witten and Coleman to
conclude that multiquark states should not exist in QCD, at least
as confined states \cite{Witten:1979kh,Coleman:1985rnk}.
Actually, as was emphasized by Weinberg \cite{Weinberg:2013cfa}, the
situation may be more complex. Dynamical effects may still be at
work, preventing the multiquark states from being dissociated into
their elementary mesonic or baryonic  clusters. In such a case, the
multiquark state will appear as a pole, or a narrow resonance, in
nonleading terms of the $1/N_c$ expansion. Contrary to the case
of ordinary mesons and baryons, in order to detect the conditions
in which multiquark states may appear, one has to go, in the
$1/N_c$ expansion of correlation functions of currents, beyond the
leading orders. Studies in this line of approach can be found in
Refs.~\cite{Knecht:2013yqa,Cohen:2014tga,Maiani:2016hxw,
Maiani:2018pef,Lucha:2017mof,Lucha:2017gqq,Lucha:2018dzq,
Lucha:2020vgf}.
\par
Since the tetraquark state can couple, in the present representation,
to two mesonic currents, it can naturally be probed in meson-meson
scattering amplitudes, appearing as a possible pole or a resonance.
We therefore shall study the tetraquark properties at large $N_c$
through the $N_c$-leading or subleading typical Feynman diagrams
that may contribute to its emergence.
We shall first concentrate, in the following, on the case of fully
exotic tetraquarks, containing four different quark flavors. This
has the advantage of excluding mixings with ordinary mesons, which
often may prevent one from drawing a clear conclusion. The case
of cryptoexotic tetraquarks will be considered afterwards. 
\par
In searching for tetraquark poles, the following features should
also be taken into account. Tetraquark poles in the $s$-channel
may emerge in two different ways. The first possibility may take
place within a class of diagrams having the same $N_c$-leading
behavior, that is, belonging to the same color-planarity, an example
of which is the class of planar diagrams. Here one sums a series of
gluon propagators which might produce the expected pole. Such a
mechanism would be representative of the production of a
bound-state spectrum through the confining forces, as is
already at work for the case of ordinary mesons. One important
property of this mechanism is that, once the global $N_c$-dependence
of the diagrams has been factorized, the bound-state equation
becomes $N_c$-independent (at leading order), implying
$N_c$-independence of the corresponding masses.
\par
The second type of mechanism is related to the summation of
diagrams not having the same $N_c$-leading behaviors. This happens
when one considers, as a starting point, diagrams not having
four-quark $s$-channel singularities, eventually having $t$- and
$u$-channel singularities, representing ordinary-meson exchanges,
or, in the simplest cases, four-meson contact terms. Examples of
these cases are presented in the forthcoming Figs.~\rf{5f1}
and \rf{5f2}. Such diagrams play the role of kernels in
bound-state equations and generate, with the aid of the four quark
propagators, a series of diagrams whose summation may lead to
the emergence of poles. There are, however, two main differences
with respect to the first type of mechanism outlined above: First,
these kernels, as can be checked from Figs.~\rf{5f1} and \rf{5f2},
are of order $O(N_c^{-2})$ or $O(N_c^{-1})$; therefore, their
iteration generates diagrams with lower $N_c$-dependences. An
immediate consequence of this is that the resulting bound-state
equation would have an $N_c$-dependent interaction kernel
(generally vanishing when $N_c\rightarrow \infty$), whose effect
would be that the mass of the possibly existing bound state or
resonance either approaches the two-meson threshold  or disappears
at infinity with increasing $N_c$. The second difference is that
the above kernels are of short-range type, since they involve
either meson contact terms or meson exchanges; therefore, this
mechanism is typical of the production of bound states or
resonances in the molecular-type scheme, where the effective
degrees of freedom are ultimately reduced to those of mesons.
\par
In our subsequent analyses of Secs.~\rf{s51}, \rf{s52} and \rf{s53}
we shall be primarily interested in the first type of mechanism,
searching for the conditions of emergence of tetraquark poles from
confining forces. The second type of mechanism, in relation with
QCD Feynman diagrams, will be considered in Sec.~\rf{s54} and, in
a more general framework, in Sec.~\rf{s6}.
\par 

\subsection{Fully exotic tetraquarks} \lb{s51}

Most of the material needed for this study has been already introduced
in Sec.~\rf{s4}. We consider two pairs of quarks and antiquarks, with
four different flavors, which we distinguish by the labels 1 and 3
for the quarks and $\overline 2$ and $\overline 4$ for the
antiquarks. We then consider the four-current correlation functions
(\rf{4e4}) and (\rf{4e5}), describing two direct channels and two
quark-recombination channels, designated by $D1$, $D2$, $R1$ and
$R2$, respectively.
\par
We first consider the direct channels. The corresponding leading
disconnected and connected diagrams, for channel $D1$, have been
given in Fig.~\rf{4f1}. It is understood that each such diagram is
accompanied by an infinite number of other diagrams with many-gluon
exchanges belonging to the same color-topology class (here, planar).
It is only the connected part of the correlation function that may
provide information about the corresponding scattering amplitude.
To isolate the latter, one has to factorize in the connected part
of the correlation function the external meson propagators, together
with the meson couplings [Eqs.~(\rf{2e12}) and (\rf{2e13})].
These diagrams, as has been shown by Eqs.~(\rf{4e6})--(\rf{4e8}),
have four-quark singularities in the $s$-channel and hence may
participate in the formation, as intermediate states, of 
two-meson states, as well as of possible tetraquark states, the latter
henceforth being designated by $T$.
\par
One then obtains the leading large-$N_c$ behaviors for the two-meson
scattering amplitudes in channels $D1$ and $D2$ and the corresponding
transition amplitudes through two-meson and tetraquark intermediate
states, respectively:
\bea
\lb{5e1}
& &A(M_{\overline 21}^{}M_{\overline 43}^{}\rightarrow
M_{\overline 21}^{}M_{\overline 43}^{})
\sim A(M_{\overline 23}^{}M_{\overline 41}^{}\rightarrow 
M_{\overline 23}^{}M_{\overline 41}^{})=O(N_c^{-2}),\\
\lb{5e2}
& &A(M_{\overline 21}^{}M_{\overline 43}^{}\rightarrow
MM\rightarrow M_{\overline 21}^{}M_{\overline 43}^{})\sim
A(M_{\overline 23}^{}M_{\overline 41}^{}\rightarrow MM\rightarrow
M_{\overline 23}^{}M_{\overline 41}^{})\nonumber \\
& &\ \ \ \ \ \ =O(N_c^{-2}),\\
\lb{5e3}
& &A(M_{\overline 21}^{}M_{\overline 43}^{}\rightarrow
T\rightarrow M_{\overline 21}^{}M_{\overline 43}^{})
\sim A(M_{\overline 23}^{}M_{\overline 41}^{}\rightarrow T\rightarrow
M_{\overline 23}^{}M_{\overline 41}^{})\nonumber \\
& &\ \ \ \ \ \ =O(N_c^{-2}).
\eea
\par
We next consider the recombination channels (\rf{4e5}).
Typical $N_c$-leading and -subleading diagrams have been shown in
Figs.~\rf{4f2} and \rf{4f4}. Only diagrams of the type of Fig.~\rf{4f4}
do have $s$-channel four-quark singularities and hence may participate
in the formation of two-meson and tetraquark states.
On the other hand, the $N_c$-leading diagrams, such as those of
Fig.~\rf{4f2}, contribute to parts of the scattering amplitude that
do not have $s$-channel singularities. One obtains the following
large-$N_c$ behaviors of the scattering amplitudes in channels $R1$
and $R2$ and the corresponding transition amplitudes through two-meson
and tetraquark intermediate states, respectively:
\bea
\lb{5e4}
& &A(M_{\overline 41}^{}M_{\overline 23}^{}\rightarrow
M_{\overline 21}^{}M_{\overline 43}^{})
\sim A(M_{\overline 21}^{}M_{\overline 43}^{}\rightarrow 
M_{\overline 41}^{}M_{\overline 23}^{}) = O(N_c^{-1}), \\
\lb{5e5}
& &A(M_{\overline 41}^{}M_{\overline 23}^{}\rightarrow
MM\rightarrow M_{\overline 21}^{}M_{\overline 43}^{})
\sim A(M_{\overline 21}^{}M_{\overline 43}^{}\rightarrow 
MM\rightarrow M_{\overline 41}^{}M_{\overline 23}^{})\nonumber \\
& &\ \ \ \ \ \ \ \ =O(N_c^{-3}),\\
\lb{5e6}
& &A(M_{\overline 41}^{}M_{\overline 23}^{}\rightarrow
T\rightarrow M_{\overline 21}^{}M_{\overline 43}^{})
\sim A(M_{\overline 21}^{}M_{\overline 43}^{}\rightarrow 
T\rightarrow M_{\overline 41}^{}M_{\overline 23}^{})\nonumber \\
& &\ \ \ \ \ \ \ \ =O(N_c^{-3}).
\eea
\par
One can analyze Eqs.~(\rf{5e1}) and (\rf{5e4}) in terms of effective
meson vertices.
Four-meson vertices of the direct type appear as being of order
$N_c^{-2}$, while those of the recombination type of order $N_c^{-1}$
(Figs.~\rf{5f1}a and \rf{5f2}a):
\bea 
\lb{5e7}
& &g(M_{\overline 21}^{}M_{\overline 43}^{}
M_{\overline 21}^{}M_{\overline 43}^{})
\sim g(M_{\overline 41}^{}M_{\overline 23}^{}
M_{\overline 41}^{}M_{\overline 23}^{})=O(N_c^{-2}), \\
\lb{5e8}
& &g(M_{\overline 41}^{}M_{\overline 23}^{}
M_{\overline 21}^{}M_{\overline 43}^{})=O(N_{\mathrm{c}}^{-1}).
\eea
Four-meson contact terms are also accompanied by glueball-exchange
and one-meson-exchange terms (Figs.~\rf{5f1}b and \rf{5f2}b).
\par
\bfg
\bc
\epsfig{file=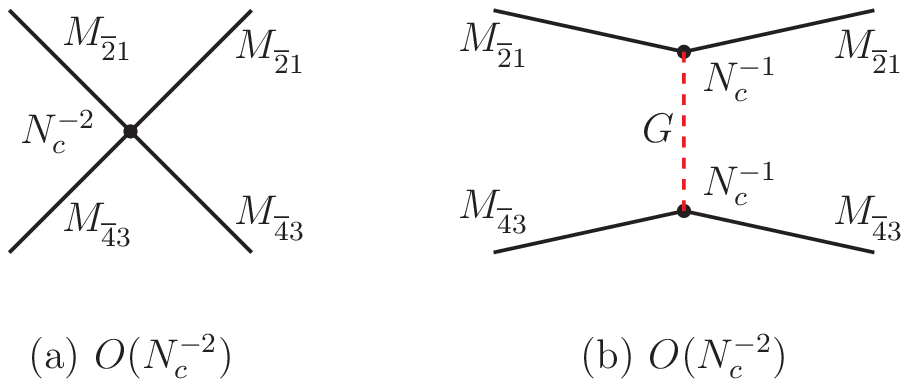,scale=0.8}
\caption{(a) Four-meson vertex in the direct channel $D1$
[Eq.~(\rf{4e4})]. (b) Glueball exchange in the same channel.
Similar diagrams also exist in the direct channel $D2$.}
\lb{5f1}
\ec
\efg
\par
\bfg
\bc
\epsfig{file=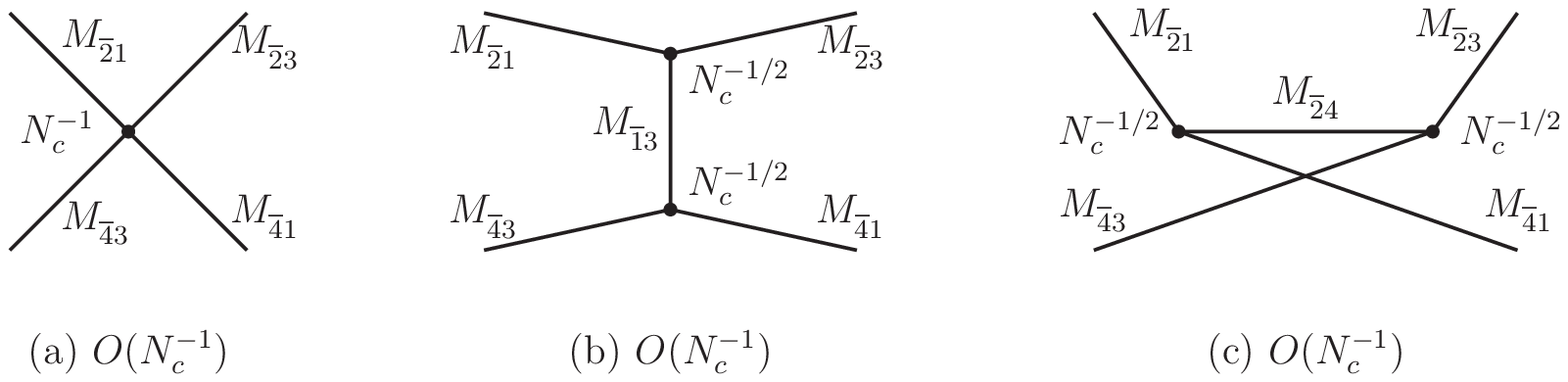,scale=0.8}
\caption{(a) Four-meson vertex in the recombination channel $R1$
[Eq.~(\rf{4e5})]. (b,c) One-meson exchanges in the same channel.
Similar diagrams also exist in the recombination channel $R2$.}
\lb{5f2}
\ec
\efg
\par
The determination of the behaviors of four-meson vertices, including
contact terms and meson exchanges, allows
us to evaluate the contributions of two-meson intermediate states
in the above processes. They are summarized in Fig.~\rf{5f3}, where
we have kept, for simplicity, only contact-type interactions. They
consistently reproduce the behaviors expected from Eqs.~(\rf{5e2}) and
(\rf{5e5}). Similar conclusions could also be obtained from the
glueball- and one-meson-exchange diagrams.
\bfg
\bc
\epsfig{file=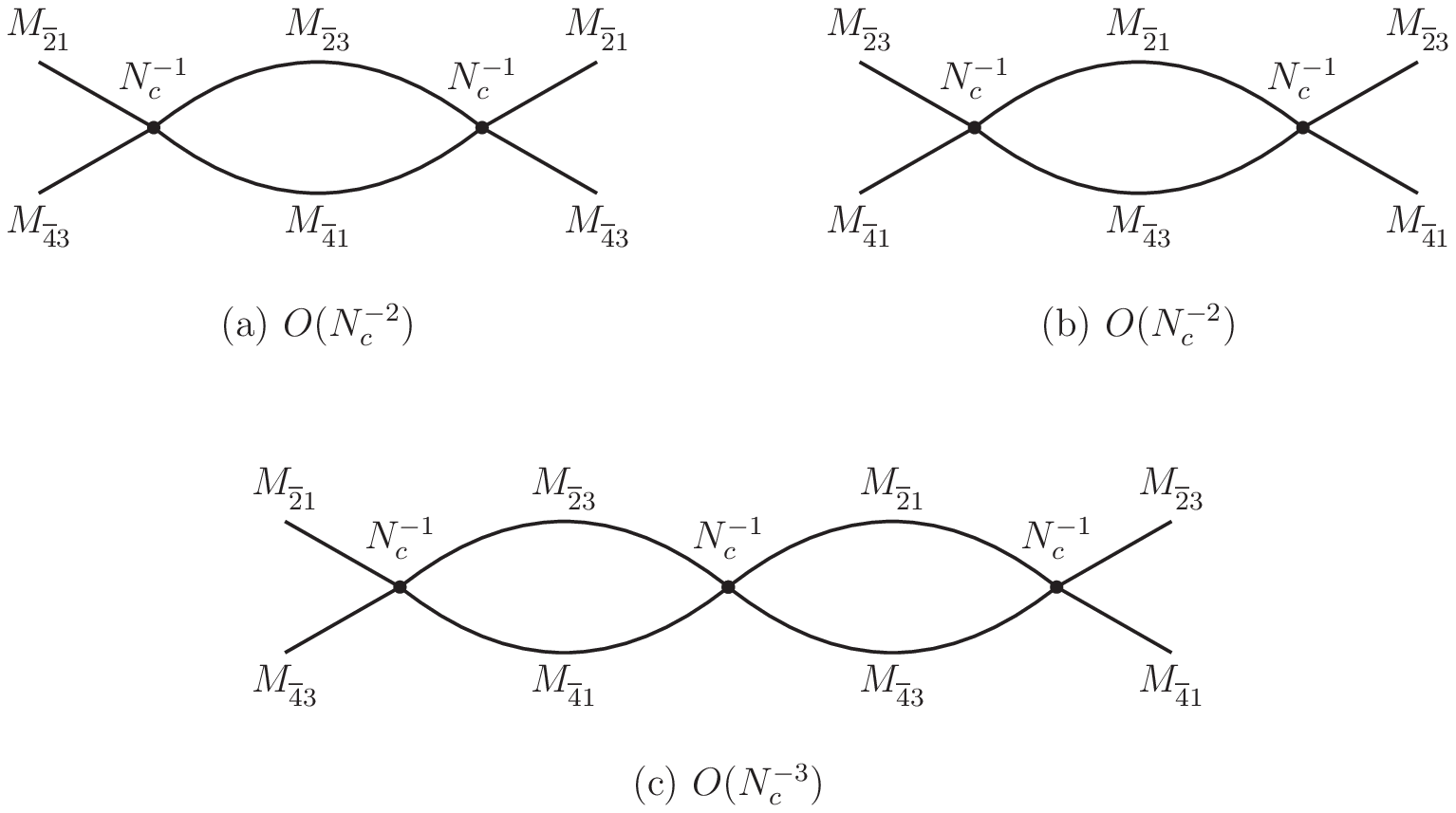,scale=0.7}
\caption{Leading-order contributions of two-meson intermediate states
to the direct, (a) and (b), and recombination, (c), channels.}
\lb{5f3}
\ec
\efg
\par
The validity of the behaviors displayed in Fig.~\rf{5f3} can also be
verified on individual Feynman diagrams with gluon exchanges, using
the Landau equations and recognizing the type of intermediate state
that can be obtained from the summation, with respect to multigluon
exchanges, of such types of diagrams. Two examples are displayed in
Fig.~\rf{5f4}.
\bfg
\bc
\epsfig{file=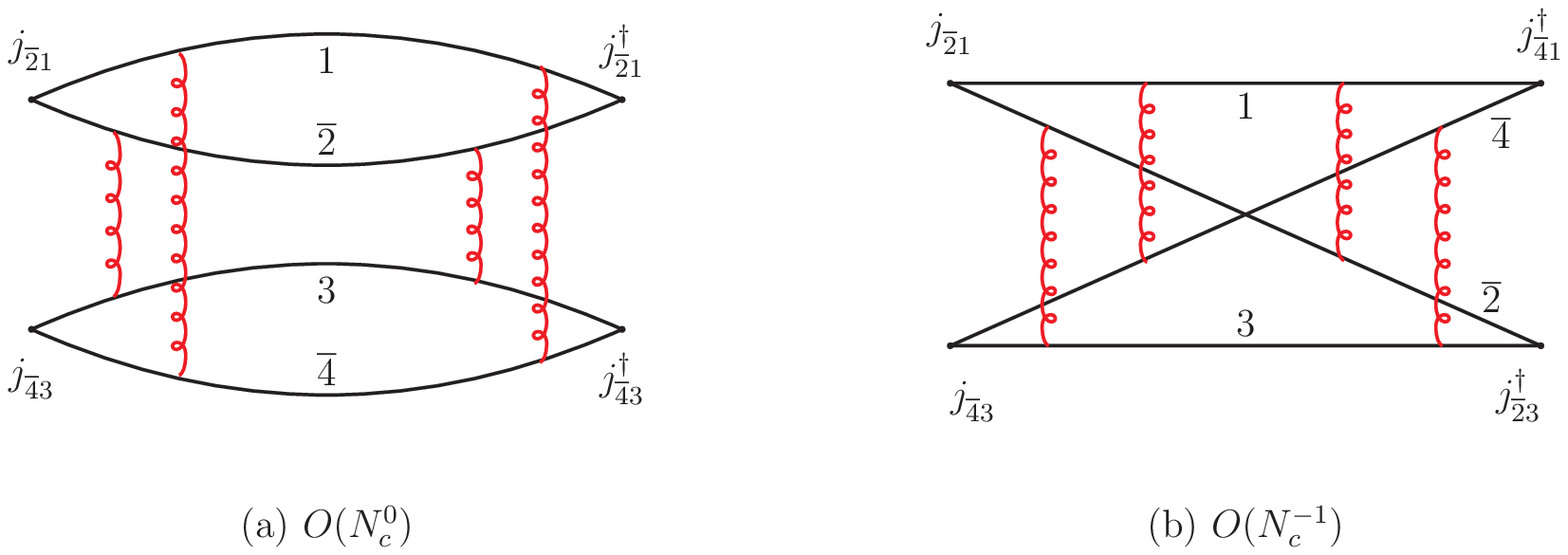,scale=0.7}
\caption{(a) Feynman diagram contributing to the reconstruction
of the two-meson intermediate state
$M_{\overline 41}M_{\overline 23}$ in the direct channel $D1$
[Fig.~\rf{5f3}a]. (b) Feynman diagram contributing to the
reconstruction of the two-meson intermediate states in the
recombination channel $R1$ [Fig.~\rf{5f3}c].}
\lb{5f4}
\ec
\efg
In diagram (a), contributing to the direct channel $D1$,
the intermediate state is manifestly the two-meson state
$M_{\overline 41}M_{\overline 23}$, while the external mesons are
$M_{\overline 21}M_{\overline 43}$. This corresponds to Fig.~\rf{5f3}a.
In diagram (b) of Fig.~\rf{5f4}, contributing to the recombination
channel $R1$, the intermediate state is composed of
$M_{\overline 21}M_{\overline 43}$ on the right and of
$M_{\overline 41}M_{\overline 23}$ on the left. This corresponds to
Fig.~\rf{5f3}c.
(Also, one should not forget that in the right and left corners of the
diagrams of Fig.~\rf{5f4}, near the vertices of the currents $j$, one
still has planar multigluon exchanges, whose infinite sum
reconstitutes the external mesons.)
\par
The properties of possibly existing tetraquark states can be extracted
from Eqs.~(\rf{5e3}) and (\rf{5e6}). One observes that a single
tetraquark alone cannot satisfy these two equations. At least two
different tetraquarks, which we designate by $T_A^{}$ and $T_B^{}$,
are needed to fulfill the conditions imposed by these equations.
The results for tetraquark to two-meson-state transition amplitudes
are the following (see Fig.~\rf{5f5}):
\bea
\lb{5e9}
& &A(T_A^{}\rightarrow M_{\overline 21}^{}M_{\overline 43}^{})
\sim O(N_c^{-1}), \ \ \ 
A(T_A^{}\rightarrow M_{\overline 23}^{}M_{\overline 41}^{})
\sim O(N_c^{-2}),\\
\lb{5e10}
& &A(T_B^{}\rightarrow M_{\overline 23}^{}M_{\overline 41}^{})
\sim O(N_c^{-1}), \ \ \ 
A(T_B^{}\rightarrow M_{\overline 21}^{}M_{\overline 43}^{})
\sim O(N_c^{-2}).
\eea
\par
\bfg
\bc
\epsfig{file=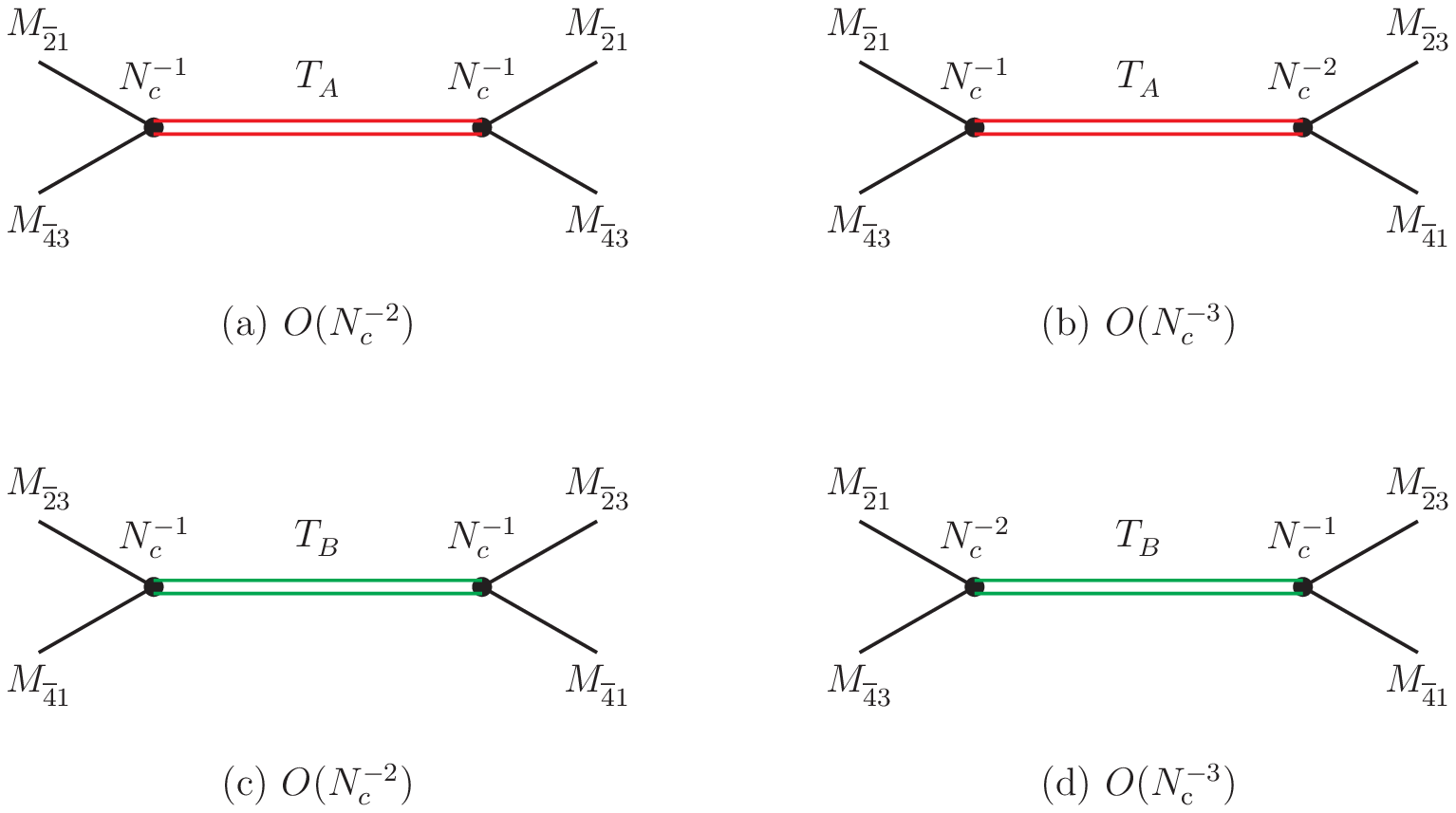,scale=0.7}
\caption{Leading-order contributions of tetraquarks $T_A^{}$ and
$T_B^{}$ to the direct (a,c) and recombination (b,d) channels.}
\lb{5f5}
\ec
\efg
\par
If the tetraquarks lie above the two-meson thresholds, the decay
widths into two mesons are
\be \lb{5e11}
\Gamma(T_A) \sim\ \Gamma(T_B)\ =\ O(N_c^{-2}),
\ee
which are smaller than those of the ordinary mesons 
[$\Gamma=O(N_c^{-1})$] by one power of $N_c^{}$. 
\par 
To have an insight into the internal structure of the two
tetraquark candidates, one can transcribe the information about
the four-meson couplings coming from Eqs.~(\rf{5e7})--(\rf{5e8})
into an effective Lagrangian, expressed through quark color-singlet
bilinears:
\bea \lb{5e12}
\mathcal{L}_{\mathrm{eff,int}}&=&-\frac{\lambda_1^{}}{N_c^{}}
[(\overline q_2^{}q_1^{})(\overline q_4^{}q_3^{})
(\overline q_3^{}q_2^{})(\overline q_1^{}q_4^{})
+(\overline q_4^{}q_1^{})(\overline q_2^{}q_3^{})
(\overline q_3^{}q_4^{})(\overline q_1^{}q_2^{})]\nonumber \\
& &-\frac{\lambda_2^{}}{N_c^{2}}
[(\overline q_2^{}q_1^{})(\overline q_4^{}q_3^{})
(\overline q_3^{}q_4^{})(\overline q_1^{}q_2^{})
+(\overline q_2^{}q_3^{})(\overline q_4^{}q_1^{})
(\overline q_1^{}q_4^{})(\overline q_3^{}q_2^{})],
\eea
where we have explicitly factored out the $N_c^{}$-dependence 
of the coupling constants. One then deduces from Eqs.~(\rf{5e9}) and 
(\rf{5e10}) that the tetraquark fields $T_A^{}$ and $T_B^{}$ should
have, at leading order at large $N_c^{}$, the following structure in
terms of the quark color-singlet bilinears:
\be \lb{5e13}
T_A^{}\ \sim\ (\overline q_2^{}q_3^{})(\overline q_4^{}q_1^{}),
\ \ \ \ \ \   
T_B^{}\ \sim\ (\overline q_2^{}q_1^{})(\overline q_4^{}q_3^{}). 
\ee
\par
This result favors a color singlet-sing\-let structure
of the tetraquarks in the exotic case. We notice that, according
to Eqs.~(\rf{5e9}) and (\rf{5e10}), the main decay channels of
the tetraquarks are not of the dissociative type, but rather of
the quark rearrangement type.
\par
It is worth emphasizing here that the two-meson contributions,
found in Eqs.~(\rf{5e2}) and (\rf{5e5}) [Fig.~\rf{5f3}],
saturate by themselves, at large $N_c^{}$, the singularity
structure emerging from the Feynman diagrams. Contrary to the
ordinary-meson case, they are in
competition with the contributions of possibly existing tetraquarks.
Therefore, the presence of the latter does not appear as mandatory
for the saturation of the large-$N_c$ equations. The results found
above about the tetraquark couplings to two mesons and about their
decay widths have, therefore, the meaning of upper bounds.
Eventually, one might encounter an intermediate situation, where one
of the tetraquarks, $T_B^{}$, say, is absent from the spectrum for some
dynamical reason. In that case, one tetraquark, $T_A^{}$, would exist
and, if it lies above the two-meson threshold, it would be observed 
through its preferred decay channel [Eq.~(\rf{5e9})]. 
\par

\subsection{Cryptoexotic states} \lb{s52}

We next consider cryptoexotic channels, involving three different
quark flavors, designated by 1, 2 and 3. As in Eqs.~(\rf{4e4})
and (\rf{4e5}), we consider correlation functions describing
two direct and two recombination channels:
\bea
\lb{5e14}
\lefteqn{\hspace{-0.3 cm}\Gamma_{D1}^{}\equiv
\langle j_{\overline 23}^{}(x)j_{\overline 31}^{}(y)
j_{\overline 31}^{\dagger}(z)j_{\overline 23}^{\dagger}(0)\rangle,
\ \ \ \Gamma_{D2}^{}\equiv
\langle j_{\overline 21}^{}(x)j_{\overline 33}^{}(y)
j_{\overline 33}^{\dagger}(z)j_{\overline 21}^{\dagger}(0)\rangle,}\\
\lb{5e15}
& &{\hspace{-1 cm}\Gamma_{R1}^{}\equiv
\langle j_{\overline 23}^{}(x)j_{\overline 31}^{}(y)
j_{\overline 33}^{\dagger}(z)j_{\overline 21}^{\dagger}(0)\rangle,
\ \ \ \Gamma_{R2}^{}\equiv
\langle j_{\overline 21}^{}(x)j_{\overline 33}^{}(y)
j_{\overline 31}^{\dagger}(z)j_{\overline 23}^{\dagger}(0)\rangle.}
\eea
\par 
Leading and subleading diagrams of the direct channel $D1$
are represented in Fig.~\rf{5f6}.
\par
\bfg
\bc
\epsfig{file=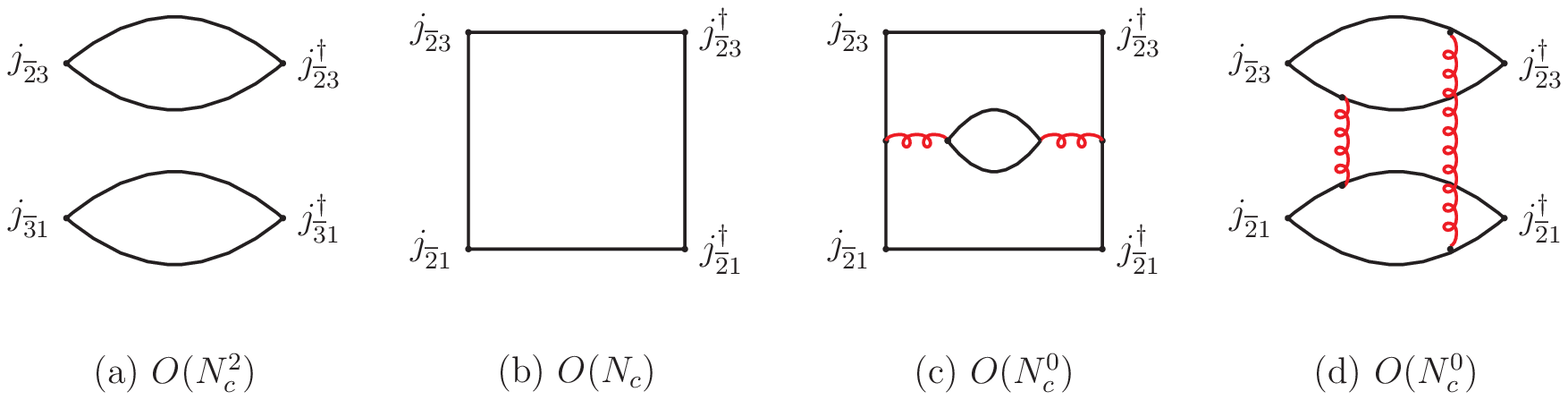,scale=0.7}
\caption{Leading- and subleading-order diagrams of the direct 
channel $D1$ of Eq.~(\rf{5e14}).}
\lb{5f6} 
\ec
\efg
\par
Diagram (b) of Fig.~\rf{5f6} corresponds to the leading-order
contribution to the meson-meson scattering amplitude. It has only
a two-quark singularity in the $s$-channel, which represents
the contribution of a single-meson intermediate state. 
This diagram has also a $t$-channel one-meson singularity, as well
as parts representing four-meson contact terms (cf.~Figs.~\rf{5f9}b
and c); by iteration, they may generate poles in two-meson
intermediate states. This type of mechanism of producing poles was
previously advocated just before Sec.~\rf{s51}, to which we refer
the reader. General aspects of it will be considered in Secs.~\rf{s54}
and \rf{s6}. We are focusing in Secs.~\rf{s51}, \rf{s52} and \rf{s53}
on the pole production mechanism that occurs in diagrams having
the same $N_c$-leading behaviors.
\par
Diagram (c) represents contributions from radiative corrections to
the previous diagram. In the space of meson states, the first part
of the intermediate states contributes to the formation of a
single-meson state, which then emits two virtual mesons, or a
tetraquark, and reabsorbs them later. This diagram may also describe
a mixing between a single-meson state and a tetraquark state, having
the same quantum numbers.
\par
Diagram (d) represents a direct contribution of two-meson states and/or 
of a tetraquark state.
\par
For the direct channel $D2$, the structure of the diagrams is similar to
that of Fig.~\rf{4f1} and is represented in Fig. \rf{5f7}.
\bfg
\bc
\epsfig{file=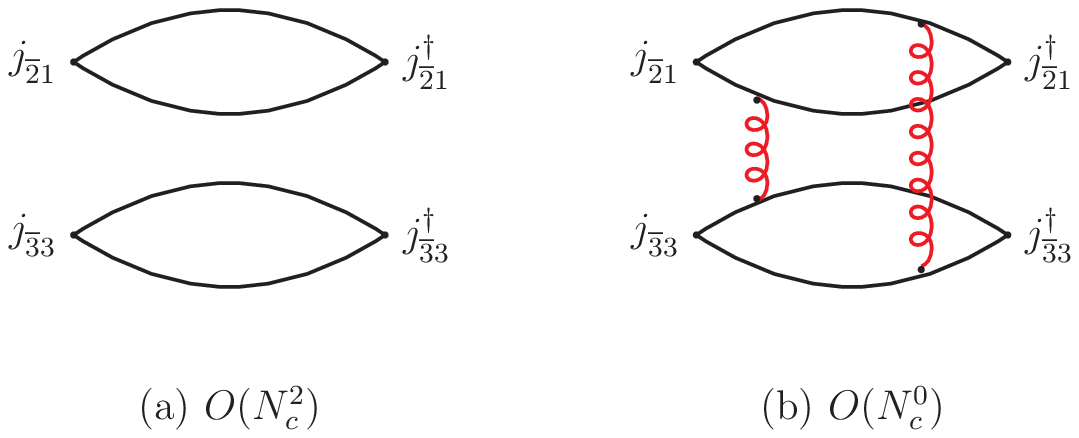,scale=0.8}
\caption{Leading- and subleading-order diagrams of the
direct channel $D2$ of Eq. (\rf{5e14}).}
\lb{5f7}
\ec
\efg
\par
For the recombination channel $R1$ of Eq.~(\rf{5e15}), the main
leading and subleading diagrams are shown in Fig.~\rf{5f8}.
Diagram (a) does not have $s$-channel singularities
[cf.~Fig.~\rf{4f2}a and related comment], while diagrams
(b) and (c) receive contributions from four-quark intermediate
states in the $s$-channel. Similar conclusions also hold for the
recombination channel $R2$.
\bfg
\bc
\epsfig{file=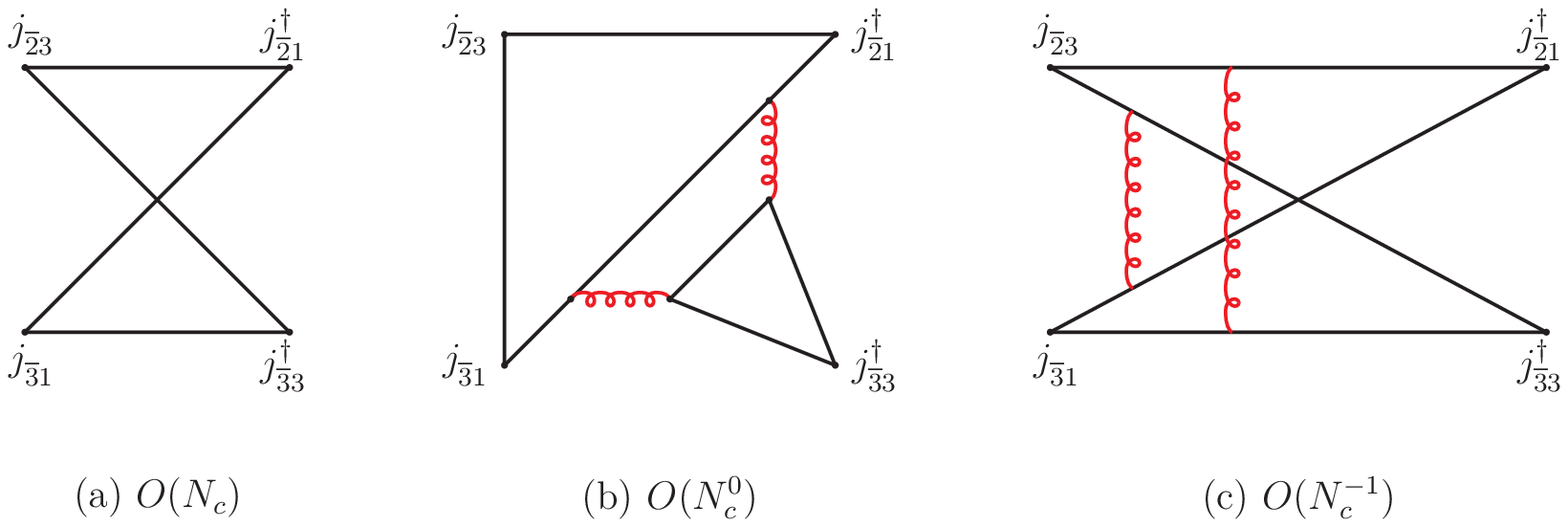,scale=0.7}
\caption{Leading- and typical subleading-order diagrams of the
recombination channel $R1$ of Eq.~(\rf{5e15}). Similar diagrams also
occur in the channel $R2$.}
\lb{5f8}
\ec
\efg
\par
From the previous results, one may obtain information about the
various scattering and transition amplitudes and effective meson
couplings, as in Eqs.~(\rf{5e1})--(\rf{5e8}). The effective meson-meson
interactions at the vertex level are summarized in Fig.~\rf{5f9}.
\bfg
\bc
\epsfig{file=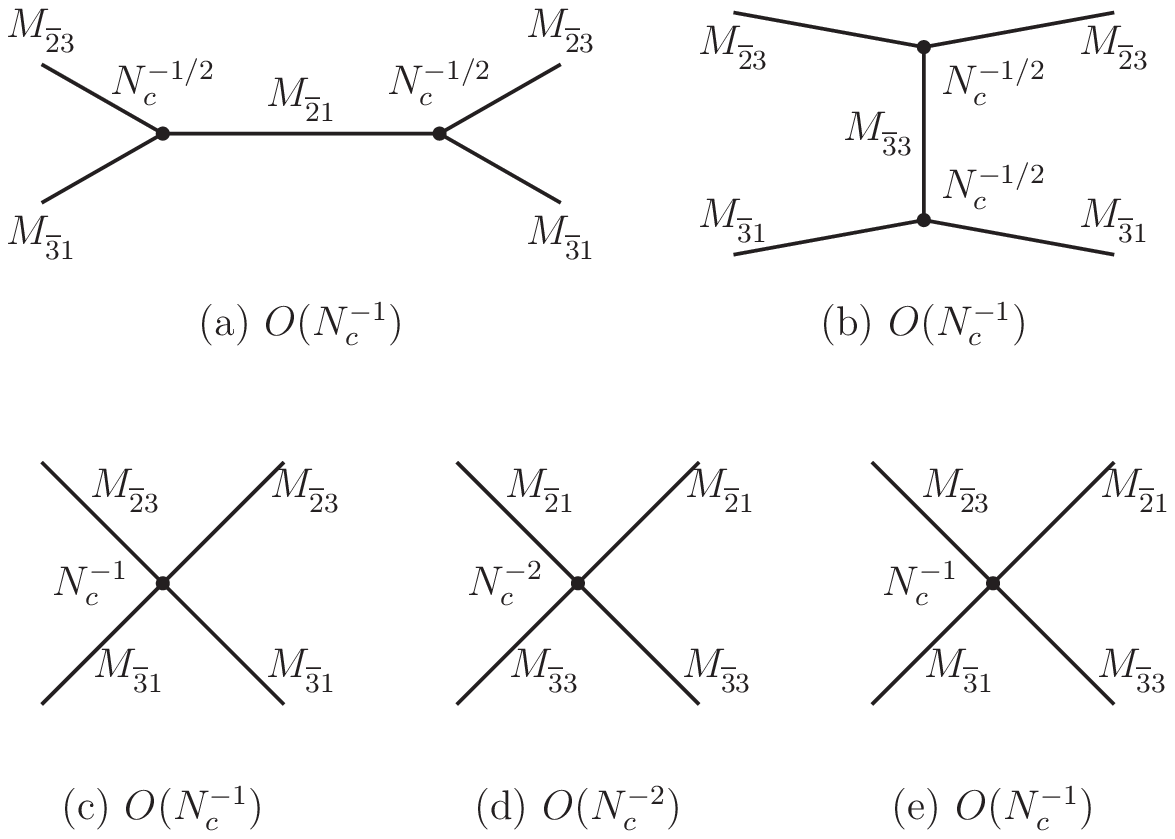,scale=0.7}
\caption{Tree-level vertex diagrams with meson propagators in
the direct channel $D1$ (a,b,c), the direct channel $D2$ (d) and
the recombination channel $R1$ (e). ($u$-channel analog of
diagram (b) not drawn.)}
\lb{5f9}
\ec
\efg
\par
The contributions of two-meson intermediate states are represented
gra\-phically in Fig.~\rf{5f10}.
\par
\bfg
\bc
\epsfig{file=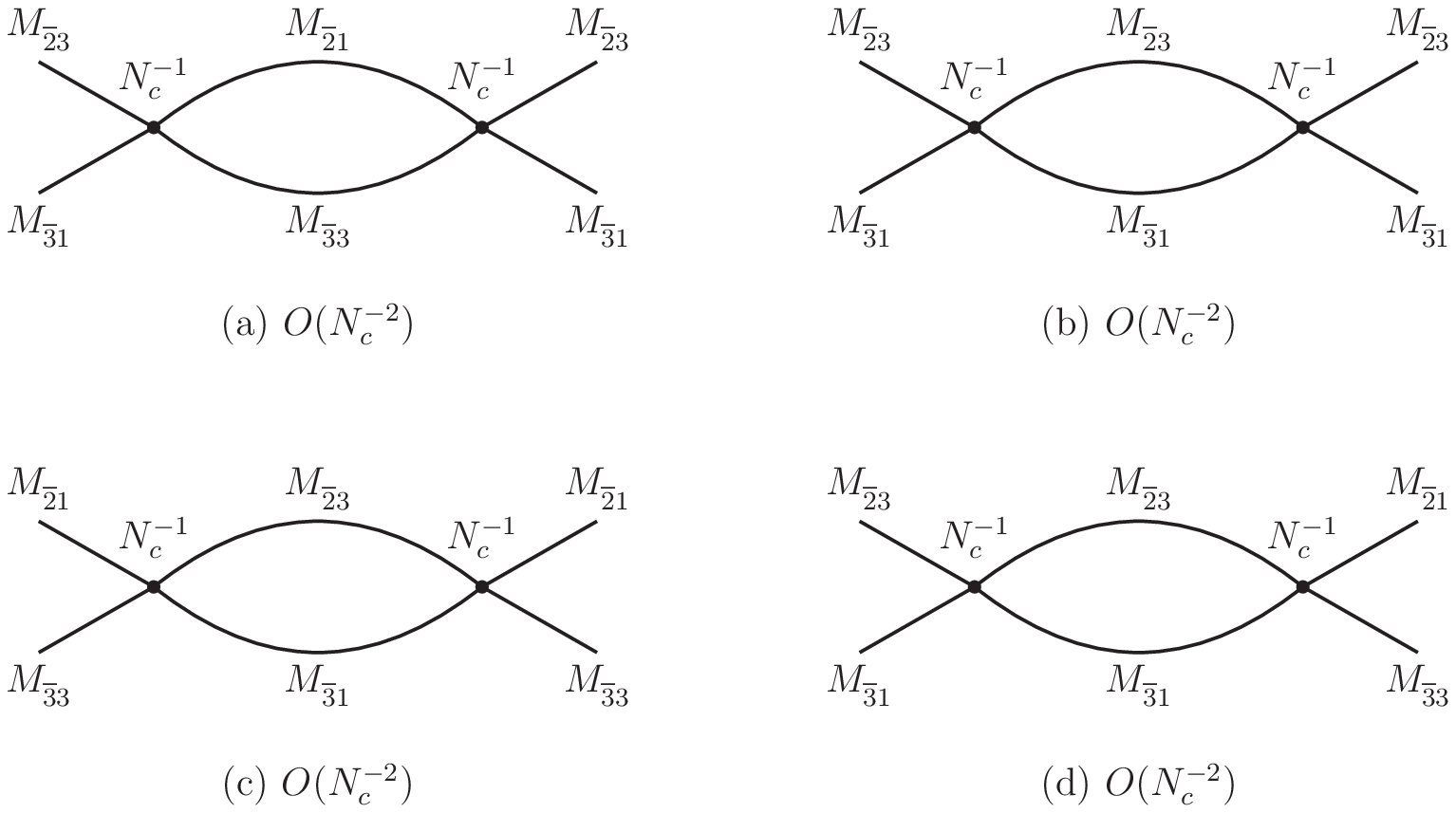,scale=0.7}
\caption{Two-meson intermediate-state contributions to the direct
channel $D1$ (a,b), the direct channel $D2$ (c) and
the recombination channel $R1$ (d).}
\lb{5f10}
\ec
\efg
\par
The tetraquark contributions are extracted in the same way as for 
the exotic channels. Because of the presence of the additional diagram
(b) of Fig.~\rf{5f8}, they are of the same order in all four channels
and hence a single tetraquark $T$ might accommodate all the
corresponding constraints. However, a more detailed analysis of the
mechanism of formation of two-meson and tetraquark intermediate
states suggests that we again are in the presence of two different
tetraquarks $T_A^{}$ and $T_B^{}$, whose field structures, in terms of
valence quarks and antiquarks, are  
\be \lb{5e16}
T_A^{}\ \sim\ (\overline q_2^{}q_3^{})(\overline q_3^{}q_1^{}),
\ \ \ \ \ \   
T_B^{}\ \sim\ (\overline q_2^{}q_1^{})(\overline q_3^{}q_3^{}). 
\ee
This conclusion is based on calculations similar to those presented
for the exotic case in Sec.~\rf{s54}, where, now, the
cryptoexotic case gives rise to additional diagrams.
The results are graphically summarized in Fig.~\rf{5f11}.
Actually, diagrams (a) and (d) of that figure may not exist,
but could be represented by the mixing-mechanism diagrams (b) of
Figs. \rf{5f13} and \rf{5f12}, respectively (see below). This issue
depends more sensitively on the formation mechanism of the tetraquarks.
\par
\bfg
\bc
\epsfig{file=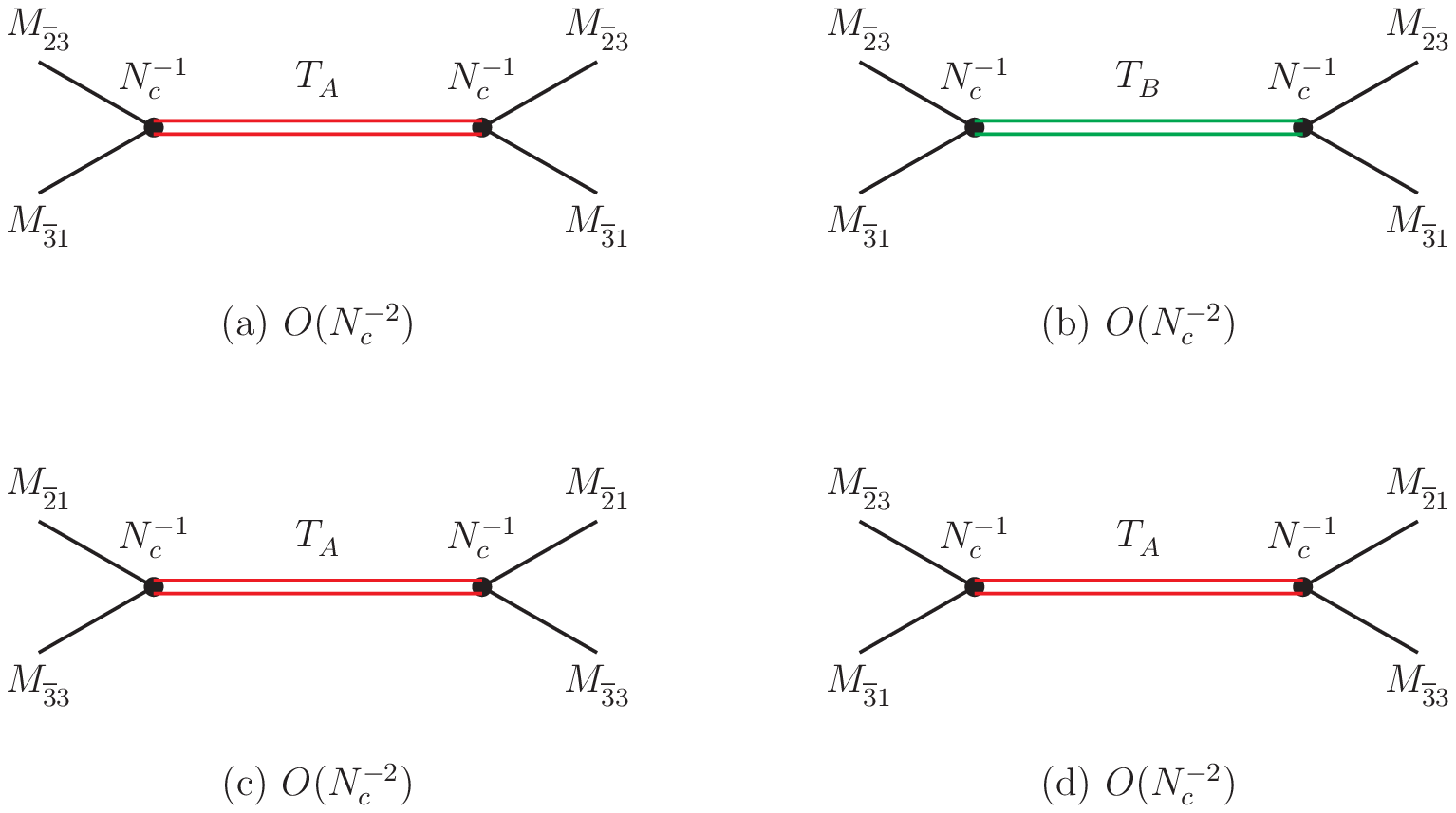,scale=0.7}
\caption{Tetraquark-state contributions to the direct
channel $D1$, (a) and (b), the direct channel $D2$, (c), and
the recombination channel $R1$, (d).}
\lb{5f11}
\ec
\efg
\par
The decay widths of the tetraquarks into two mesons are again of
order $N_{\mathrm{c}}^{-2}$ [Eq.~(\rf{5e11})].
\par
Diagram (b) of Fig.~\rf{5f8} may also describe mixings of two-meson
or tetra\-quark states with a single-meson state that appears in the
left part of the diagram. Figure \rf{5f12} graphically describes
this phenomenon.
\par
\bfg
\bc
\epsfig{file=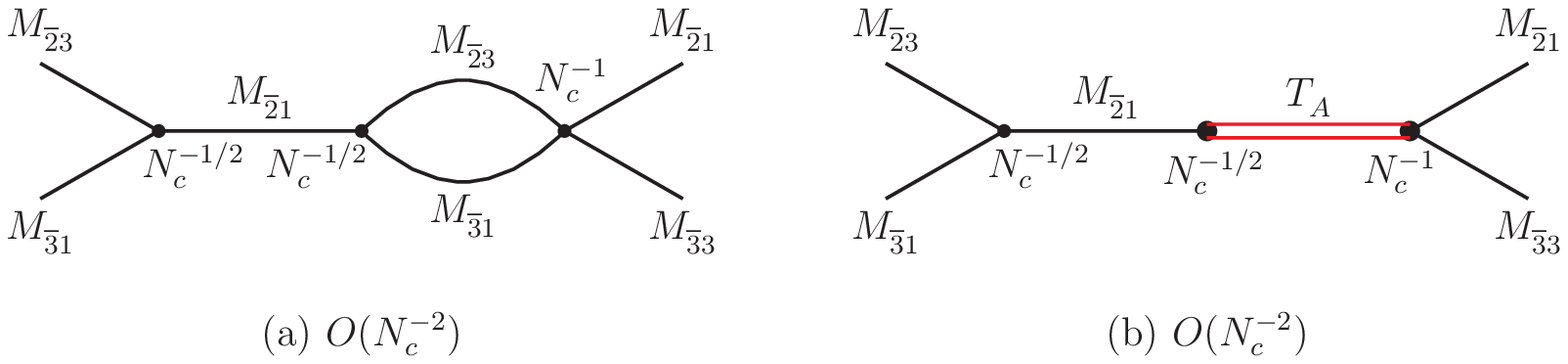,scale=0.7}
\caption{Mixings, in the recombination channel $R1$, of a
single-meson state with two-meson (a) and tetraquark (b) states.}
\lb{5f12}
\ec
\efg
\par
Mixings of a single-meson state with two-meson and tetraquark states
also exist in the direct channel $D1$, as was previously mentioned,
emerging from diagrams of the type of Fig.~\rf{5f6}c. Since in the
quark loop the quark and the antiquark can have any flavor, the
resulting two-meson and tetraquark states may belong to another
class of cryptoexotic states. Figure \rf{5f13} graphically describes
this phenomenon.
\bfg
\bc
\epsfig{file=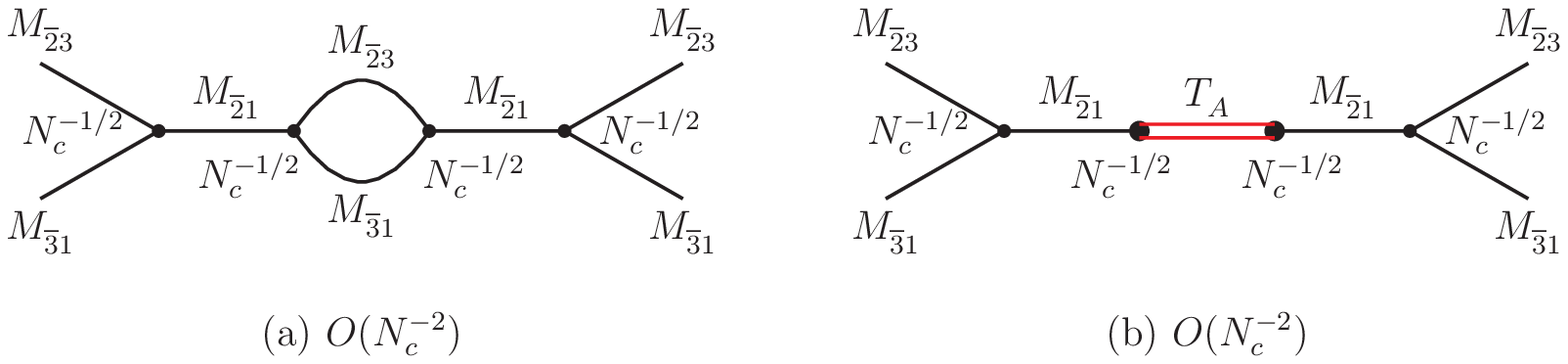,scale=0.7}
\caption{Mixings, in the direct channel $D1$, of a
single-meson state with two-meson (a) and tetraquark (b) states.}
\lb{5f13}
\ec
\efg
Cryptoexotic tetraquarks may therefore decay into two mesons
either through a direct coupling or through a mixing with single-meson
states. In both cases, the transition amplitude is of order
$O(N_c^{-1})$.
\par
Cryptoexotic channels with two quark flavors can be treated in the
same way as before. Here, the correlation functions to be considered
are
\bea
\lb{5e17}
\lefteqn{\hspace{-0.3 cm}\Gamma_{D1}^{}\equiv
\langle j_{\overline 21}^{}(x)j_{\overline 12}^{}(y)
j_{\overline 12}^{\dagger}(z)j_{\overline 21}^{\dagger}(0)\rangle,
\ \ \ \Gamma_{D2}^{}\equiv
\langle j_{\overline 22}^{}(x)j_{\overline 11}^{}(y)
j_{\overline 11}^{\dagger}(z)j_{\overline 22}^{\dagger}(0)\rangle,}\\
\lb{5e18}
& &{\hspace{-1 cm}\Gamma_{R1}^{}\equiv
\langle j_{\overline 21}^{}(x)j_{\overline 12}^{}(y)
j_{\overline 11}^{\dagger}(z)j_{\overline 22}^{\dagger}(0)\rangle,
\ \ \ \Gamma_{R2}^{}\equiv
\langle j_{\overline 22}^{}(x)j_{\overline 11}^{}(y)
j_{\overline 12}^{\dagger}(z)j_{\overline 21}^{\dagger}(0)\rangle.}
\eea
\par 
Most of the leading and subleading diagrams are similar to those
found in the three-flavor case. In addition, one finds, in the
direct channel $D1$, anni\-hi\-lation-type diagrams involving at
least two gluon lines, which produce, as intermediate states in
the $s$-channel, glueballs (cf.~Figs.~\rf{5f14} and \rf{2f13}).
Mixings of tetraquarks with glueball states are of subleading order.
Therefore, the main conclusions about tetraquark decay widths and
two-meson intermediate states remain unchanged. 
\bfg
\bc
\epsfig{file=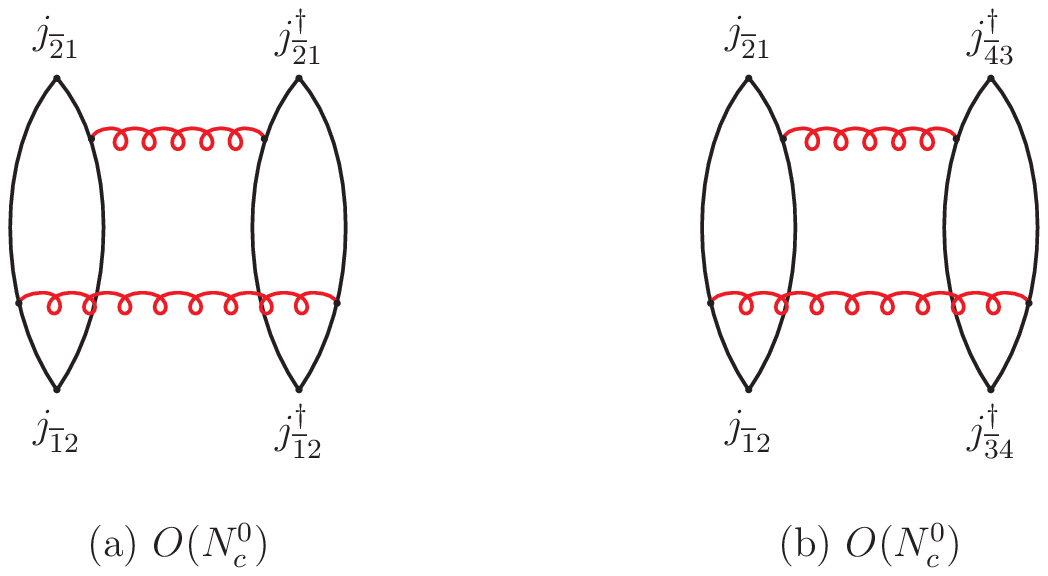,scale=0.7}
\caption{(a) Glueball appearance in the intermediate states of the
$s$-channel of the scattering process
$M_{\overline 21}^{}M_{\overline 12}^{}\rightarrow
M_{\overline 21}^{}M_{\overline 12}^{}$. (b) Similar phenomenon
in the scattering process
$M_{\overline 43}^{}M_{\overline 34}^{}\rightarrow
M_{\overline 21}^{}M_{\overline 12}^{}$. This diagram also
contributes in the $t$-channel of the scattering process
$M_{\overline 21}^{}M_{\overline 43}^{}\rightarrow
M_{\overline 21}^{}M_{\overline 43}^{}$ (cf.~Fig.~\rf{5f1}b).}
\lb{5f14}
\ec
\efg
\par


\subsection{Open-flavor-type states} \lb{s53}

We now consider the case of an open flavor, where two quark
fields have the same flavor. The corresponding four-point
correlation function is
\be \lb{5e19}
\Gamma\equiv\langle j_{\overline 23}^{}(x)j_{\overline 13}^{}(y)
j_{\overline 13}^{\dagger}(z)j_{\overline 23}^{\dagger}(0)\rangle.
\ee
Here, the direct and the recombination channels are identical,
with the common scattering process
$M_{\overline 23}^{}M_{\overline 13}^{}\rightarrow
M_{\overline 23}^{}M_{\overline 13}^{}$.
The corresponding leading and main subleading diagrams are
represented in Fig.~\rf{5f15}. 
\bfg
\bc
\epsfig{file=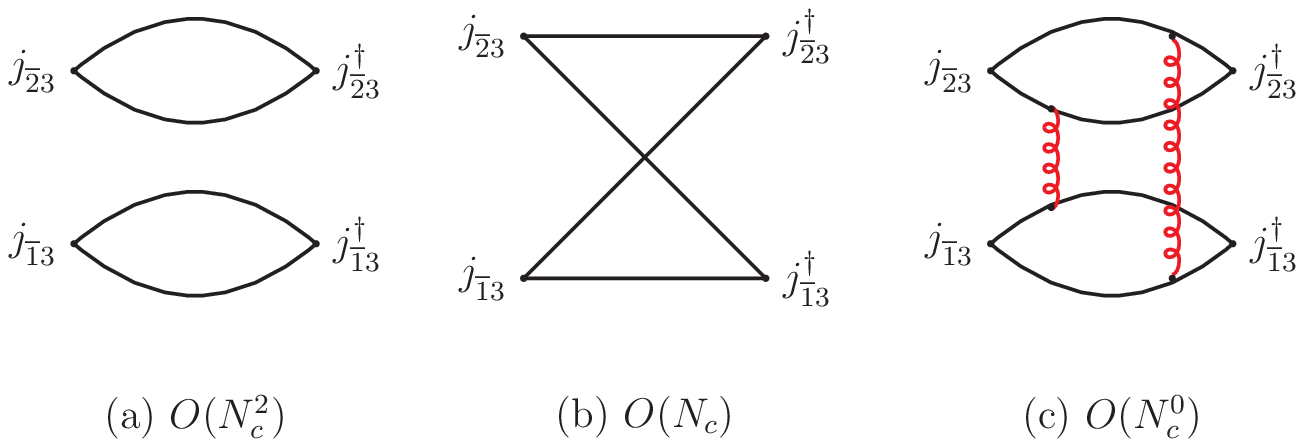,scale=0.7}
\caption{Leading and subleading diagrams in the open-flavor
channel of the correlation function (\rf{5e19}).}
\lb{5f15}
\ec
\efg
\par
The effective meson-meson interaction vertex, the two-meson
interme\-diate-state contribution and the tetraquark state are
graphically represented in Fig.~\rf{5f16}. The decay width of
the tetraquark into two mesons is of order $N_c^{-2}$. The tetraquark
state couples to the local current 
$(\overline q_2^{}q_3^{})(\overline q_1^{}q_3^{})$, which should be
antisymmetrized with respect to the quark field $q_3^{}$, taking
into account its spin degrees of freedom.
\bfg
\bc
\epsfig{file=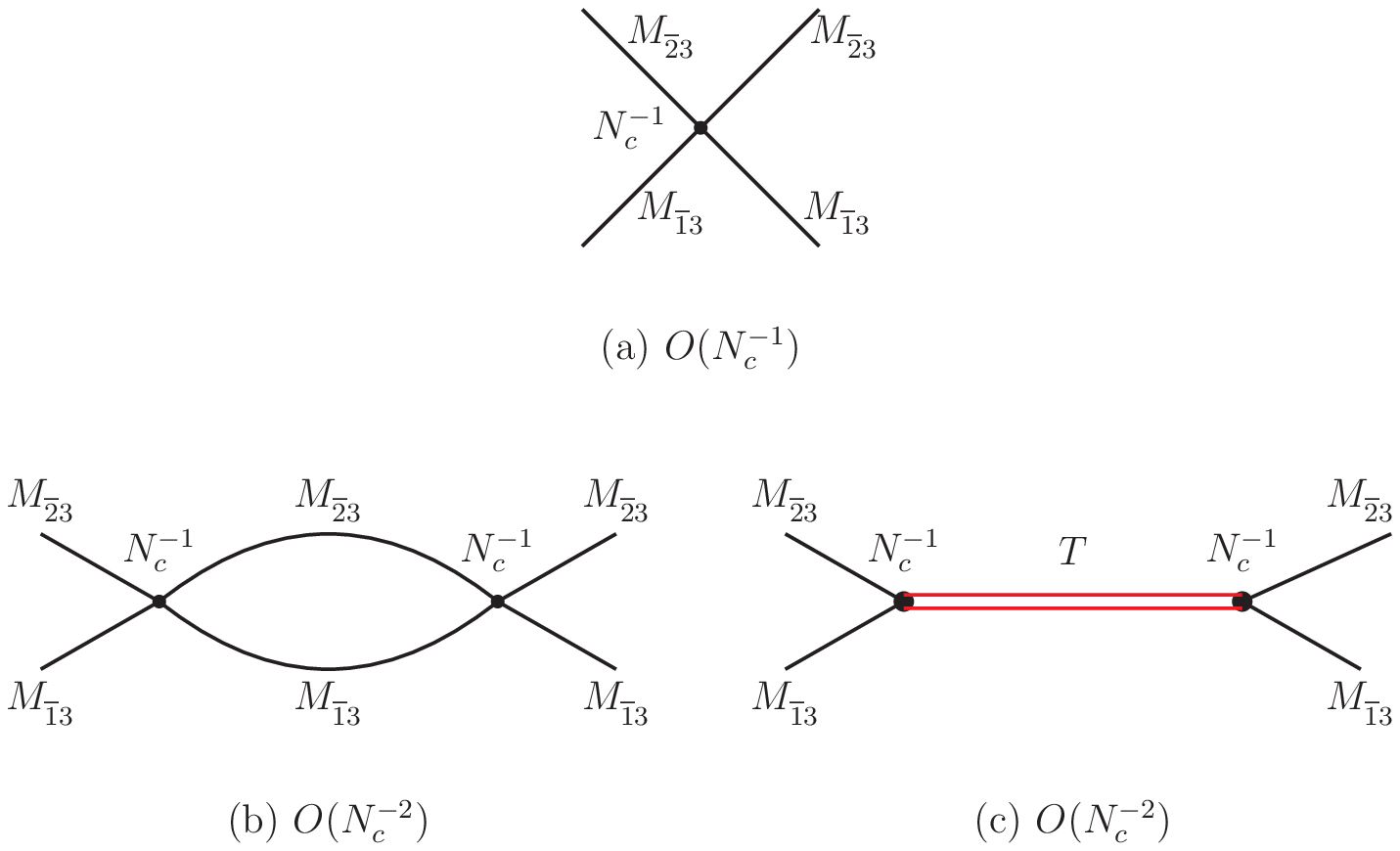,scale=0.7}
\caption{Four-meson effective vertex (a), two-meson intermediate
states (b), and tetraquark intermediate state (c), corresponding to
the open-flavor scattering process   
$M_{\overline 23}^{}M_{\overline 13}^{}\rightarrow
M_{\overline 23}^{}M_{\overline 13}^{}$. (Meson-exchange diagrams, as
in Fig.~\rf{5f2}, are not represented.)}
\lb{5f16}
\ec
\efg
\par

\subsection{Are there tetraquarks at large $N_c^{}$?} \lb{s54}

The fact that, in the exotic and cryptoexotic channels, the possibly
existing tetraquarks should have, at $N_c^{}$-leading order of the
connected diagrams, an internal structure made of two mesonic clusters
[Eqs.~(\rf{5e13}) and (\rf{5e16})] requires further clarification for
the understanding of such a result. The main observation is that
these solutions do not correspond to the expected diquark-antidiquark
structure, which would result from a confining mechanism of the
four-constituent system.
\par
The reason for this can be traced back to the different behaviors of
quark-antiquark and diquark systems at large $N_c^{}$.
In quark-antiquark systems, ladder-type gluon-exchange diagrams are
planar and therefore the infinite number of such diagrams contributes
with equal power of $N_c^{}$ to the whole sum (cf.~Fig.~\rf{2f5}).
This is a necessary condition to possibly form, at that order of
$N_c^{}$, a bound-state pole in the corresponding scattering amplitude.
\par
This is not the case for diquark (or antidiquark) systems. Ladder-type
diagrams are damped by factors of $N_c^{-1}$ at each inclusion of
a gluon line (cf.~Fig.~\rf{2f15}). Here, it is the completely crossed
diagrams which play the role of planar diagrams. However, the sum of
such diagrams is of no help for the formation of bound states, since
completely crossed diagrams do not have $s$-channel singularities and
are parts of the definition of the irreducible kernel of any integral
or bound-state equation. The formation of a bound state (or of a
quasi-bound state) of a diquark necessitates the summation of ladder
diagrams. One cannot obtain, with such type of summation, a bound state
which would be stable with respect to a given order of $N_c^{}$.
\par
The above features can also be formulated in terms of the
quark-antiquark and quark-quark scattering amplitudes, 
designated by $\mathcal{T}$, the first being in the color-singlet
representation and the second in the antisymmetric representation.
Designating by $K$ the one-gluon-exchange kernel, wherein the
coupling constant has been redefined according to 't Hooft's limit,
Eq.~(\rf{2e3}), and from which color indices have been either factorized
or summed together with nearby other color-tensor contributions, the
two scattering amplitudes satisfy, after summation of all ladder
diagrams, the integral equations
\bea
\lb{5e20}
& &\mathcal{T}_{\overline qq}^{}=\frac{1}{N_c^{}}K+K*G_0^{}*
\mathcal{T}_{\overline qq}^{},\\
\lb{5e21}
& &\mathcal{T}_{qq}^{}=\frac{1}{N_c^{}}K+\frac{1}{N_c^{}}K*G_0^{}*
\mathcal{T}_{qq}^{},
\eea
where $G_0^{}$ represents the two quark propagators, the star
operation takes into account the eventual integrations with respect
to the momenta, and the leading behaviors in $N_c$ have been
factorized.
\par
In the quark-antiquark case, Eq.~(\rf{5e20}), a rescaling of
$\mathcal{T}$ in the form $\mathcal{T}/N_c$ removes the factor
$1/N_c$ from the equation and transforms the latter into an
$N_c$-independent equation. If the latter equation is assumed,
with an appropriate form of the gluon propagator, to be valid in the
confining regime, then the resulting meson bound states will have,
at leading order of $N_c$, $N_c$-independent masses, thus confirming
the soft behavior of the latter under variations of $N_c$ from infinity
down to $N_c=3$.
\par
This is not the case for the diquark system. The previous rescaling of 
$\mathcal{T}$ does not remove the $1/N_c$ dependence of the kernel
of the integral equation (\rf{5e21}).
To have a qualitative understanding of the consequences of the
$1/N_c{}$ dependence of the kernel, which we assume being transmitted
to the confining interaction, we consider the illustrative example
of two heavy quarks, with reduced mass $\mu$, satisfying the
Schr\"odinger equation with a linearly confining potential:
\be \lb{5e22}
\Big[\ E-\frac{\mathbf{p}^2}{2\mu}-\frac{1}{N_c^{}}\sigma r\ \Big]
\phi=0,  
\ee
where $\sigma$ is the string tension and the $1/N_c{}$ dependence
of the potential has been explicitly factorized.
This equation leads to the formation of confined bound states,
with bound state energies and mean spatial sizes scaling with
respect to $N_c^{}$ in the following way:
\be \lb{5e23}
E\ \sim\ \Big(\frac{N_c^{}\sqrt{\sigma}}{2\mu}\Big)^{1/3}
\frac{\sqrt{\sigma}}{N_c^{}},\ \ \ \ \ \
\langle r\rangle\ \sim\ \Big(\frac{N_c^{}\sqrt{\sigma}}{2\mu}\Big)^{1/3}
\frac{1}{\sqrt{\sigma}}.
\ee
At large $N_c^{}$, the bound-state energies decrease and the diquark
masses tend to the two-quark mass threshold. On the other hand, the
mean spatial sizes of the diquarks increase, although weakly, with
$N_c^{}$ and the diquarks cease to be compact objects, in
contradiction with the initial objective of finding diquark and
antidiquark systems with compact sizes.
In this situation, the system might easily switch to the $N_c$-dominant
configuration made of two mesonic clusters.
\par
The above example can be completed by considering, for the case of
heavy quarks, instead of the confining potential, the color-Coulomb
potential, which governs the dynamics at short distances and which
is assumed to dominate for heavy-quark systems. 
Here, the interaction potential is $V(r)=-\lambda/(2N_cr)$, where
$\lambda$ is the redefined coupling constant squared at large $N_c$
[Eq.~(\rf{2f3})]. Then, when $N_c$ increases, the bound-state spectrum
shrinks rapidly, like $1/N_c^2$, to the two-quark threshold from below,
while the mean spatial sizes of the bound states increase like $N_c$,
much more rapidly than in the purely confining-potential example. In
the present case, when the diquark system reaches the confining region,
it already has ceased to be compact.
\par
The above results can also be understood with the aid of the
quadratic Casimirs of the various representations. Assuming that
the confining interaction kernel has the same color-representation
property as the gluon propagator, one can evaluate the relative
strengths of the various two-body potentials (for more details,
cf. Ref. \cite{Esposito:2016noz}, Appendix B). For the
quark-antiquark system in the singlet representation, the potential
is proportional to $-(N_c^2-1)/(2N_c)$, where the minus sign reflects
the attractive nature of the potential. For the same system in the
adjoint representation, it is proportional to $+1/(2N_c)$. For the
diquark system in the antisymmetric representation $[2]$
(Sec.~\rf{s32}), it is proportional to $-(N_c^{}+1)/(2N_c)$, while in
the symmetric representation $[1,1]$, it is proportional to
$+(N_c^{}-1)/(2N_c)$.
We observe that, at large $N_c$, the ratio between the diquark
potential in the antisymmetric representation and the quark-antiquark
potential in the singlet representation decreases like $1/(N_c-1)$,
displaying the dominance of the latter potential in the course of
the formation of bound systems. For $N_c=3$, however, the latter
ratio is only 1/2, which might make possible the formation of diquark
systems inside tetraquarks.
\par
To remedy the above difficulties of the diquark scheme, several
dynamical mechanisms have been advocated, either at the
experimental production level \cite{Brodsky:2014xia,Lebed:2017min},
or at the inner interaction level \cite{Maiani:2017kyi}.
\par
On theoretical grounds, an alternative viewpoint has been advocated
in Ref.~\cite{Maiani:2018pef}. It was argued that the planar diagrams
of the type of Figs.~\rf{4f1}b and \rf{5f4}a do not represent, in
spite of gluon exchanges and $s$-channel cuts, genuine interactions
between mesons, but rather depict different ways of representing
color or momentum flows in a system of two noninteracting mesons, and
therefore tetraquark formation graphs should begin from nonplanar
diagrams. This argument, if accepted, brings the contribution of the
direct-channel $D1$ and $D2$ scattering amplitudes to order $N_c^{-4}$,
instead of $N_c^{-2}$. However, one still has a discrepancy with
the recombination channels, which remain at order $N_c^{-3}$. To
remain at the end with one type of tetraquark, in the
diquark-antidiquark antisymmetric representation, one is obliged to
impose an additional selection rule, according to which tetraquarks
may appear only in direct \emph{or} recombination channels.
\par
The main point of the above argument, the noninteracting feature of
the two mesons, does not seem, however, well-founded. Diagrams of
the types of Figs.~\rf{4f1}b and \rf{5f4}a are representatives
of an infinite set of planar diagrams containing at the right and left
corners of the quark loops (near the currents $j$) planar gluons
exchanged between antiquark 2 and quark 1, and between antiquark 4
and quark 3, which means that the initial and final parts of the
diagrams already contain mesons $M_{\overline 21}^{}$ and
$M_{\overline 43}^{}$; any gluon exchanged between these mesons
represents a genuine interaction and not merely a color- or
momentum-flow artifact. It is only these kinds of diagram that 
can describe two-meson-loop formation, as depicted in Fig.~\rf{5f3},
which is of order $N_c^{-2}$. We shall describe below in more detail
the mechanism of the two-meson interaction. The question as to
whether the planar diagrams may produce by themselves tetraquark poles
is more involved and requires further analysis, which we shall also
present hereafter.
\par
We come back to the $N_c$-dominant structure of the four-quark
system, made of two mesonic clusters [Eqs.~(\rf{5e13}) and (\rf{5e16})].
The principal question that remains to be answered is whether such
solutions are compatible with the formation of tetraquarks.
Considering, for definiteness, the direct channel $D1$ of the 
flavor-exotic case [Eq.~(\rf{4e4})], the candidate tetraquark would
have, according to Eq.~(\rf{5e13}), the structure
$T_A^{}\sim(\overline q_2^{}q_3^{})(\overline q_4^{}q_1^{})$,
which is generated by means of sums, with respect to ladder-gluon
lines, of diagrams of the type of Fig.~\rf{5f4}a. The gluon lines
between quark 1 and antiquark 4 generate the scattering amplitude
$\mathcal{T}_{\overline 41}$. Similarly, the gluon lines between
antiquark 2 and quark 3 generate the scattering amplitude 
$\mathcal{T}_{\overline 23}$. The two scattering amplitudes are
disconnected from each other. Nevertheless, they are embedded into
the structure of the meson-meson scattering amplitude
$\mathcal{T}_{[(\overline 21)(\overline 43),
(\overline 21)(\overline 43)]}\equiv
\mathcal{T}(M_{\overline 21}^{}M_{\overline 43}^{}
\rightarrow M_{\overline 21}^{}M_{\overline 43}^{})$ and are
subjected to additional loop integrations, providing a connected
structure. A typical contribution is graphically represented in
Fig. \rf{5f17}. Other contributions involve either
$\mathcal{T}_{\overline 41}$ alone or $\mathcal{T}_{\overline 23}$
alone.
\bfg
\bc
\epsfig{file=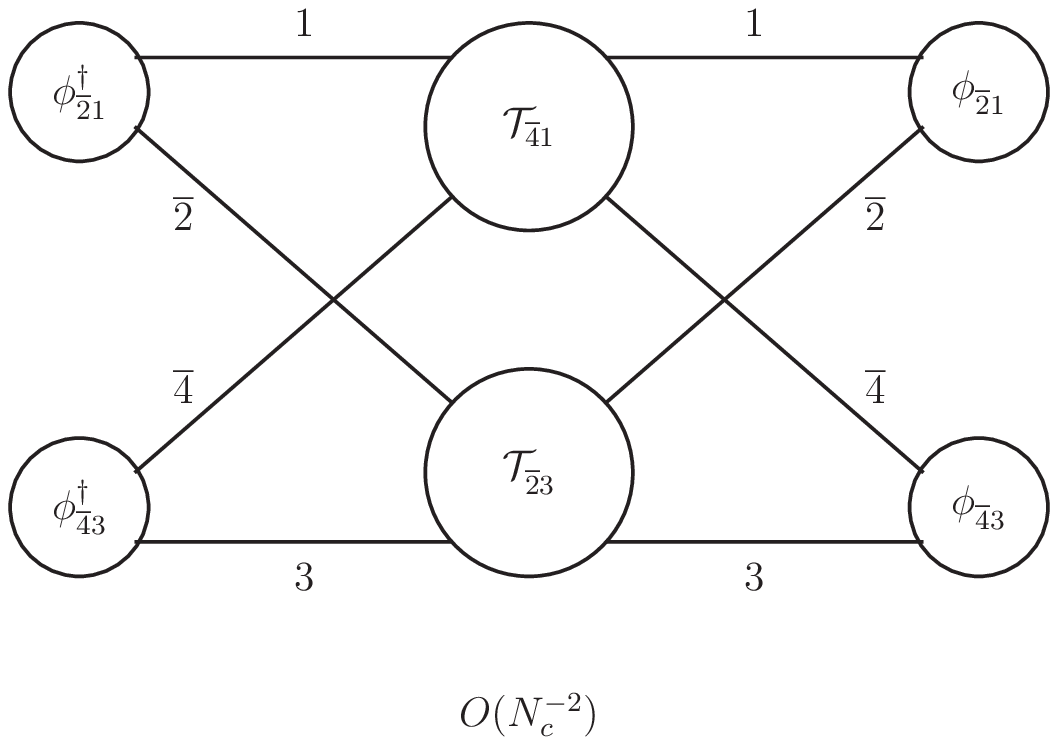,scale=0.7}
\caption{Typical contribution of the quark-antiquark scattering
amplitudes $\mathcal{T}_{\overline 41}^{}$ and
$\mathcal{T}_{\overline 23}^{}$ to the meson-meson scattering amplitude
of the direct channel $D1$ [Eq.~(\rf{4e4})]. Full lines represent quark
propagators. $\phi_{\overline 21}^{}$ and $\phi_{\overline 43}^{}$
are the wave functions of the external mesons. Other contributions
involve either $\mathcal{T}_{\overline 41}$ alone or
$\mathcal{T}_{\overline 23}$ alone. The $N_c$-counting rules have
been detailed in the text.}
\lb{5f17}
\ec
\efg
\par
At leading order in $N_c$, the scattering amplitudes
$\mathcal{T}_{\overline qq}^{}$ are saturated by an infinite sum of
stable mesons [Sec.~\rf{s23}], which we label by a global increasing
index $n$, the value $n=0$ corresponding to the ground state.
They have then the following structure:
\be \lb{5e24}
\mathcal{T}_{\overline qq}^{}(P_{\overline qq}^{},\ldots)=
\frac{1}{N_c}K-\frac{i}{N_c}\sum_{n=0}^{\infty}
\frac{\phi_{\overline qq,n}^{}\phi_{\overline qq,n}^{\dagger}}
{P_{\overline qq}^2-P_n^2},
\ee
where $P_{\overline qq}^{}$ is the total momentum of the quark-antiquark
system and $P_n^2$ is the mass squared of the $n$th meson, with the
$\phi_{\overline qq,n}^{}$s being the corresponding wave
functions\footnote{The wave functions $\phi$ are equal, up to
normalization factors, to the Bethe--Salpeter wave functions multiplied
by the inverse of the two quark propagators.}.
The one-gluon-exchange contribution (or its equivalent one in a
confining scheme) remains outside the sum, since it cannot contribute
simultaneously to the product of the meson wave functions.
An explicit realization of this structure is provided in
two-dimensional QCD (cf. Eq. (18) of Ref.~\cite{Callan:1975ps}).
\par
The $N_c$-counting rules in diagrams of the type of Fig.~\rf{5f17}
and of the next two figures (\rf{5f18} and \rf{5f19}) are the
following: according to Eq.~(\rf{5e24}), each appearance of a
scattering amplitude $\mathcal{T}_{\overline qq}^{}$ is accompanied
by a factor $1/N_c$, while the appearance of a meson wave function
$\phi$ is accompanied by a factor $1/N_c^{1/2}$; every (four-dimensional)
loop momentum integration is accompanied by a factor $N_c$.
\par
The $s$-channel singularities of the object
of Fig.~\rf{5f17} are governed by those coming from the product
$\mathcal{T}_{\overline 41}\mathcal{T}_{\overline 23}$.
Each of these scattering amplitudes has meson poles as 
singularities and hence behaves as a sum of effective meson
propagators. The $s$-channel singularities that result from the
integrations are therefore those of two-meson scattering amplitudes.
They do not involve, however, any pole-type singularity which might
signal the possible presence of a tetraquark state. It might, however,
happen that the infinite sum of the two-meson contributions, which all
have the same type of singularities but are located at different
positions, produces a pole-type singularity through a divergence
occurring in the vicinity of some particular point. Such a possibility,
which closely depends on the behaviors of the various overlappings of
wave functions, is the only one which might exist within the class of
$N_c$-leading diagrams. It would no longer be the result of purely
confining interactions, since the latter have been absorbed by the
meson formations, but rather would be the result of a residual effect
of them, coming from the existence of the tower of an infinite number
of meson states. The possibility of such a mechanism needs, however,
further detailed investigations.
\par
In case the above mechanism does not produce a bound state, a second
possibility of producing a tetraquark pole remains: an 
iteration mechanism of diagrams of the type of Fig.~\rf{5f17}.
\par
\bfg
\bc
\epsfig{file=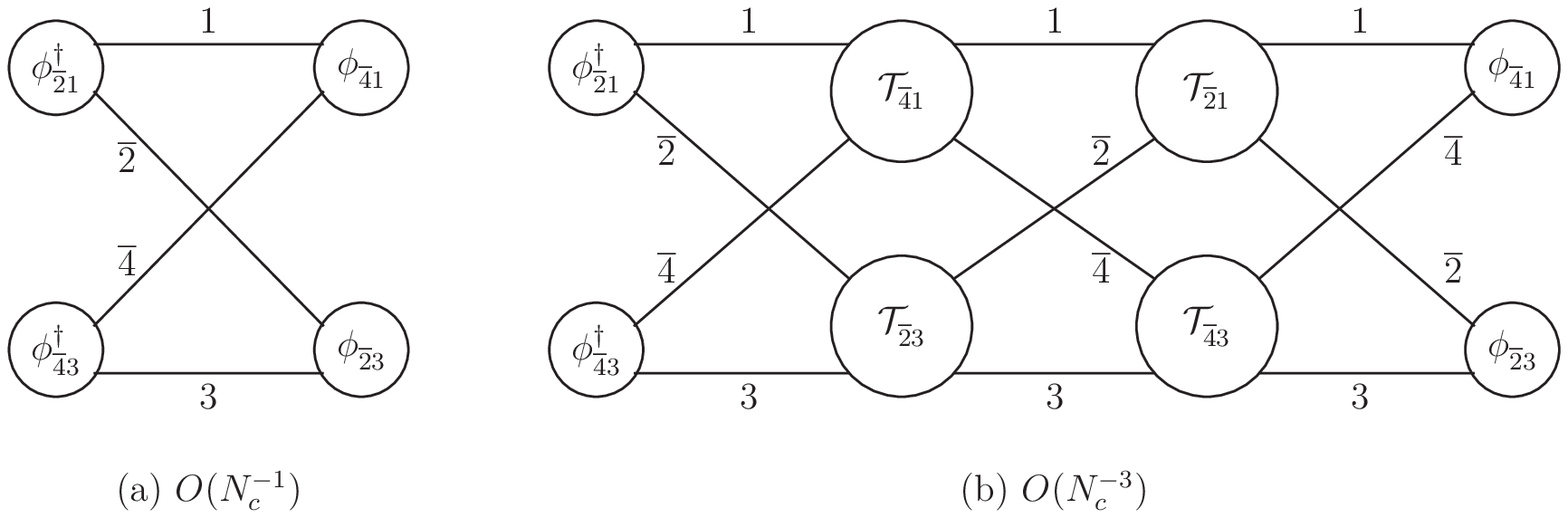,scale=0.7}
\caption{(a) The generic diagram corresponding to Fig.~\rf{4f2}a, of
the recombination channel $R1$ [Eq.~(\rf{4e5})]. (b) First iteration
with the contributions of the quark-antiquark scattering amplitudes
$\mathcal{T}_{\overline 41}^{}$ and $\mathcal{T}_{\overline 23}^{}$,
on the one hand, and 
$\mathcal{T}_{\overline 21}^{}$ and $\mathcal{T}_{\overline 43}^{}$,
on the other, to the same scattering amplitude. 
Same notations as in Fig.~\rf{5f17}. A typical Feynman
diagram participating in this process is presented in Fig.~\rf{5f4}b.
Because of the structure of the quark-antiquark scattering amplitude
(\rf{5e24}) in terms of meson wave functions, this diagram involves
the succession of the recombination channels $R1-R2-R1$.
Insertion of the kernel of the above diagram, made of the product
$R1-R2$, on the left of the kernel of the diagram of Fig.~\rf{5f17}
generates the first iteration of the latter, introducing at the same
time a factor of $N_c^{-2}$.}
\lb{5f18}
\ec
\efg
\bfg
\hspace{0.25 cm}
\epsfig{file=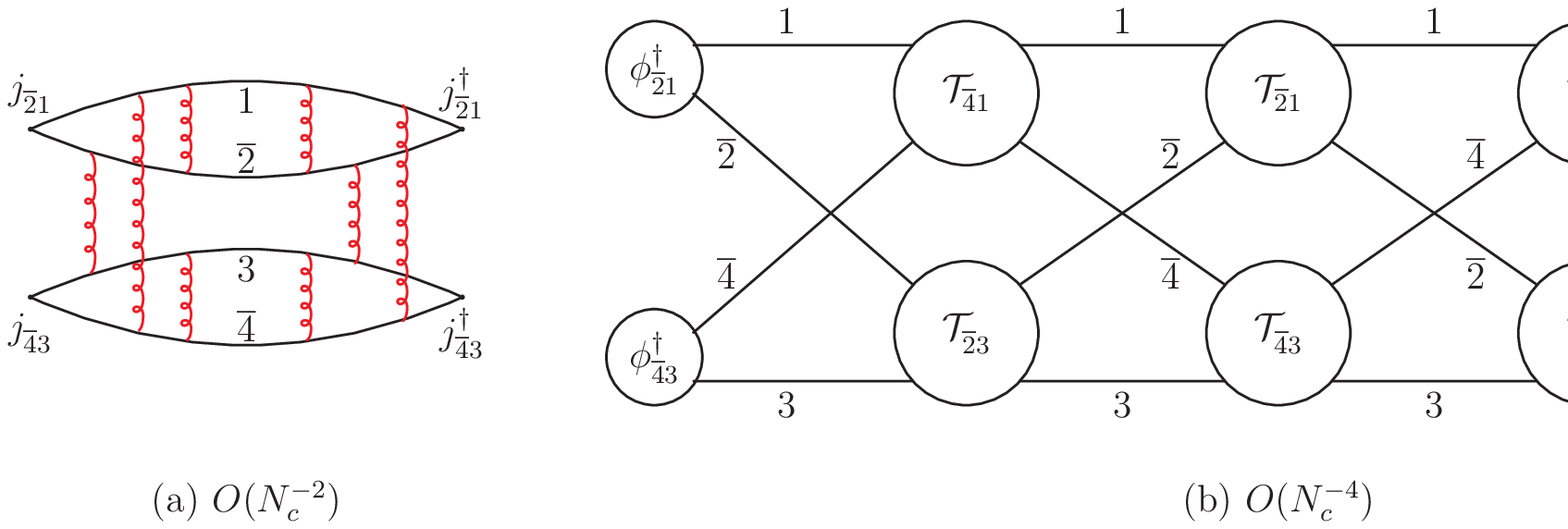,scale=0.6}
\caption{First iteration of the scattering amplitude of the direct
channel $D1$ of Fig.~\rf{5f17} with the insertion of the product
$R1-R2$ of the two recombination channels.
Each insertion introduces a factor of $N_c^{-2}$.
(a) Typical Feynman diagram of the first iteration.
(b) General structure of the iteration, which involves the succession
of the recombination channels $R1-R2-R1-R2$.}
\lb{5f19}
\efg
\par
The effective iteration kernel is made of the succession of the
recombination channels $R1$ and $R2$, examples of which are presented
in Fig.~\rf{5f18}, for the recombination channel $R1$, and in
Fig.~\rf{5f19}, for the direct channel $D1$. Because
of the structure of the quark-antiquark scattering amplitude in
terms of meson wave functions [Eq.~(\rf{5e24})], the kernel of
the iterations in the recombination as well as in the direct
channels has the structure of a product of the two recombination
channels $R1$ and $R2$.
An infinite series of such iterations, which act in convolution in
momentum space, might generate, after summation, a tetraquark pole in
the scattering amplitude.
However, each iteration introduces here a factor of
$N_c^{-2}$, which means that the effective interaction kernel of the
possibly resulting integral equation is of order $N_c^{-2}$ and hence
vanishes when $N_c^{}$ tends to infinity.
Since the confining interactions are explicitly absent between
the meson clusters, one faces here two possibilities.
\par
In the first one, the resulting interaction is of the residual
long-range type (van der Waals, or another hidden mechanism). If it
is globally attractive, then a bound state might exist, approaching
the lowest two-meson threshold when $N_c$ goes to infinity. However,
the existence of such types of forces has not been observed up to now
in the domain of light quarks. Whether they exist and are operative
with heavy quarks remains a key issue of this mechanism.
\par
Van der Waals forces have been considered in the realization of
``hadrocharmonium'' (or, more generally, of hadroquarkonium) states
\cite{Dubynskiy:2008mq,Voloshin:2007dx,Voloshin:2018vym}. Here, one
considers four-quark systems made of a heavy-quark $Q\bar Q$ pair and
a light-quark pair. The heavy-quark pair will have the tendency to
form a heavy quarkonium hadron, while the light quarks form a kind
of hadron matter, interacting with the heavy-quark core by means of
QCD gluon-exchange van der Waals forces. The possibility of global
attractive forces and of existence of bound states has been shown. 
The difference of this system from conventional molecular states
comes from the fact that the resulting bound state (or resonance)
may lie relatively far from the two-meson threshold and can have
distinctive decay channels.
\par
In the second possibility, one rather is in the presence of effective
short-range molecular-type interactions between mesons. The properties
of these types of interaction will be considered in Sec. \rf{s6}.
The main property that concerns us here is that when the strength
of the interaction tends to zero, and more generally becomes lower
than a critical strength, no bound states exist, but resonances might
appear; in the strength-vanishing limit, the resonance masses are
pushed towards infinity. Therefore, when $N_c$ tends to infinity,
no tetraquark bound states or low-mass resonances are expected to
occur. It is only for finite (possibly large) values of $N_c$
that tetraquark bound states might appear in the spectrum, in the
vicinity of the two-meson threshold.
\par
The detailed prediction of the above possibilities depends on the
related integral equation that governs the pole production.
The exact expression of the latter takes, however, a
rather intricate form, since it actually involves, because of the
infinite number of meson states, an infinite number of coupled
equations, with off-mass-shell scattering amplitudes. This problem
is not yet analyzed in the literature.
\par
Bound-state equations for four-quark states, based on the
Bethe--Salpeter and Dyson--Schwinger equations framework, have been
considered in
Refs. \cite{Heupel:2012ua,Eichmann:2015cra,Wallbott:2019dng,
Eichmann:2020oqt},
where a detailed study of the various mechanisms of formation of
tetraquark states has been undertaken. The results confirm the dominant
role of meson clusters inside the tetraquark bound states.
\par
The same mechanism as described above could also be
applied to the other direct channel, $D2$, of the meson-meson
scattering amplitude [Eq.~(\rf{4e4})] by exchanging
the roles of the antiquarks $\overline 2$ and $\overline 4$.
Here, however, the quark flavors and masses having been changed,
the structure of the kernel of the iteration is different from that
of channel $D1$ and hence one should expect to find a tetraquark bound
state different from that of the $D1$ channel. This is why
the existence of two different direct channels might induce the
existence of two different tetraquarks. Each of these, if they exist,
would have a priviledged decay channel, according to the structure
of their couplings to two-meson states.
\par
The previous analysis could also be applied to the case of
cryptoexotic channels, as discussed in Sec.~\rf{s52}, with appropriate
adaptations, taking into account the additional diagrammatic
possibilities, which are mainly related to mixing possibilities with
ordinary mesons. One also encounters here two different structures
in the iteration kernels, leading to two different tetraquarks, as was
emphasized in Sec.~\rf{s52}. 
In the case of a molecular-type mechanism, an additional contribution
comes from diagrams of the types of Figs.~\rf{5f9}b, \rf{5f9}c and
\rf{5f9}e, which, by iteration, may participate in the formation of
the bound state; hence, the possible existence of diagrams \rf{5f11}a
and \rf{5f11}d.
\par
In conclusion, the tetraquark formation mechanism, as resulting from
the large-$N_c$ analysis of Feynman diagrams, seems to be dominated
by the meson-meson interactions through a collaborative iteration of
direct and recombination sectors\footnote{Quark recombination or
interchange channels in meson-meson scattering have been considered
in Ref. \cite{Barnes:1991em}.}. From that viewpoint, we are
rather close to a molecular-type structure, in which the role of
confining interactions has been absorbed by the formation of mesons.
However, residual long-range type interactions may still survive
and contribute in a more specific way.
\par
We shall come back, through another perspective, to the comparison of
the diquark-formation and the mesonic-cluster-formation mechanisms
in Sec.~\rf{s7}.
\par

\section{Molecular states} \lb{s6}

Molecular structure of exotic hadrons had been considered since
the early days of the charmonium discovery
\cite{Voloshin:1976ap,DeRujula:1976zlg,Tornqvist:1993ng}. 
Taking the analogy of the formation of atomic molecules or of nuclei,
it is natural to consider the possibility of the existence of bound
states or resonances resulting from the direct interaction of
ordinary hadrons. Since the latter mutually interact by means of
short-range forces, generated by meson exchanges, one is entitled
to use effective field theories
\cite{Weinberg:1978kz,Gasser:1983yg,Gasser:1984gg,Manohar:1996cq,
Georgi:1990um,Wise:1992hn,Isgur:1991wq,Neubert:1993mb,
Brambilla:2004jw,Gasser:1987rb,Weinberg:1991um,Ordonez:1995rz,
Kaplan:1996xu,Epelbaum:1998ka,Epelbaum:1999dj,Bedaque:2002mn,
Hammer:2019poc}, adapted to the range of energies and masses that
are involved
\cite{Close:2003sg,Wong:2003xk,Swanson:2003tb,Amsler:2004ps,
Tornqvist:2004qy,AlFiky:2005jd,Fleming:2007rp,Dong:2009yp,
Gutsche:2010jf,Molina:2010tx,Aceti:2012cb,Valderrama:2012jv,
Karliner:2016ith,Sakai:2017avl,Baru:2018qkb,Kalashnikova:2018vkv,
Valderrama:2019sid,Habashi:2020ofb,Dong:2021juy,Chen:2021erj}.
In particular, the presence, in the exotic hadrons, of heavy quarks
allows the use, at least partially, of a nonrelativistic formalism,
which considerably facilitates the analysis of the problem.
\par

\subsection{Effective-theory matching} \lb{s61}

The main idea of the formulation of effective field theories is
that the description of the dynamics of physical systems depends
on the energy scale or the distance scale at which one evaluates
physical observables. At low energies, or at large distances,
the dynamics should be insensitive to the details of the dynamics
at high energies or at short distances. It is then advantageous to
integrate out the degrees of freedom that describe the
short-distance dynamics and keep from them only overall effects,
thus reducing the number of degrees of freedom needed for the
description of the system at large distances. In this respect,
a composite particle, which, at the microscopic level, is made of
more elementary constituents, could be approximated, when observed
from a sufficiently large distance, by a pointlike particle, with
appropriate attributes.
\par
The effective theory, resulting from the above reduction of degrees
of freedom, should, however, reproduce the same results than the
full, or microscopic, theory from which it is deduced. This is
generally ensured by imposing matching conditions between the two
theories concerning several observables, like the scattering
amplitudes or form factors.
The structure of the effective theory is organized according to
counting rules involving the energy scale that validates its
existence. In principle, the effective-theory Lagrangian should
contain an infinite number of terms, ranked according to their
dimensionality (in energy scale), the most important terms having
the lowest dimensionality. Truncation of the series after the first
few terms depends on the precision of the calculation that is
required and on the energy scale to which one wishes to extend the
predictions.
The matching conditions allow the determination of the
parameters of the effective theory in terms of quantities known
from the full theory. It is worth emphasizing that, in general,
the effective-theory reduction may involve several energy scales,
for example, when the system contains particles with masses belonging
to different energy scales, in which case the matching conditions
become more involved.
\par
Since most of the experimentally observed tetraquark candidates are
located in the vicinity of two-meson thresholds, the issue of interest
in the molecular scheme is the understanding of the conditions in
which such a situation emerges, both in the effective-theory and in
the full-theory frameworks.
\par
In this respect, Luke and Manohar \cite{Luke:1996hj} have made, by
means of a simplified model, which we briefly sketch below, a thorough
analysis of the various aspects of the problem one meets. The model 
considers, as the full theory, a theory of nonrelativistic fermions
with mass $M$, interacting by means of a Yukawa-type coupling
with a scalar field of mass $m$ and coupling constant $g$. The effective
theory is obtained by integrating out the scalar field and keeping
only the fermion field. In the latter theory, the leading-order
interaction is represented by a four-fermion contact term, with
coupling constant $h$.
Nonleading interactions are represented by
higher-dimensional operators. Matching conditions are implemented
by considering the elastic (off-energy) two-fermion scattering
amplitude. In the full theory, the most important contributions come
from the series of ladder diagrams, while in the effective theory,
the equivalent contributions come from the series of chains of bubble
diagrams, generated by the four-fermion contact term. These diagrams
are represented in Fig.~\rf{6f1}, where the main parameters of the two
theories are also displayed.
This model has also direct connection with the evaluation of the
nucleon-nucleon scattering amplitude at low energies, considered in
\cite{Weinberg:1991um,Ordonez:1995rz,Kaplan:1996xu}.
\par
\bfg
\bc
\epsfig{file=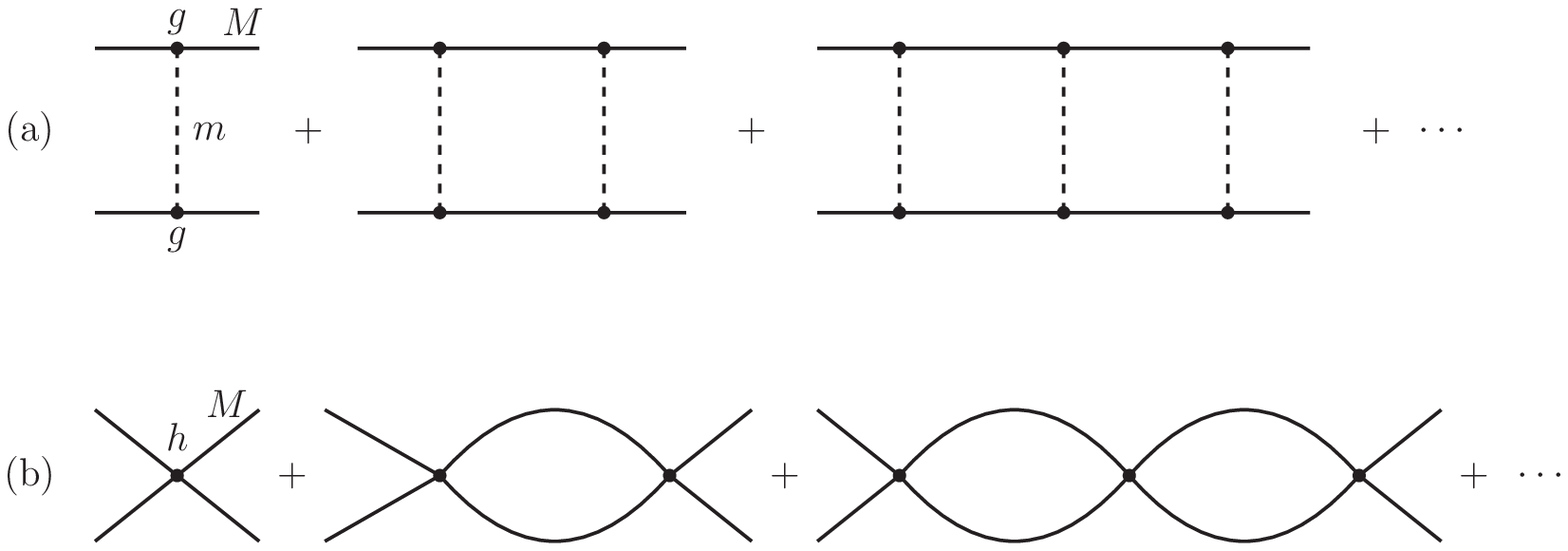,scale=0.7}
\caption{(a) The series of ladder diagrams in the full theory.
(b) The series of chains of bubble diagrams in the effective theory.
The masses and coupling constants are displayed.}
\lb{6f1}
\ec
\efg
\par
In the full theory, the bound-state problem is governed by the
Schr\"odinger equation with the (attractive) Yukawa potential. Bound
states exist only when the coupling constant (squared) is greater than
some critical value:
\be \lb{6e1}
g^2\geq g_{\mathrm{cr.}}^2=1.7\times \Big(\frac{4\pi m}{M}\Big).
\ee
At the critical value, the bound state appears at the two-fermion
threshold (zero binding energy) and when $g^2$ is gradually increased,
the binding energy increases and the bound state goes down in the
potential well. For higher values of $g^2$, new bound states
may appear, representing excited states.
When $g^2< g_{\mathrm{cr.}}^2$, the bound states disappear, having been
transformed into resonances. For values of $g^2$ approaching zero, the
lowest-energy resonance has an energy (more precisely its real part)
that increases up to infinity.
\par
Nonrelativistic or semirelativistic theories are generally expanded
in inverse powers of the heavy-fermion mass $M$. A kinematic quantity
that is adequate for such expansions is the ratio of the
c.m.~momentum $Q$ of the fermions to the mass $M$, $Q/M$.
Introducing the velocity $v$ of the fermions through the relation
$Q=Mv$, the above expansion is therefore an expansion with respect to
$v$; small velocities ensure rapid convergence of the corresponding
series. In some cases, infrared singularities of loop diagrams are
represented by negative powers of $v$ and demand a separate treatment
or isolation of such terms.
\par
Considering now the effective theory, the matching condition with the
full theory, up to some order in the loop counting, implies a
redefinition of the effective coupling $h$. The latter then takes the
following form:
\be \lb{6e2}
h=\sum_{n=0}^{\infty}h^{(n)},
\ee
where $h^{(n)}$ represents the contribution coming from the $n$-loop
calculation. The calculations, up to two
loops, provide the following expansion of $h$:
\be \lb{6e3}
h=h^{(0)}\Big[\ 1+\frac{1}{2}\Big(\frac{g^2M}{4\pi m}\Big)+
\ln\Big(\frac{4}{3}\Big)\,\Big(\frac{g^2M}{4\pi m}\Big)^2+\cdots\ \Big],
\ \ \ \ \ \ h^{(0)}=\frac{g^2}{m^2}.
\ee    
\par
The chain diagrams of Fig.~\rf{6f1}b can be summed using the full
$h$ as the effective coupling. The bubble diagram is ultraviolet
divergent and requires renormalization. After the integration of the
temporal component of the loop variable is done (the effective theory
being formulated in the nonrelativistic regime), the three-dimensional
part is regularized by dimensional regularization, in which case no
subtraction is needed, the ultraviolet divergence being linear (odd
power of the momentum). The series is simply a geometric series and is
easily summed. One finds for the off-energy scattering amplitude
\be \lb{6e4}
\mathcal{A}=\frac{h}{1+hM^{3/2}(-E)^{1/2}/(4\pi)},
\ee
where $E$ is the total (nonrelativistic) energy of the two fermions.
For $h$ negative, there is a bound state with energy (cf. also
\cite{Weinberg:1991um})
\be \lb{6e5}
E=-\frac{16\pi^2}{h^2M^3}.
\ee
(The binding energy is $B=-E$.)
When $|h|\rightarrow \infty$, the bound-state energy tends to zero
and the bound state approaches the threshold. We have seen that in
the full theory this situation occurs when $g^2$ tends to
$g_{\mathrm{cr.}}^2$ from above [Eq.~(\rf{6e1})]. This means that
the above limit of $h$ occurs for a finite value of $g^2$, which is
$g_{\mathrm{cr.}}^2$; therefore, the series (\rf{6e2}) diverges for
that value of $g^2$, which signals the fact that the correspondence
between the full and the effective theories is no longer
perturbative. In this domain of the coupling constant, the
higher-dimensional operators, which have been neglected in the above
evaluations, become relevant to all orders and may signal the
breakdown of the effective theory. On the other
hand, when $h$ tends to zero in its negative domain, the bound-state
energy tends to $-\infty$ and the bound state disappears from the
bottom of the energy domain. In the full theory, this happens when
$g^2$ tends to $+\infty$. However, when $g^2$ gradually increases, new
bound states appear in the spectrum of the full theory, representing
excited states; these are not reproduced in the effective theory.
In the domain of positive values of $h$, bound states do not exist.
This, therefore, corresponds to the domain
$0\leq g^2\leq g_{\mathrm{cr.}}^2$; here, the full theory displays
resonances, but in the effective theory they are absent. (The effective
theory might display a resonance in the case of derivative-type
couplings, a situation that occurs in chiral perturbation theory
(cf. Sec.~\rf{s63}).) A perturbative matching between the full and the
effective theories occurs only in the weak-coupling regime of the
former, i.e., $g^2\simeq 0$, which entails $h\simeq 0$ and $h>0$. 
\par
From the above comparisons, one may deduce a schematic qualitative
correspondence between the couplings of the full and the effective
theories, which is represented in Fig. \rf{6f2}.
\bfg
\bc
\epsfig{file=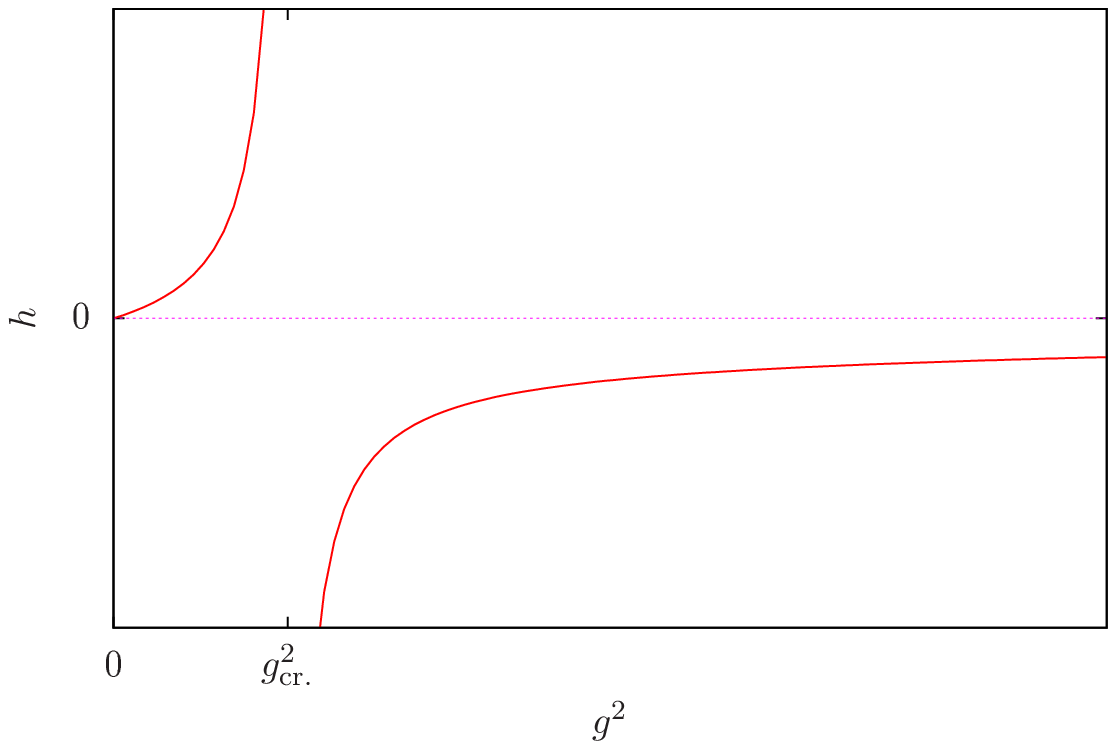,scale=0.8}
\caption{A schematic behavior of the effective-theory coupling
constant $h$ with respect to variations of the full-theory coupling
constant squared $g^2$.}  
\lb{6f2}
\ec
\efg
The more singular
behavior of the effective theory is a consequence of the fact that
the bubble diagram is ultraviolet divergent, while the box diagram
in the full theory is finite and therefore reflects smoothly the
variations of the coupling constant. In $x$-space, the bound-state
equation of the effective theory is governed by the three-dimensional
$\delta$-function potential, which requires renormalization of the
coupling constant \cite{Jackiw:1991je}.
\par
In spite of the breakdown of perturbation theory in the region
$g^2\sim g_{\mathrm{cr.}}^2$ and the lack of explicit correspondence
between $g^2$ and $h$, it is possible to deduce further information
through parameters that have direct connection with experimental
measurements. These are the $S$-wave scattering length $a$ and the
effective range $r_e^{}$. When the bound state approaches the
threshold, the scattering length increases and tends to $\infty$
at threshold, while the effective range remains finite. It is then
shown that $h$ is related to the exact scattering length of the
Yukawa theory
\cite{Weinberg:1991um,Kaplan:1996xu}:
\be \lb{6e6}
h=-\frac{4\pi}{M}\,a,
\ee
whereas higher-dimensional derivative terms are suppressed by
powers of $(4\pi r_e^{}/a)$. Therefore, the bound-state energy
(\rf{6e5}) remains a valid result for large values of the
scattering length. However, the radius of convergence of
the effective theory is much reduced and is given by values of the
momentum $k$ of the order of $\sqrt{2/(ar_e^{})}$ and not by
$r_e^{-1}\sim m^{}$.
\par
The above considerations allow us to foresee the implications of
the large-$N_c^{}$ limit on the possibility of formation of bound
states. Here, the full theory is QCD, while the effective theory
is a meson theory, where mesons mutually interact either by meson
exchanges, characterized by some effective generic coupling $g^2$,
or by contact terms, characterized by an effective coupling $h$.
We have seen in Sec.~\rf{s23} that three-meson couplings
and four-meson couplings generally scale like $1/N_c^{1/2}$ and
$1/N_c$, respectively (cf.~Figs.~\rf{2f10} and \rf{2f12}). 
Therefore, in the large-$N_c$ limit, the meson-interaction couplings
tend to zero and one reaches the situation where the mesons become
free non\-inter\-acting particles. For large finite values of $N_c$,
the effective theory is then in the weak-coupling regime, which
corresponds to the phase $g^2\simeq 0$, $h\simeq 0$ with $h>0$.
In this situation, one does not expect to find bound states. Rather,
one should have resonances located far from the two-meson thresholds.
\par
When $N_c$ is decreased down to finite values, close to the physical
value 3, two types of evolution might be expected, depending on the
quark masses that are involved and also on the detailed quantum
numbers of the system that is considered. 
For light quarks, the evolution probably remains in the phase
$g^2\leq g_{\mathrm{cr.}}^2$, $h>0$, in which case the lowest-mass
resonance approaches the two-meson threshold, but still remains
sufficiently far from it. This is corroborated by results obtained
within the framework of chiral perturbation theory (cf.~Sec.~\rf{s63}).
For systems involving heavy quarks, it seems that the evolution
reaches the critical region of the vicinity of the two-meson
threshold, characterized by $g^2\sim g_{\mathrm{cr.}}^2$ and
$|h|\sim \infty$, where a bound state or a resonance might appear.
It is worth noticing that $g_{\mathrm{cr.}}^2$ is inversely proportional
to the external-meson masses [Eq.~(\rf{6e1})], which feel the heavy
quark masses, and therefore has a smaller value for such systems,
making easier the appearance of near-threshold structures.
This is corroborated by the experimental observations of exotic
hadrons involving heavy quarks. A similar conclusion has also been
reached, from a different point of view, in Ref.~\cite{Dong:2020hxe}.
The above descriptions, while remaining at the level of observations,
require, however, more dynamical justifications at the QCD level. 
\par

\subsection{Compositeness} \lb{s62}

The description of the internal structure of a bound state depends
upon the scale at which the latter is probed. The mean size of the
bound state is one of the criteria that can be used to distinguish
two situations: (i) a large size, characterizing a loosely bound state,
in which one may distinguish the existence of two or several clusters;
(ii) a compact size, characterizing an undecomposable or elementary
object. It is evident, however, that probing the latter object with
higher precision, one may discover that, in turn, it is also
decomposable into more elementary clusters. The most natural example
of this situation comes from the nuclei, which, in first
approximation, can be described as made of nucleons, considered as
pointlike objects. However, observing the nucleons at shorter distances,
one realizes that the latter are themselves made of more elementary
particles, which are the quarks and the gluons.
\par
The same criterion also applies to the case of exotic hadrons.
Molecular-type exotic hadrons would be described as composed of
clusters of ordinary hadrons, interacting by means of effective forces,
while compact exotic hadrons would be described by the direct
interactions of quarks and gluons, without leading to the appearance
of hadronic clusters. The solution of bound-state equations and
comparison of their predictions with experimental data would be
sufficient to settle the question of the nature of a bound state;
however, in many cases, the interactions that are at work have
a non\-per\-tur\-ba\-tive character and do not allow for a deductive
solution of the problem. In this case, the knowledge of external
criteria, related to experimental data, could bring a complementary
view to the efforts of understanding the problem.
\par
In this respect, Weinberg proposed in the past, in the framework of
nonrelativistic quantum mechanics, an independent criterion for the
probe of compositeness of the deuteron, whose binding energy is much
smaller than the scale of the strong interactions that govern
nuclear physics \cite{Weinberg:1962hj,Weinberg:1965zz}. Defining
$Z$ as the probability of having the deuteron as an elementary particle
($0\leq Z\leq 1$), the probability of having it as made of a proton and
a neutron is $(1-Z)$. The vicinity of the deuteron state to the
two-nucleon threshold allows one to use, for the scattering phase
shift $\delta$ of the two nucleons, the effective range expansion in
terms of the scattering length $a$ and the effective-range $r_e^{}$
\be \lb{6e7}
k\cot\delta\simeq -\frac{1}{a}+\frac{1}{2}r_e^{}k^2,
\ee
and to relate the latter quantities to the binding energy and the
parameter $Z$. Weinberg finds
\bea \lb{6e8}
& &a=[2(1-Z)/(2-Z)]\,R+O(m_{\pi}^{-1}),\ \ \ 
r_e^{}=[-Z/(1-Z)]\,R+O(m_{\pi}^{-1}),\nonumber \\
& &R=(2\mu B)^{-1/2},
\eea
where $B$ is the deuteron binding energy, $B=2.22$ MeV, $\mu$ the
proton-neutron reduced mass, $m_{\pi}^{}$ the pion mass, and $R$
represents the deuteron radius.
If the deuteron is composite, $Z\simeq 0$, $a$ takes its maximum value,
while $r_e^{}\simeq 0$; in the opposite case, if the deuteron is
elementary, $Z\simeq 1$, $a\simeq 0$ and $r_e^{}\rightarrow -\infty$. 
Experimental data, $a=+5.41$ fm and $r_e^{}=1.75$ fm
\cite{Weinberg:1965zz,Stoks:1988zz}, clearly favor the composite
nature of the deuteron, made of a proton and a neutron. Notice that,
for $Z\simeq 0$, the relationship between the binding energy and
the scattering length of Eq.~(\rf{6e8}) reduces to Eq.~(\rf{6e5}),
using (\rf{6e6}).
\par
Weinberg's criterion has been investigated by many authors and extended
to a wider range of applicability, such as to resonances,
multi-channel processes and relativistic cases
\cite{Baru:2003qq,Cleven:2011gp,Hanhart:2011jz,Hyodo:2011qc,
Sekihara:2014kya,Guo:2015daa,Meissner:2015mza,Guo:2020vmu}.
Oller has proposed a new criterion for compositeness, based on
the use of the number operators of free particles
\cite{Oller:2017alp}. The compositeness criterion thus brings 
a complementary constraint for the analysis of the internal structure
of bound states and resonances.
\par

\subsection{Resonances with light quarks} \lb{s63}

Molecular systems made of light quarks ($u,~d,~s$) need the use of a
relativistic formalism. Since the hadronic clusters interact by
means of short-range forces and the light quarks have a relativistic
motion, one guesses that the mean forces experienced by the
clusters are weaker than in the case of heavy quarks, which tend to
stabilize the general motion. Therefore, the effective theory is
expected here to be in its weak-coupling regime, where no bound
states can be produced. Rather, one expects the appearance of
resonances, generally located, mostly for the case of the quarks $u$
and $d$, far from the two-hadron threshold.
\par
The light-quark effective field theory which describes QCD at low
energies is chiral perturbation theory (ChPT)
\cite{Weinberg:1978kz,Gasser:1983yg,Gasser:1984gg,Manohar:1996cq}.
The case of SU(2)$_L\times$SU(2)$_R$ chiral symmetry, involving
pions, has been widely studied in the literature and precise
experimental tests have confirmed its validity \cite{Colangelo:2001df,
Ananthanarayan:2000ht}.
\par
The domain of validity of ChPT, concerning the meson momenta, extends
over a few hundreds of MeV. Since the perturbative expansion is done
polynomially (up to logarithms) in the momenta, it is not expected to
find poles with the first few terms and to be able to probe directly
the properties of resonances. To extend the domain of predictivity of 
ChPT, it has been combined with dispersion relations, using
analyticity and crossing-symmetry properties 
\cite{Colangelo:2001df,Ananthanarayan:2000ht}. 
Extending the above setup to the complex plane, it has been shown
that the $\pi\pi$ scattering amplitude in its partial $S$ wave, with
isospin 0, possesses a pole in the second Riemann sheet at the
complex mass value $M=(441-i272)$ MeV \cite{Caprini:2005zr}. This
solution has been identified with the $f_0(500)/\sigma$ meson, whose
existence had been controversial for several decades, but which later
had re-emerged through new experimental results as a wide resonance.
Confirmation of the above result has been obtained in
\cite{Mennessier:2010xg,GarciaMartin:2011jx,Moussallam:2011zg}.
\par
It had also been suggested that properties of resonances could be
more directly probed by using elastic unitarity in its full form,
by means of the inverse amplitude method 
\cite{Truong:1988zp,Dobado:1989qm,Dobado:1996ps,Boglione:1996uz,
Nieves:1999bx}.
A thorough study of the scalar resonance $f_0^{}(500)/\sigma$, based
on the combined frameworks of ChPT, dispersion relations and
unitarization, has been undertaken by Pel{\'a}ez \textit{et al.}
\cite{Pelaez:2006nj,Pelaez:2015qba,Pelaez:2021dak}. Their analysis
also leads to a pole position in the second Riemann sheet of the
complex plane, at the mass value $M=(449-i275)$ MeV, thus confirming
the result of \cite{Caprini:2005zr}.
\par
To appreciate the power of the unitarization scheme
(cf. also \cite{Oller:1997ti,Oller:1997ng,Oller:2019opk,Oller:2020guq,
Yao:2020bxx}), one may consider
the partial $S$-wave isospin-0 $\pi\pi$ scattering amplitude in
its leading order, $O(p^2)$, of ChPT. The latter reads
\be \lb{6e9}
t_{\ell=0}^{I=0}(s)\equiv t(s)=\frac{2s-M_{\pi}^2}{32\pi F_{\pi}^2},
\ee
where $s$ is the Mandelstam variable and $F_{\pi}^{}$ is the pion
decay constant, defined as in Eq.~(\rf{2e12}) with an axial-vector
current ($F_{\pi}^{}\simeq 92.3$ MeV) and $M_{\pi}^{}$ is the pion mass.
Notice that in ChPT the
meson-meson interactions begin with derivative couplings, this is
why an $s$-dependence appears at leading order. The elastic unitarity
condition reads
\be \lb{6e10}
\mathrm{Im}\,t(s)=\sigma(s)|t(s)|^2,\ \ \ \ \ \
\sigma(s)=\Big(1-4M_{\pi}^2/s\Big)^{1/2}.
\ee
This shows that the imaginary part of the scattering amplitude is
of higher order, $O(p^4)$, than the real part.
In terms of the inverse amplitude it takes the form
\be \lb{6e11}
\mathrm{Im}\,\frac{1}{t(s)}=-\frac{\mathrm{Im}\,t(s)}{|t(s)|^2}
=-\sigma(s).
\ee
Therefore, the imaginary part of the inverse of the amplitude is
explicitly known and reduces to a kinematic factor. This allows one
to complete expression (\rf{6e9}), by incorporating in it information
(\rf{6e11}), and considering (\rf{6e9}) as the real part of the
amplitude:
\be \lb{6e12}
t(s)=\frac{\mathrm{Re}\,t(s)}{1-i\sigma(s)\mathrm{Re}\,t(s)}.
\ee
\par
In searching for poles in the second Riemann sheet, one considers
the complex conjugate of the amplitude of the first Riemann sheet.
Expression (\rf{6e12}) becomes
\be \lb{6e13}
t^{\mathrm{II}}(s)=\frac{\mathrm{Re}\,t(s)}{1+
i\sigma^{\mathrm{II}}(s)\mathrm{Re}\,t(s)}.
\ee
Notice that this equation is the analog of Eq.~(\rf{6e4}), obtained
by summing a series of bubble diagrams. In the lower $s$-plane,
$\sigma(s)$ undergoes the change
$\sigma^{\mathrm{II}}(s)=-(\sigma(s^*))^*$.
The pole position is obtained from the zero of the denominator,
\be \lb{6e14}
\sigma(s_{\sigma})(2s_{\sigma}-M_{\pi}^2)=i32\pi F_{\pi}^2,
\ee
which gives $\sqrt{s_{\sigma}}=(493-i441)$ MeV, of the same order
of magnitude as the precise values obtained above. This means that
the leading-order ChPT scattering amplitude, together with the
unitarization condition, drags the solution to its exact value.
\par
One of the advantages of the inverse amplitude method is that it
displays more explicitly the large-$N_c$ behavior of the various
contributions. At large $N_c$, the pion mass remains unaffected
at leading order, $M_{\pi}^{}=O(N_c^0)$, while the decay constant
scales like $N_c^{1/2}$, $F_{\pi}^{}=O(N_c^{1/2})$ [Eq.~(\rf{2e13})].
Taking into account these behaviors, one finds that Eq.~(\rf{6e14})
reduces to
\be \lb{6e15}
\sqrt{s_{\sigma}}=(1-i)\sqrt{8\pi}F_{\pi}^{}=O(N_c^{1/2}),
\ee
which shows that the mass and width of the $f_0(500)/\sigma$ meson
increase like $\sqrt{N_c^{}}$ at large $N_c$. This is in accordance
with the weak-coupling regime of molecular effective theories that
we have met in Sec.~\rf{s61}. This shows that the
$f_0(500)/\sigma$ meson is mainly made of two pions, rather than
of a pair of quark and antiquark, in which case the mass should
remain stable under changes of $N_c$ and the width would decrease
\cite{Pelaez:2006nj,Pelaez:2015qba} (cf. also
\cite{Sun:2005uk,Nieves:2011gb}). Nevertheless, because of
mixing possibilities, the $f_0(500)/\sigma$ seems to have a small
component of $\overline qq$ state.
\par
The detailed analysis applied to the case of $f_0(500)/\sigma$ has
also been applied to the case of the $\rho$ meson, which appears
as a resonance in the $P$ wave. 
Here, however, the behaviors of the mass and the width under variations
of $N_c$ confirm the fact that the $\rho$ meson is mainly made of
a $\overline qq$ pair \cite{Pelaez:2006nj,Pelaez:2015qba}.
\par
In conclusion, the molecular scheme, considered as an effective
theory, provides a systematic tool of investigation of the
properties of many exotic-type states, either in the domain of
heavy quarks, or that of light quarks.
\par

\section{The cluster reducibility problem} \lb{s7}

One salient feature of the multiquark currents, met in Secs.~\rf{s31}
and \rf{s32}, is their decomposition property into combinations of
products of meson and/or baryon currents, typical examples of which are
Eqs.~(\rf{3e6}) and (\rf{3e10}). This suggests that multiquark
states are not color-irreducible, unlike ordinary hadrons, and,
therefore, could not be put on the same footing as the latter states.
The consequences of this fact are easily conceivable. If, within a
multiquark state, clusters of ordinary hadrons may be formed, and since
the mutual interactions of the latter are not confining, the
multiquark state will have the tendancy to be dissociated into its
hadronic components or to be transformed into a loosely bound state
of hadrons.
\par
One might still think that the above color-reducibility property
concerns only couplings to local operators which involve a few
moments of the corresponding bound state wave function. Actually,
the property is very general and concerns also the couplings to
multilocal operators \cite{Lucha:2019cdc}. 

\subsection{Cluster reducibility of multilocal operators} \lb{s71}

We shall briefly sketch, in this subsection, the case of multilocal
operators.
\par
The proof of cluster reducibility of multilocal operators is
based on two properties of the gauge links (\rf{3e24}): they are
elements of the gauge group SU(3) (and, more generally, of SU($N_c$)),
and have a determinant equal to 1. Hence, they satisfy the group
composition law
\be \lb{7e1}
U_{\ b}^a(C_{zy})U_{\ c}^b(C_{yx})=U_{\ c}^a(C_{zyx}),
\ee
where $C_{zyx}$ is the line composed of the union of the two lines
$C_{zy}$ and $C_{yx}$, with a junction point at $y$. The expression
of the determinant of $U$ is (cf.~Ref.~\cite{Migdal:1984gj},
Appendix C)
\be \lb{7e2}
\mathrm{det}(U(C_{yx}))=1=\frac{1}{3!}\epsilon_{a_1a_2a_3}
\epsilon^{b_1b_2b_3}U_{\ b_1}^{a_1}(C_{yx})U_{\ b_2}^{a_2}(C_{yx})
U_{\ b_3}^{a_3}(C_{yx}).
\ee
Considering, for definiteness, the tetraquark operator of
Fig.~\rf{3f2}a, multiplying it with the determinant of the gauge
link of the line $C_{yx}$ and using the contraction property of two
$\epsilon$ tensors,
\be \lb{7e3}
\epsilon^{c_1c_2c_3}\epsilon_{d_1d_2d_3}=\delta_{\ d_1}^{c_1}
\delta_{\ d_2}^{c_2}\delta_{\ d_3}^{c_3}+\sum_{k_i}(-1)^P
\delta_{\ d_1}^{c_{k_1}}\delta_{\ d_2}^{c_{k_2}}
\delta_{\ d_3}^{c_{k_3}},
\ee
where the sum runs over all permutations of the indices $k_i$, with
the sign of the parity of the permutation represented by $(-1)^P$,
one ends up with a decomposition of the operator into a sum of
products of two meson operators of the type (\rf{3e25}), where
the gauge link lines have a polygonal structure. Further
simplification occurs, using the backtracking condition
\cite{Makeenko:1999hq}
\be \lb{7e4}
U_{\ b}^a(C_{yx})U_{\ c}^b(C_{xy})=\delta_{\ c}^a,
\ee
which expresses the unitarity property of the parallel transport
operation, leading to the decomposition displayed graphically in
Fig.~\rf{7f1}, which is the multilocal form of the first of
Eqs.~(\rf{3e6}). 
\bfg
\bc
\epsfig{file=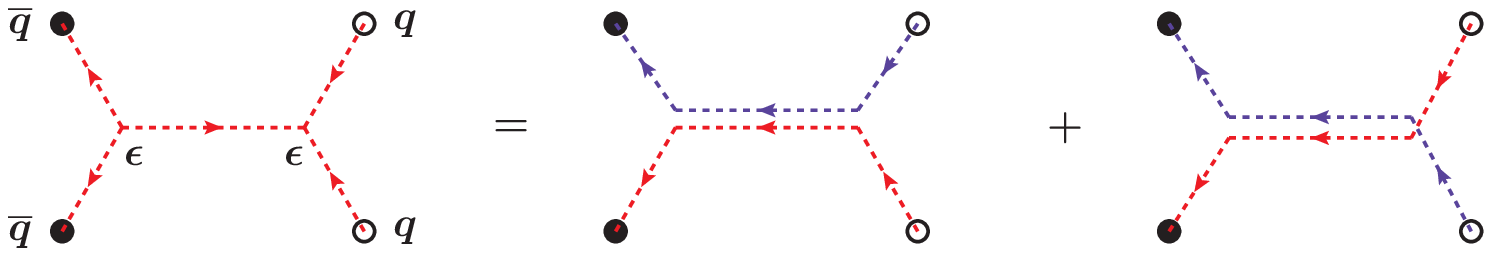,scale=0.8}
\caption{Decomposition of the tetraquark operator into a
combination of products of two meson operators.}
\lb{7f1}
\ec
\efg
The meson operators that appear on the right-hand side of the
equation come out with polygonal phase-factor lines.   
\par
The same procedure can be applied to the cases of pentaquark
and hexaquark operators. The pentaquark operator of Fig.~\rf{3f2}b
is decomposed into a combination of products of a mesonic and a
baryonic operator (cf. Fig.~\rf{7f2}). 
\bfg
\bc
\epsfig{file=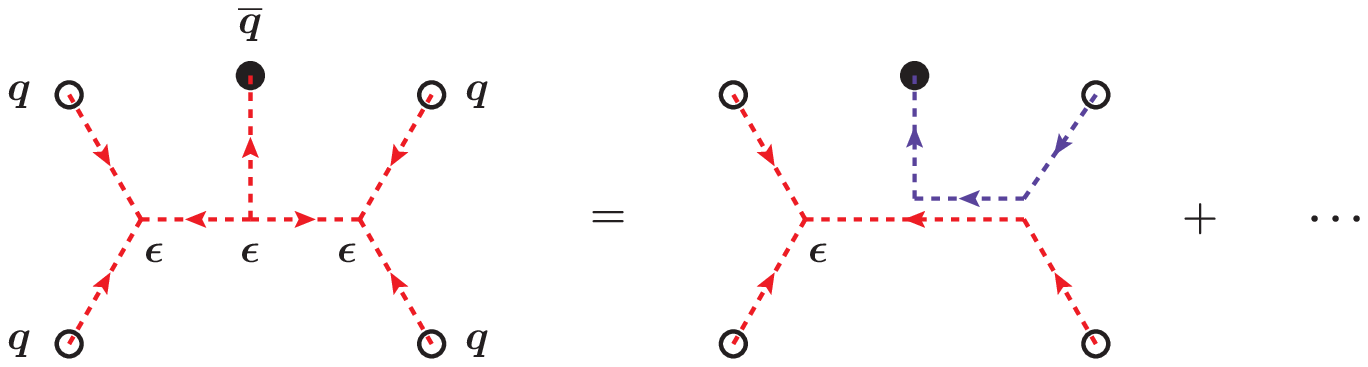,scale=0.8}
\caption{Decomposition of the pentaquark operator into a
combination of products of a meson and a baryon operator.
The ellipsis indicates the remaining three other products.}
\lb{7f2}
\ec
\efg
\par
The hexaquark operator of Fig.~\rf{3f2}c is decomposed into a
combination of products of two baryonic operators (cf. Fig.~\rf{7f3}).
\bfg
\bc
\epsfig{file=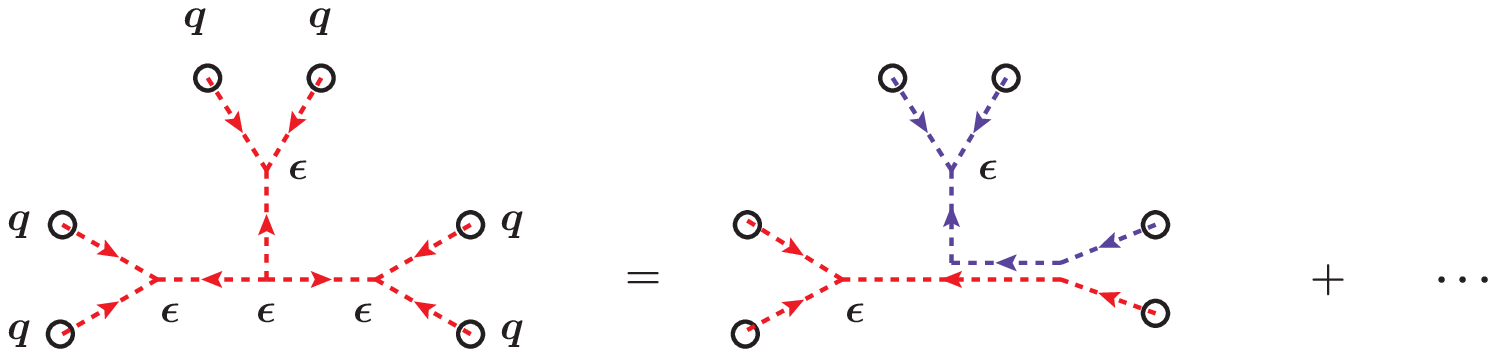,scale=0.8}
\caption{Decomposition of the hexaquark operator into a
combination of products of two baryon operators.
The ellipsis indicates the remaining other products.}
\lb{7f3}
\ec
\efg
\par
The cluster reducibility property remains also valid in the
more general SU($N_c$) case \cite{Lucha:2019cdc}, by appropriately
generalizing the $\epsilon$-tensor properties \cite{Migdal:1984gj}. 
Thus, for tetraquarks, the first operator in Fig.~\rf{3f4} decomposes
into $(N_c-1)!$ combinations of $(N_c-1)$ mesonic operators. The last
operator in that figure decomposes into combinations of two mesonic
operators and of Wilson loops, according to the types of paths used
between the two junction points. (In the case of straight lines, the
Wilson loops disappear.) Similar decompositions occur for the
other operators (not drawn) of that figure denoted by the ellipses
and corresponding to intermediate-type representations.
\par
For pentaquarks, the first operator in Fig.~\rf{3f5} decomposes into
a combination of products of $(N_c-2)$ mesonic operators and one
baryonic operator. The last operator of the figure decomposes
into a combination of products of one mesonic operator and one
baryonic operator, and possibly of Wilson loops.
\par
For hexaquarks, the first operator in Fig.~\rf{3f6} decomposes into
a combination of $(N_c-1)$ baryonic operators, while the last
operator of that figure decomposes into a combination of products
of two baryonic operators, and possibly of Wilson loops.
\par

\subsection{Energy balance} \lb{s72}

The main question that remains to be clarified concerning the cluster
reducibility property of multiquark operators is whether it survives
in the quantized theory. This is not a trivial question, because
the equivalence relations, which were established on formal grounds
between $Y$-shaped operators and combinations of products
of ordinary hadronic operators, involve products of some phase
factors along the same lines, but belonging to different global
paths (cf. Figs.~\rf{7f1}, \rf{7f2} and \rf{7f3}, right-hand sides
of the equalities).
The disentanglement of such phase factors, in order to introduce
them inside different hadronic states, corresponding to the
clusters, might not be a neutral operation and might imply energy
loss or gain. We shall show below, on simple examples, that this
is indeed the case and that the result depends on the geometrical
properties, in coordinate space, of the representations that are
considered, favoring only one of the two sides of the equivalence
relations.
\par
General energy considerations are much more transparent in the case
of static quarks, corresponding to the infinite mass limit of
heavy quarks, kinematic effects of the motion of quarks being
then suppressed; actually, the latter, for finite masses, do not
affect the main property of confinement of the theory and
introduce only nonleading terms in the confining regime.
Therefore, the static limit is expected to provide, in the
confining regime, the main qualitative properties that are
searched for.
\par
To extract the interaction energy properties of a static system,
one generally considers correlation functions of appropriate
operators having couplings to such systems. The calculation
involves, among others, the propagators of the static quarks
in the presence of gluon fields, which are essentially proportional
to the gluon-field path-ordered phase factors along the time direction
\cite{Brown:1979ya,Kogut:1982ds,Kaku:1993ym}. Considering
in the correlation functions gauge-invariant multilocal operators,
which involve in their definitions gluon-field phase factors,
one ends up with the vacuum expectation value of an expression
involving the color trace of phase factors along a closed contour
(or traces along closed contours in the case of non-connected
operators), which defines the vacuum average of the Wilson loop.
On the other hand, inserting in the correlation function a complete
set of intermediate hadronic states and taking the total evolution
time $T$ to infinity, one selects the ground-state hadron, which
yields a factor $e^{-iET}$, where $E$ is the corresponding
interaction energy, the quark-mass contributions having been
factorized. Then the comparison of this term with the Wilson-loop
contribution allows for the determination of $E$.
\par
We first consider the simplest example, corresponding to a mesonic
operator [Eq.~(\rf{3e25})] with a phase factor along a
straight-line segment [Fig.~\rf{3f1}a] of length $R$, taken in the
three-dimensional space orthogonal to the time axis; $R$ is equal
to the distance between the quark and the antiquark, considered
at equal times. The resulting Wilson-loop contour, for an evolution
of the system during a time $T$, is a rectangle of length $T$ and
width $R$, represented in Fig.~\rf{7f4}.
\bfg
\bc
\epsfig{file=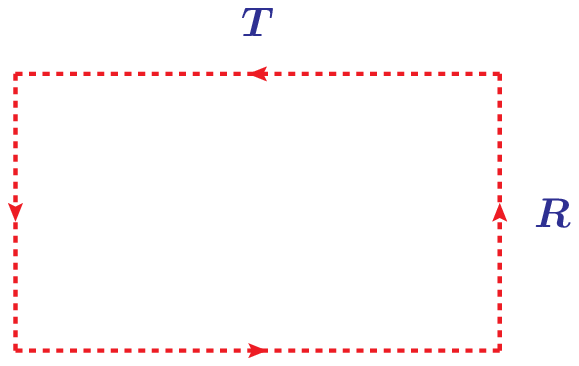,scale=0.8}
\caption{Wilson-loop contour resulting from the time evolution of
a meson containing a static quark-antiquark pair.}
\lb{7f4}
\ec
\efg
\par
The evaluation of the vacuum average of Wilson loops, in the absence
of a completely analytic solution of QCD in its confining regime,
can be numerically done in lattice gauge theory, where one works in
Euclidean spacetime. Furthermore, making there the strong-coupling
expansion leads to analytic predictions, which are generally confirmed
by experimental data \cite{Wilson:1974sk,Kogut:1982ds,Kaku:1993ym}.
Generally, in the evaluation of the energy balance, one uses the
fact that, in Euclidean space, the generating functional of
Green's functions is dominated by field configurations which minimize
the energy.
\par
At leading order of the strong-coupling expansion, the Wilson-loop
vacuum average is obtained by paving the minimal area enclosed by
the contour with the lattice plaquettes. In the case of the rectangle
above, the area is simply the product $RT$; the Wilson-loop
vacuum average is then proportional to the factor $e^{-\sigma RT}$,
where $\sigma$ is the ``string tension'', here defined as
$\sigma=\frac{1}{a^2}\ln(g^2)$, where $a$ is the lattice spacing and
$g$ the QCD coupling constant. Upon comparing this factor with
$e^{-ET}$ (the Euclidean version of $e^{-iET}$), one obtains
\be \lb{7e5}
E=\sigma R,
\ee
which means that the static quark-antiquark interaction energy
increases linearly with the separation distance. This result
analytically establishes the confinement of quarks. It has also
been confirmed by direct numerical evaluations in lattice theory
\cite{Bali:2000gf}. Corrections, coming from finite mass and spin
effects, have been evaluated in Ref. \cite{Eichten:1980mw}.
\par
We next examine the question of the possible influence on the
energy of the state coming from a deformation of the phase-factor
line in the definition of the mesonic operator (\rf{3e25}). We
choose, as a simple example, a rectangular line in position space,
orthogonal to the time direction (Fig.~\rf{7f5}a). The Wilson-loop
contour, generated by this operator during a time evolution
$T$, is represented in Fig.~\rf{7f5}b by the dashed oriented
line. The area mapped by the lattice plaquettes in the
strong-coupling approximation is equal to $R(T+2d)$. It is only
the factor multiplying the variable $T$ that contributes to the
interaction energy; from this, one deduces the same relation
as in Eq.~(\rf{7e5}), which shows that the phase-factor line
deformation in the mesonic operator does not modify the energy of
the state. The resulting change in the Wilson-loop vacuum average
comes from the factor $e^{-\sigma 2dR}$, which is absorbed in the
meson wave-function expressions. Also, line deformations having
components on the time axis, with finite size $\Delta t_0$, say, cannot
modify the energy of the state, since the size $\Delta t_0$
becomes negligible in front of $T$, when the latter goes to infinity.
\par
\bfg
\bc
\epsfig{file=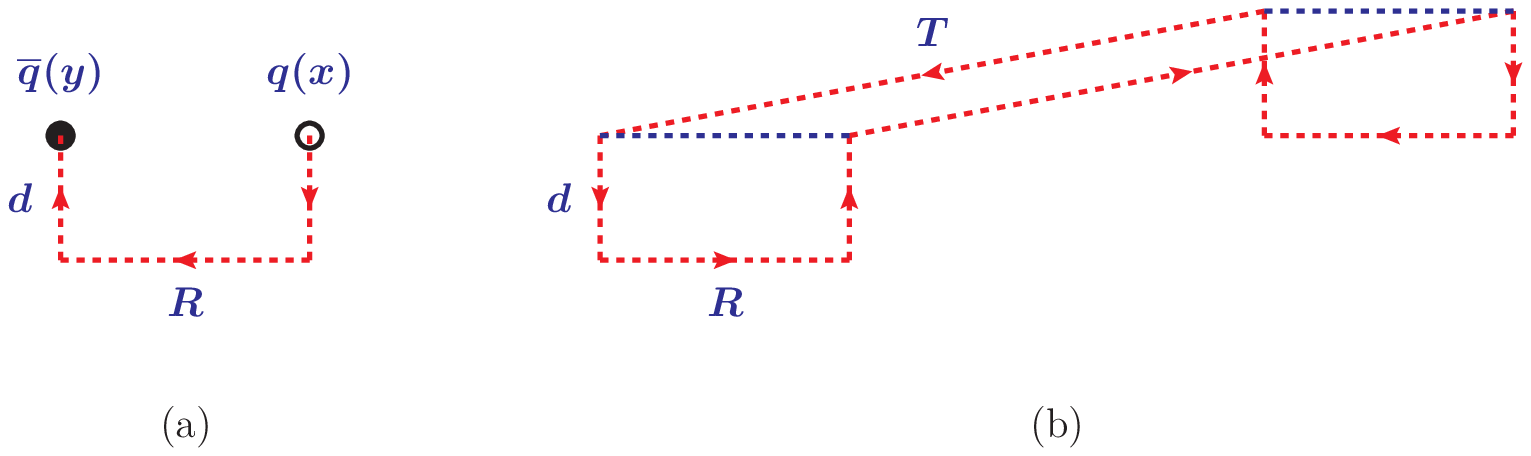,scale=0.8}
\caption{(a) A mesonic operator with a rectangular phase-factor line;
(b) the Wilson-loop contour, generated after a time evolution $T$,
represented by the dashed oriented line.}
\lb{7f5}
\ec
\efg
\par
It is worth noticing that, in the continuum theory, it is expected
that, for large contours, the Wilson-loop vacuum average is saturated
by the minimal surface enclosed by the contour
\cite{Wilson:1974sk,Makeenko:1980wr,Jugeau:2003df}. The minimal surface
corresponding to the contour of Fig.~\rf{7f5}b does not coincide with
the union of the three rectangular areas delineated by the contour
and paved by the plaquettes of the lattice. According to the defining
equation of the minimal surfaces, it lies outside these areas
\cite{Jugeau:2003df}. However, when the limit $T\rightarrow\infty$ is
taken, it shrinks to the above areas and thus provides the same result
as that obtained on the lattice.
\par
We now consider the problem of the tetraquark operator in the
diquark-antidiquark antisymmetric representation together with
its reduction into two mesonic operators, as represented in
Fig.~\rf{7f1}. As a consequence of the independence of the energy
of a system of the types of phase-factor lines, as shown above, one
can deform, in the mesonic operators of the right-hand side of the
equality in Fig.~\rf{7f1}, the phase-factor lines to transform them
into straight line segments joining the quark to the antiquark.
In the configuration adopted in that figure, the first diagram of
the two-meson system involves smaller lengths for the distances
between the quark and the antiquark inside the mesons with respect
to the second diagram. According to Eq.~(\rf{7e5}), the energy of
the first diagram being thus smaller than that of the second diagram, 
one can drop for the present study the contribution of the latter,
which will give negligible contributions compared to the first
one in the Wilson-loop evaluations. (We recall that the quarks are
static.) Since we are interested in qualitative aspects, we make
further simplifications in the geometric configurations of each
representation. We choose equal distances $\ell$ between the
quark and the antiquark in each meson. The two mesonic operators
are placed in the same plane, along parallel directions, the two
quarks and the two antiquarks being aligned along vertical lines,
separated by a distance $d$ (see Fig.~\rf{7f6}a, top). The same
configuration of the quarks and the antiquarks is also chosen for
the tetraquark operator (Fig.~\rf{7f6}b, top).
\bfg
\vspace*{0.5 cm}
\bc
\epsfig{file=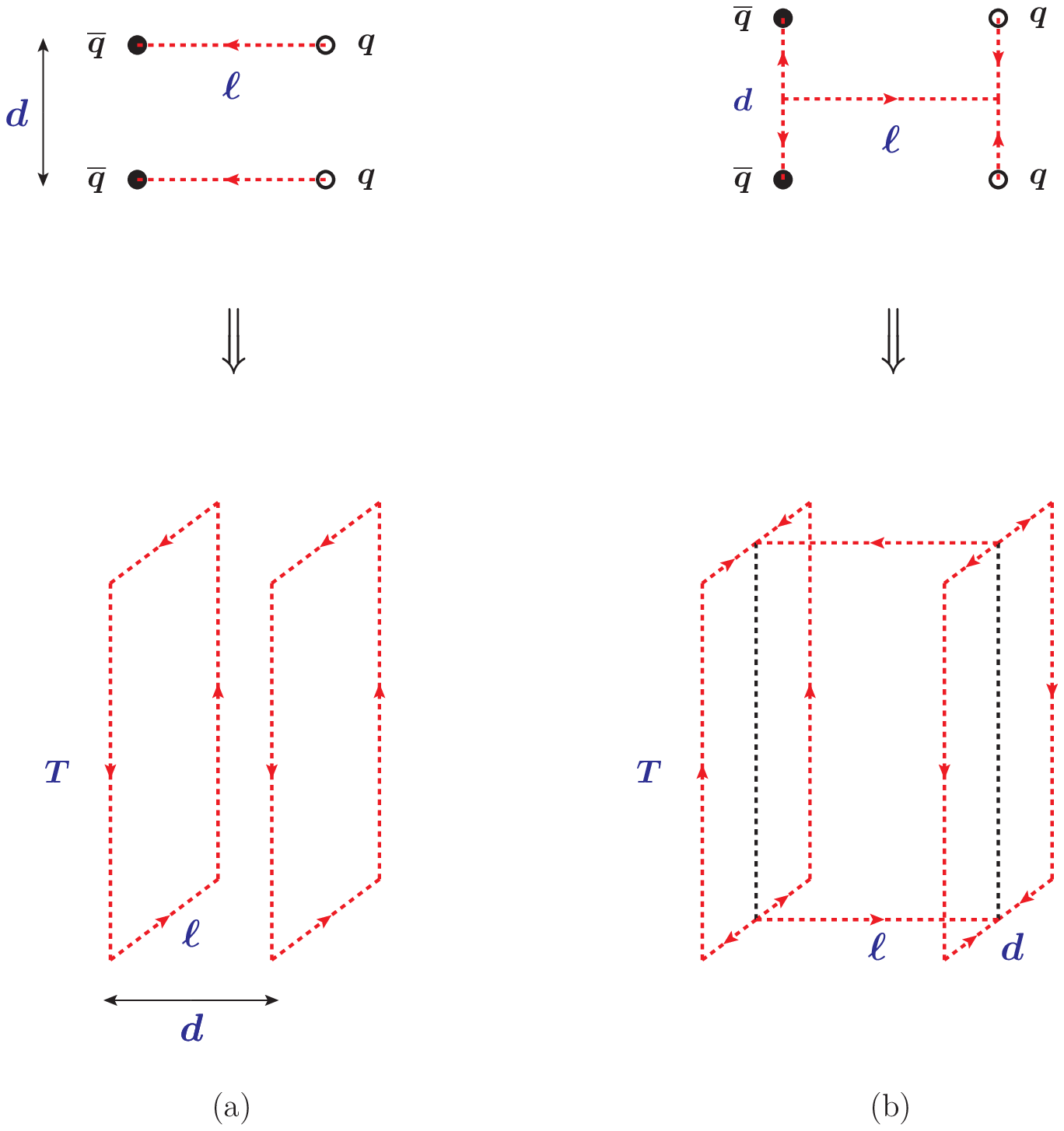,scale=0.8}
\caption{(a) Two mesonic operators and the Wilson loop contours
generated after a time evolution $T$. (b) Tetraquark operator in
the diquark-antidiquark antisymmetric representation and the
Wilson-loop contour generated after a time evolution $T$,
represented by the dashed oriented lines.}
\lb{7f6}
\ec
\efg
\par
The evolution of the two systems during a time $T$ generates
Wilson loops whose contours are represented in Fig.~\rf{7f6}.
The system of two mesons generates two independent factorized
Wilson loops, each with a rectangular contour. (In the continuum
theory, factorization of Wilson loops occurs at leading order of
large $N_c$ \cite{Witten:1979pi}; nonfactorizable contributions
are of the OZI-rule violating type, cf. Fig.~\rf{2f13}.) The total
area enclosed by the contours is $2\ell T$. The tetraquark operator
generates a single Wilson loop, whose contour is composed of a
central rectangle of width $\ell$ and of four wings, each having
a width equal to $d/2$. The total area enclosed in the contour
is $(\ell+2d)T$. One then obtains for the two-meson system and
for the tetraquark system the following interaction energies:
\be \lb{7e6}
E_{\mathrm{2mes.}}=\sigma (2\ell),\ \ \ \ \ \
E_{\mathrm{tetrq.}}=\sigma(\ell+2d).
\ee
The two energies are not equal. The system which will dominate
in the generating functional of Green's functions is the one that
has the smallest energy. A fair dominance of the tetraquark
system thus requires
\be \lb{7e7}
\sigma(\ell+2d)\ \ll\ \sigma(2\ell)\ \ \ \ \Longrightarrow\ \ \ \ \
d\ \ll\ \ell/2.
\ee
\par
The meaning of Eq.~(\rf{7e7}) is that, when the distance between the
two mesons is much smaller compared to their mean size, the
tetraquark operator may be considered as the representative of the
system. More generally, one might also have situations where
the two mesons are overlapping each other. Therefore, condition
(\rf{7e7}) depicts situations where the two quark-antiquark pairs
are located in a small volume, in which the two mesons are not
sufficiently separated from each other. When a clear separation of
the two mesons is realized, such that $d\gg \ell/2$, then the latter
system becomes the dominant one.
\par
The analysis presented above can easily be extended to the case of
the general group SU($N_c$). We consider here the two extreme cases of
Fig.~\rf{3f4}. For the left diagram, one obtains, in similar
geometric configurations as before,
$E_{\mathrm{tetrq.}}=\sigma(\ell+(N_c-1)d)$ and 
$E_{(N_c-1)\mathrm{mes.}}=\sigma (N_c-1)\ell$. At large $N_c$, the
dominance of the tetraquark system occurs when $d\ll \ell$, a condition
that is qualitatively similar to Eq.~(\rf{7e7}). For the right
diagram of Fig.~\rf{3f4}, one obtains\footnote{For the tetraquark
system, one has $(N_c-2)$ independent Wilson lines.}  
$E_{\mathrm{tetrq.}}=\sigma((N_c-2)\ell+2d)$ and 
$E_{2\mathrm{mes.}}=\sigma (2\ell)$. At large $N_c$, the two-meson
system is dominant for any situation; this is a corollary to a
similar property arising from the comparison of the quadratic Casimirs
of the diquark and quark-antiquark systems (Sec.~\rf{s54}).
\par
The previous analyses can also be applied to other
multiquark systems, like pentaquarks and hexaquarks, reaching
similar conclusions: the multiquark operator, constructed in the
diquark-type antisymmetric representation (or, equivalently,
in the string-junction-type or $Y$-shaped-type representations)
is representative of the system only when all quarks and antiquarks
are positioned in a small volume of space, where the mesonic or
baryonic clusters are overlapping each other or are very close
to each other; outside such a volume, the mesonic and baryonic
clusters become more faithful representatives of the system under
study. 
\par
It is evident that when the static approximation is relaxed, the
energy spectrum of the four-quark system will be determined from
the solution of the corresponding bound-state equation with its
energy eigenvalues, which implicitly takes into account the underlying
gauge-field configurations and the probabilities of their realization
in the corresponding dynamical situation. The quantum-mechanical
outcome of the previous analysis is the sharp dominance of the
two-meson description for the tetraquark case, which is corroborated
by the analysis based on the comparison of the quadratic Casimirs of
the diquark and quark-antiquark systems (Sec.~\rf{s54}; cf.~also
Refs.~\cite{Heupel:2012ua,Eichmann:2015cra,Wallbott:2019dng,
Eichmann:2020oqt} and Sec.~\rf{s73}).
\par
The strong-coupling approximation in lattice gauge theories for
four-quark systems has been first considered by Dosch
\cite{Dosch:1982ep}.
The results presented above have been confirmed by direct numerical
calculations on the lattice for the SU(3) case
\cite{Alexandrou:2004ak,Okiharu:2004wy,
Okiharu:2004ve,Suganuma:2011ci,Cardoso:2011fq,Bicudo:2017usw}.
\par
In conclusion, the cluster separability property of multiquark
operators, obtained on formal grounds, has a weaker significance when
energy balance is considered in the static limit, where the quarks
stand at fixed spatial positions.
Although the string-junction-type representation does not
survive in all space, it may still dominate in small volumes, from
which it may influence, by continuity on the frontier of the volume,
the properties of the system in the external volume. Nevertheless,
for the general case of moving quarks, one should expect a stronger
dominance of the mesonic or baryonic clusters. We shall study, in the
next subsection, in more detail, the contributions of each type of
description to the interaction energy of the system.
\par

\subsection{Geometric partitioning} \lb{s73}

A general feature of the static interaction energies is that they
continue representing the dragging guide of the system under
consideration even when the constituents are in motion, after, of
course, taking into account the kinematic modifications. We shall
now consider the case of moving quarks and antiquarks in the
nonrelativistic approximation, corresponding, in practice, to heavy
quarks and antiquarks. This generalization is sufficient to deduce
the essential qualitative aspects of the problem.
\par
We denote henceforth the interaction energies by $V$. The system
that is considered is the tetraquark system in its $Y$-shaped
representation (Fig.~\rf{7f7}a) and in its mesonic-cluster-type
representations (Figs.~\rf{7f7}b and c), as deduced from the cluster
reducibility relation of Fig.~\rf{7f1}. The two quarks are designated
by 1 and 3, the antiquarks by $\overline 2$ and $\overline 4$, and the
two junction points of the $Y$-shaped representation by $k$ and $\ell$.
The quarks and antiquarks being in motion, the lengths of the various
segments of Fig.~\rf{7f7} are now variables of the problem. On the
other hand, the positions of the junction points $k$ and $\ell$ are
not predetermined; they are obtained after a minimization of the
interaction energy $V$ of the $Y$-shaped representation with respect
to these points. The latter are called ``Steiner points'' in the
literature. In general, for a configuration of the type of Fig.
\rf{7f7}a, the point $k$ corresponds to the position from which the
pairs of points $(1,3)$, $(3,\ell)$ and $(\ell,1)$ are seen under
$120^{\circ}$ and similarly for the point $\ell$. In principle, the
minimization program should be applied continuously for every
configuration of the quark and antiquark positions; this, however,
is a lengthy time consuming task and generally one is satisfied with
simple geometric configurations, which are proved as introducing
only tiny quantitative errors.
\bfg
\bc
\epsfig{file=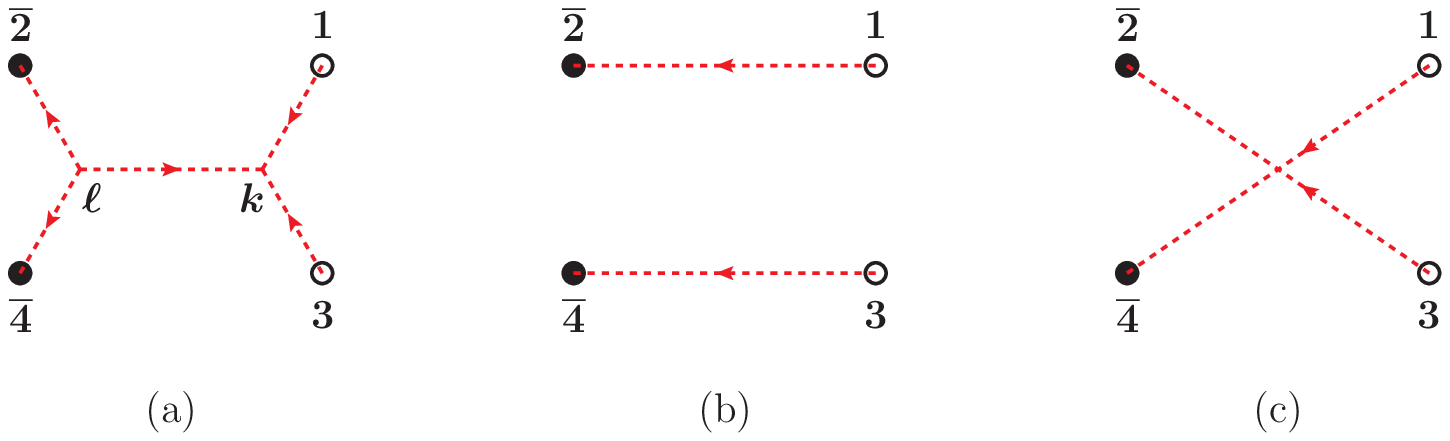,scale=0.8}
\caption{The tetraquark system in (a) the $Y$-shaped representation,
(b) and (c) two-mesonic cluster representations.}
\lb{7f7}
\ec
\efg
\par
The interaction or potential energy of the two-mesonic clusters is
composed of the contributions of Figs.~\rf{7f7}b and c. According to
Eq.~(\rf{7e5}), they are, respectively,
\be \lb{7e8}
V_{\overline 21,\overline 43}^{}=
\sigma(r_{\overline 21}^{}+r_{\overline 43}^{}),\ \ \ \ \
V_{\overline 41,\overline 23}^{}=
\sigma(r_{\overline 41}^{}+r_{\overline 23}^{}),
\ee
where we have designated by $r_{ij}^{}$ the three-dimensional distance
between positions $i$ and $j$.
\par
According to the positions of the quarks and antiquarks, the
interaction energy that prevails is the one that is minimal:
\be \lb{7e9}
V_{\mathrm{2mes.}}^{}=\mathrm{min}(V_{\overline 21,\overline 43}^{},
V_{\overline 41,\overline 23}^{}),
\ee
which can also be rewritten as
\bea \lb{7e10}
V_{\mathrm{2mes.}}^{}&=&\sigma(r_{\overline 21}^{}
+r_{\overline 43}^{})\,\theta((r_{\overline 41}^{}
+r_{\overline 23}^{})-(r_{\overline 21}^{}
+r_{\overline 43}^{}))\nonumber \\
& &\ \ \ \ \ \ \ \ +\sigma(r_{\overline 41}^{}+r_{\overline 23}^{})\,
\theta((r_{\overline 21}^{}+r_{\overline 43}^{})-
(r_{\overline 41}^{}+r_{\overline 23}^{})).
\eea
\par
On the other hand, the $Y$-shaped potential takes the form
\be \lb{7e11}
V_Y^{}=\sigma(r_{k1}^{}+r_{k3}^{}+r_{\overline 2\ell}^{}
+r_{\overline 4\ell}^{}+r_{k\ell}^{}).
\ee
\par
The final potential is then the minimum of $V_{\mathrm{2mes.}}^{}$
and $V_Y^{}$:
\be \lb{7e12}
V_{\mathrm{tetrq.}}=\mathrm{min}(V_Y^{},V_{\mathrm{2mes.}}^{}),
\ee
which is explicitly dependent on the positions of the quarks and
the antiquarks. The $Y$-shaped potential will dominate in small
volumes, where the quarks and the antiquarks are close to each other,
while the two-meson potential will dominate when the two mesonic
clusters are well separated.
\par
The two-meson potential (\rf{7e10}), which is composed of the
contributions of two different clusters, exclusive to each other,
is based on the quark rearrangement mechanism when two quarks or two
antiquarks come close to each other. This potential is known in
the literature under the name of ``flip-flop''
\cite{Miyazawa:1979vx,Lenz:1985jk,Oka:1984yx,Oka:1985vg,Carlson:1991zt,
Martens:2006ac,Vijande:2007ix,Richard:2009rp,Ay:2009zp,Vijande:2011im,
Bicudo:2010mv,Bicudo:2015bra,Bicudo:2015kna}. One of its features is
that it is free of long-range van der Waals-type forces, which
unavoidably occur in additive quark models with confining potentials
\cite{Fishbane:1977ay,Appelquist:1978rt,Willey:1978fm,Matsuyama:1978hf,
Gavela:1979zu}. Actually, Ref.~\cite{Lenz:1985jk} adopted it just on
the basis of the latter property. It also introduced the concept of
``configuration-space partitioning'', due to the appearance of
different potentials according to the positions of the quarks and the
antiquarks. However, one should be aware that
potential (\rf{7e10}) is not a mere model proposition, but is the
outcome of QCD lattice calculations in the strong-coupling limit, and
is verified in full numerical calculations
\cite{Dosch:1982ep,Alexandrou:2004ak,Okiharu:2004wy,
Okiharu:2004ve,Suganuma:2011ci,Bicudo:2017usw}. In the continuum theory,
it has support from Wilson-loop calculations in the large-$N_c^{}$
limit \cite{Makeenko:1980wr,Jugeau:2003df}.
\par
In calculations with potential (\rf{7e12}), one is interested in
the possible existence of tetraquark bound states, which would be
stable under strong interactions and would thus provide a clear
experimental signal. This would happen if the bound state is located
below the two-meson thresholds. Narrow resonance-type states, which
would be located above two-meson thresholds, might also give signals
about the existence of quasi-stable tetraquark states. Detailed
calculations have been done in 
Refs.~\cite{Vijande:2007ix,Richard:2009rp,Ay:2009zp,Vijande:2011im,
Bicudo:2010mv,Bicudo:2015bra,Bicudo:2015kna}. In particular,
Ref.~\cite{Vijande:2007ix} provides instructive details about the
contributions of the various forces. Ignoring spin degrees of freedom,
systems made of quarks with two different flavors, of the types
$QQ\bar q\bar q$ and $Qq\overline Q\overline q$,
have been considered. It turns out that, for the existence of a bound
state, the $Y$-shaped potential plays a minor role; the main
role is played by the flip-flop potential. Thus the system
$QQ\bar q\bar q$ has always a bound state for any value of
the ratio of the quark masses, with a binding energy rather small
compared to the strong-interaction energy scale. The system
$Qq\overline Q\overline q$ has bound states only for comparable masses
of the two quarks.
\par
It may seem puzzling how the flip-flop potential, where the two
clusters do not directly interact, can produce a bound state.
Actually, the interaction is hidden in the quark rearrangement
mechanism, since the flip-flop potential involves transitions
between the quark configurations $(\bar 21)(\bar 43)$ and
$(\bar 41)(\bar 23)$. The dynamical mechanism that realizes
such transitions at the level of scattering amplitudes and Feynman
diagrams is based on the quark rearrangement channels $R1$ and
$R2$, the role of which, as a possible kernel in iteration
series of bound-state emergence, was stressed in Sec.~\rf{s54}.
\par
The existence of a stable tetraquark bound state with the
structure $QQ\bar q\bar q$ had also been predicted
by Manohar and Wise \cite{Manohar:1992nd} on the basis of the
heavy-quark limit. In this limit, the interaction between the
two heavy quarks is well described by the short-range component
of the confining potential and, because of its attractive nature,
it produces a deeply bound diquark state, which behaves like an
almost pointlike color-antitriplet heavy antiquark, which then forms
with the two light antiquarks an antibaryon-like bound state.
This mechanism has also been advocated in recent studies, together
with the heavy-quark symmetry, to predict, on quantitative grounds,
the tetraquark bound-state masses
\cite{Karliner:2017qjm,Eichten:2017ffp,Quigg:2018eza}.
\par
Although the predictions about the existence of tetraquark bound
states of the type $QQ\bar q\bar q$ seem to be similar in
the two approaches based on the geometric partitioning, on one hand,
and on the heavy-quark symmetry, on the other, the lines of approach
do not seem to have common features. In the heavy-quark symmetry
approach, the heavy diquark system is reduced to a pointlike antiquark,
while in the geometric partitioning approach it is just the contrary
that is used, that is, the cluster reducibility of the system into
mesonic clusters and their mutual interaction through the quark
rearrangement mechanism. 
\par
Coming back to the geometric partitioning idea, it seems to provide
a refined analysis of the conditions under which the reducibility of
the multiquark operators and states occurs. The string-junction or
$Y$-shaped junction-type description of multiquark states seems to
survive only in small volumes of space, leaving the rest of space
to the description based on the mesonic or baryonic clusters, which
continue interacting by means of the quark exchange mechanism,
producing, in turn, weakly bound multiquark states.
\par
Nevertheless, the qualitative, as well as quantitative, conclusions
reached up to now by means of the latter descriptions cannot be
considered as definitive. The main reason of this is related to the
fact that a complete description of a multiquark state necessitates,
even in the nonrelativistic limit, the use of a multichannel
interacting system, relating the independent basis states, the
number of which is not limited to one. Taking the example of the
tetraquark, we have seen, in Sec.~\rf{s31}, that there are two
independent basis states or sectors, which could be taken to be the
sectors $(\overline 2 1)(\overline 4 3)$ and
$(\overline 4 1)(\overline 2 3)$
in their color-singlet-singlet representation, respectively, or the
sector $(\overline 2 1)(\overline 4 3)$ in its singlet-singlet
and octet-octet representations, or the diquark-antidiquark
sectors in their antisymmetric and symmetric representations.
Usually, the diquark-antidiquark sector in its symmetric
representation, as well as the octet-octet sector, are
discarded on the basis that their internal interaction is repulsive
and could not lead to the formation of diquark or antidiquark
or quark-antiquark intermediate bound states. However, one should
also take into account the fact that the mutual interaction between 
these clusters to form a color-singlet state is still attractive
with a strength at least twice greater than the conventional
quark-antiquark interaction forming a color singlet. The existence
of such forces might still substantially modify the predictions
obtained up to now. Therefore, complementary studies are still
needed in this approach to reach a definitive conclusion.
\par
As a last remark, geometric partitioning, which has been formulated
in a nonrelativistic framework, cannot be considered, in general, as
an instantaneous phenomenon. It is the result of a transition process
from an energetically favorable configuration to another one. This
transition involves the quark rearrangement or interchange mechanism.
Therefore, it might be that geometric partitioning is actually a
simplified description of the more complicated quark rearrangement
mechanism, which involves, in its generality, many Feynman diagrams.
\par

\section{Summary and concluding remarks} \lb{s8}

The extension of the color gauge group SU(3) to SU($N_c^{}$), with
large $N_c^{}$, as had been proposed by 't Hooft, has been revealed
to form an efficient tool for the investigation of the nonperturbative
regime of QCD. Without solving the theory, it has clarified many
theoretical questions that had been raised with the emergence of QCD,
some of which having already appeared with the early days of the quark
model.
\par
It is in the large-$N_c^{}$ limit that the notion of confinement takes
its idealized formulation. In this limit, quark-pair creation being
suppressed, mesons appear as made of a pair of quark and antiquark,
therefore providing to the notion of valence quarks a precise meaning.
The $1/N_c$ expansion method, starting from leading terms, provides
a systematic tool for a qualitative understanding of the order of
magnitude of physical processes and observables. However, in this
limit, baryons undergo a huge transition, since the number of their
constituents increases with $N_c$ and they tend to have a solitonic
structure, necessitating a treatment different from that of mesons.
\par
It is then natural to apply the large-$N_c$ analysis to the case of
exotic hadrons, which, in the language of valence quarks, are states
containing more quarks and antiquarks than the ordinary hadrons.
Many newly experimentally discovered or observed particles fall into
this category, since their quantum numbers or decay modes do not
fit into the scheme of the ordinary quark model.
\par
The main theoretical question that arises at this level is whether
such states are color irreducible, like ordinary hadrons. The latter,
at the valence-quark level, cannot be decomposed into simpler
color-singlet states. The answer, for the exotic states, is negative.
They are decomposable into combinations of products of ordinary
hadrons. This property is true not only for local interpolating
currents, but also for multilocal operators involving gauge links.
This means that exotic states are not natural extensions of ordinary
hadrons and could not be solutions resulting from fully confining 
forces, with a spectroscopy made of towers of bound states with
increasing masses. Such a situation might be reached only with the
existence of hidden fine-tuning mechanisms that could favor the
confining forces to take place, without being destabilized by
cluster decomposition.
\par
In passing to the gauge group SU($N_c)$, new technical complications
arise. It turns out that exotic states can be probed or described
by several inequivalent representations, each containing different
numbers of valence quarks. Thus, tetraquarks, which in the case of
SU(3) are described as made of two pairs of valence quarks and
antiquarks, have now $(N_c^{}-2)$ different representations, a generic
representation having $J$ pairs of quarks and antiquarks, $J$ taking
values from 2 to $(N_c^{}-1)$. Similar generalizations occur also for
pentaquarks and hexaquarks. Concerning tetraquarks, the interpolating
operators that contain two pairs of quarks and antiquarks are the
most convenient ones, since they remain closest to the SU(3) case
and allow easy recognition of experimental outcomes. It is the latter
representation which has been adopted throughout the present work.
\par
Since tetraquarks are expected to decay or to couple to two-meson
states, it is natural to analyze their properties by means of
meson-meson scattering amplitudes. The case of fully exotic tetraquarks,
containing four different quark flavors, is the simplest one, for there
mixing channels with ordinary mesons are avoided. One is then in the
presence of a coupled-channel problem, where the contributions of all
channels shoud be taken into account. It turns out that, at large $N_c$,
the diagonal channels (or direct channels) are dominant, while
the off-diagonal channels (or quark-rearrangement or
quark-interchanging channels) are subdominant. This has as a consequence
that if there are tetraquark states, they are two in number, each one
corresponding to the diagonal-channel solution, and each of them having
a dominant coupling with the two mesons of that channel, of the order
of $1/N_c$. In case of a possibility of decay, the corresponding
decay width would be of the order of $1/N_c^{2}$. This solution
does not favor the diquark scheme, which, because of the confinement
constraint, is built on a single antisymmetric representation and thus
predicts a single tetraquark, having equal couplings with the two
mesons of each channel (spin quantum numbers having been ignored, as not
being essential for these analyses).
\par
It should be stressed that, contrary to the case of ordinary hadrons,
the large-$N_c$ analysis does not imply the existence of tetraquarks.
These are in competition with contributions of two-meson intermediate
states, which consistently can saturate the various equations.
Tetraquarks, if they exist, are additonal contributions to the
intermediate states.
Therefore, predictions obtained about tetraquarks at large $N_c$ should
be considered as upper bounds for the related quantities.
\par
The large-$N_c$ analysis also allows the study of the formation
mechanism of tetraquark states. At leading order, it is the two-meson
clusters that provide the main contributions and these are expected not
to mutually interact by means of confining forces.
In that scheme, the main formation mechanism is generated by the internal
contributions of quark-rearrangement (quark-interchange) processes. These
sum up and might produce tetraquark poles in the meson-meson
scattering amplitudes. There are two possible interpretations of the
global outcome of this mechanism: (i) The whole summation is reduced
to an effective meson-meson interaction, producing molecular-type
tetraquarks. (ii) The interaction between the meson clusters, even if
not confining, is of the residual type of confining interactions, not
reducible to meson exchanges or contact terms.
Only a more detailed investigation of the corresponding dynamics might
provide a clarification of that issue.
\par
For cryptoexotic states, with three or two quark flavors, mixing
diagrams, involving ordinary mesons as intermediate states, complicate
the extraction of the tetraquark properties. Nevertheless, the
existence of four-channel processes and the fact that the
quark-rearrangement mechanism is still at work, imply, in general,
the possibility of the existence of two different tetraquarks,
eventually having priviledged couplings with the mesons of the
diagonal-type channels.
\par
Finally, the idea of geometric partitioning, which takes into account
the energy balance of meson-cluster- and string-junction-type
configurations, has provided further clarification about the dominant
configurations which might produce tetraquark states. Except in small
volumes, where the four quarks are located, most of space is dominated
by two-meson-cluster configurations. The transition from one of
these configurations to the other implies again the quark-rearrangement
mechanism.
\par
The mechanism of formation of tetraquarks is not yet fully understood,
due to the complexity of experimental data and the lack of explicit
theoretical solutions. However, the large-$N_c$
approach provides a complementary view to the problem by establishing
a hierarchy among the various types of contributions. The general
outcome that emerges from that approach is that the tetraquark formation
is mainly dominated by the quark-rearrangement mechanism, which operates
at the internal level of the processes.
\par
The problem of exotic hadrons still remains a challenge for
all theoretical approaches.
\par



\vspace{0.5 cm}
\noindent
\textbf{Acknowledgements}
\par
\vspace{0.25 cm} 
D.~M. acknowledges support from the Austrian Science Fund (FWF),
Grant No.~P29028.
H.~S. acknowledges support from the EU research and innovation
programme Horizon 2020, under Grant agreement No.~824093.
D.~M. and H.~S. are grateful for support under
joint CNRS/RFBR Grant No.~PRC Russia/19-52-15022.
The figures (except Fig.~\rf{6f2}) were drawn with the aid of the
package Axodraw2 \cite{Collins:2016aya}.
\par


\bibliographystyle{elsarticle-num} 
\bibliography{tqlnc.bib}






\end{document}